\documentclass[a4paper,11pt]{article}
\pdfoutput=1

\usepackage{jheppub}

\usepackage{epsfig,multicol}
\usepackage{graphicx}
\usepackage{amsmath,amssymb}
\usepackage{mathrsfs}
\usepackage{multirow}

\usepackage{float}

\newcommand\fverb{\setbox\fverbbox=\hbox\bgroup\verb}
\newcommand\fverbdo{\egroup\medskip\noindent%
			\fbox{\unhbox\fverbbox}\ }
\newcommand\fverbit{\egroup\item[\fbox{\unhbox\fverbbox}]}
\newbox\fverbbox

\newcommand{\pslash}{p\kern-1ex /}
\newcommand{\qslash}{q\kern-1ex /}
\newcommand{\lslash}{l\kern-1ex /}
\newcommand{\sslash}{s\kern-1ex /}
\newcommand{\Dslash}{\mathcal{D}\kern-1.5ex /}

\newcommand{\beqa}{\begin{eqnarray}}
\newcommand{\eeqa}{\end{eqnarray}}
\newcommand{\rmO}{\mathrm{O}}

\newcommand{\ba}{\begin{eqnarray}}
\newcommand{\ea}{\end{eqnarray}}
\newcommand{\be}{\begin{equation}}

\newcommand{\mO}[1]{\text{O(#1)}}

\newcommand{\hatk}{{\hat{k}}}

\title{Wiener-Hopf solution of the free energy TBA problem and
instanton sectors in the \mO{3} sigma model
}

\author[a]{Zolt\'an Bajnok,}
\author[a]{J\'anos Balog,}
\author[a,b]{and Istv\'an Vona}

\affiliation[a]{Holographic QFT Group, Institute for Particle and Nuclear Physics, HUN-REN Wigner Research Centre for Physics,\\ H-1525 Budapest 114, P.O.B. 49, Hungary}
\affiliation[b]{MTA-ELTE "Momentum" Integrable Quantum Dynamics Research Group, Department of Theoretical Physics, E\"otv\"os Lor\'and University,\\ P\'azm\'any P\'eter s\'et\'any 1/A, 1117 Budapest, Hungary}

\emailAdd{bajnok.zoltan@wigner.hu}
\emailAdd{balog.janos@wigner.hu}
\emailAdd{vona.istvan@wigner.hu}

\abstract{
Perturbation theory in asymptotically free quantum field theories
is asymptotic. The factorially growing perturbative coefficients carry
information about non-perturbative corrections, which can be related
to renormalons and instantons. Using the Wiener-Hopf technique we
determine the full analytic solution for the free energy density in
the two dimensional \mO{$N$} sigma models. For $N>3$ there are no instantons,
and we found that the perturbative series carries all the information
about the non-perturbative corrections. However, in the \mO{3} case, we
identify several non-perturbative sectors that are not related to
the asymptotics of the perturbative series. The number of sectors
depends on the observables: for the ground-state energy density we
identify three sectors, which we attribute to instantons. For the
free energy density in the running perturbative coupling we found
infinitely many sectors. 

}

\begin{document}


\newcommand{\con}{\,\star\hspace{-3.7mm}\bigcirc\,}
\newcommand{\convu}{\,\star\hspace{-3.1mm}\bigcirc\,}
\newcommand{\Eps}{\Epsilon}
\newcommand{\gM}{\mathcal{M}}
\newcommand{\dD}{\mathcal{D}}
\newcommand{\gG}{\mathcal{G}}
\newcommand{\pa}{\partial}
\newcommand{\eps}{\epsilon}
\newcommand{\La}{\Lambda}
\newcommand{\De}{\Delta}
\newcommand{\nonu}{\nonumber}
\newcommand{\beq}{\begin{eqnarray}}
\newcommand{\eeq}{\end{eqnarray}}
\newcommand{\ka}{\kappa}
\newcommand{\ee}{\end{equation}}
\newcommand{\an}{\ensuremath{\alpha_0}}
\newcommand{\bn}{\ensuremath{\beta_0}}
\newcommand{\dn}{\ensuremath{\delta_0}}
\newcommand{\al}{\alpha}
\newcommand{\bm}{\begin{multline}}
\newcommand{\fm}{\end{multline}}
\newcommand{\de}{\delta}
\newcommand{\dpd}{\int {\rm d}^d p}
\newcommand{\dqd}{\int {\rm d}^d q}
\newcommand{\dxd}{\int {\rm d}^d x}
\newcommand{\dyd}{\int {\rm d}^d y}
\newcommand{\dud}{\int {\rm d}^d u}
\newcommand{\dzd}{\int {\rm d}^d z}
\newcommand{\dpp}{\int \frac{{\rm d}^d p}{p^2}}
\newcommand{\dqq}{\int \frac{{\rm d}^d q}{q^2}}




\maketitle

\section{Introduction}
\label{sectIntro}

Perturbation theory works very well for weakly interacting quantum
field theories including the electro-weak theory, which describes
the weak and the electromagnetic fundamental interactions \cite{Aoyama:2012wk}.
In the case of the strong interaction, however, quantum chromodynamics
(QCD) suffers from the asymptotic nature of the perturbative expansion
\cite{Dyson:1952tj,Hurst:1952zh,Lipatov:1976ny}. This manifests itself
in the factorially growing coefficients, which renders the perturbative
series naively meaningless. The path integral formulation naturally
leads to the use of Borel resummation \cite{Spada:2018iyp},
but singularities on the positive
real line imply non-perturbative corrections and prevent us to obtain
a unique result. These singularities in many cases are encoded in
the asymptotic behaviour of the perturbative series and are related
to renormalons or instantons \cite{tHooft:1977xjm,Beneke:1998ui}.
Unfortunately QCD is a very complicated theory and we do not have access
to many perturbative coefficients, nor the analytical structure
of the observables on the Borel plane, although partial results have
been already obtained from a first few coefficients \cite{Caprini:2020lff}.
It is thus advantageous to analyse these questions in simplified models,
which shares many features with QCD, but otherwise tractable. 

The two dimensional non-linear \mO{$N$} sigma models are of this type.
On the one hand, they are asymptotically free in perturbation theory
and the mass of the particles is generated dynamically. On the other
hand, they are integrable and the ground-state energy density can
be exactly calculated, via a linear TBA integral equation, which is
the thermodynamic limit of the Bethe ansatz equation. The non-linear
\mO{$N$} sigma models were providing useful toy models in which
various new phenomena were exactly calculated for the first time.
The integrable scattering matrix of the particles were calculated
in \cite{Zamolodchikov:1977nu}, the perturbative $\Lambda$ scale
was related to the massgap in \cite{Hasenfratz:1990zz,Hasenfratz:1990ab},
the \mO{6} sigma model contributed to our understanding to the AdS/CFT
duality \cite{Bajnok:2008it}, and recently resurgence theory, relating
perturbative and non-perturbative corrections, were tested in \cite{Marino:2019eym,Abbott:2020mba,Abbott:2020qnl,Marino:2021dzn,Bajnok:2021dri,Bajnok:2021zjm,Bajnok:2022rtu}.
The \mO{$N$} models and their integrable cousins have proven to be very
useful, exactly soluble examples, where resurgence theory \cite{Dingle1973:asy,Ecalle1981:res}
could be developed and tested. For modern reviews about resurgence,
see \cite{Marino:2012zq,Dorigoni:2014hea,Aniceto:2018bis}. 

Indeed, there has been significant activity and relevant
progress recently in the perturbative and non-perturbative investigations of
TBA type linear integral equations. The motivation and results came
from the statistical \cite{Marino:2019fuy,Marino:2019wra,Marino:2020dgc,Marino:2020ggm,Ristivojevic:2022vky,Ristivojevic:2022gxr,Liu:2023tft,Reichert:2020ymc} as well as from the particle
physics side \cite{Marino:2021dzn,Bajnok:2021dri,Bajnok:2021zjm,Marino:2022ykm,Bajnok:2022rtu,Reis:2022tni,Marino:2023epd,Aniceto:2015rua,
Dorigoni:2015dha,Schepers:2023dqk}.
The first step in these investigations originated from the \mO{$N$}
models, where Volin developed an efficient algebraic method to calculate
a high number of perturbative coefficients \cite{Volin:2009wr,Volin:2010cq}.
Using resurgence theory the leading exponentially suppressed corrections
were extracted from the asymptotics of the perturbative coefficients
\cite{Abbott:2020mba,Abbott:2020qnl,Bajnok:2021dri}. These are related
to instantons or renormalons, which were further confirmed in particular
cases by large $N$ calculations \cite{Marino:2021six,DiPietro:2021yxb,Marino:2024uco}
and in the \mO{3} model by introducing a $\theta$-term \cite{Marino:2022ykm}.
A systematic expansion of the Wiener-Hopf integral equation led to
the structure of all the non-perturbative corrections with explicit
results for their first few coefficients \cite{Marino:2021dzn}. The
result is a trans-series, i.e. a series in the perturbative coupling
(and in its logarithm) as well as in the non-perturbative, exponentially
suppressed contributions. Each non-perturbative correction is multiplied
by an asymptotic perturbative series. The convergence properties
of the truncated trans-series were analysed in \cite{Marino:2023epd}.
The Wiener-Hopf analysis was later extended to higher orders and the
logarithmic coupling dependence was removed by introducing the running
coupling \cite{Bajnok:2022rtu}. Finally, in \cite{Bajnok:2022xgx}
we presented a way to determine the full trans-series, i.e. all the
non-perturbative terms together with their perturbative expansions.
In the present paper we elaborate and extend the method for the \mO{$N$}
models and investigate the resurgence properties of various observables.
The main question we are asking is to what extent the perturbative
series determines the non-perturbative corrections and how many independent
non-perturbative sectors we have.

\section{Summary of the Wiener-Hopf method and the main results}
\label{sect0}

As explained in the Introduction, 2-dimensional integrable models are useful
toy models in which non-perturbative field theory features,
otherwise unavailable, can be studied in detail.
In this paper our aim is to study problems related to
resurgence theory. These include behaviour of very high
orders of perturbation theory, trans-series, Borel
resummation, cancellation of ambiguities, etc.

The particular physical quantity we will examine is the
free energy in the presence of an external field in the
2-dimesional \mO{$N$} nonlinear sigma model.
The study of this quantity provides us an almost 
unique situation: all ingredients are
explicitly calculable.
On the one hand, the free energy is exactly calculable
because it is described by a linear
TBA integral equation based on the bootstrap S-matrix.
On the other hand, it can be expanded perturbatively
to very high orders efficiently
\cite{Volin:2009wr,Volin:2010cq}.

The study of the free energy was initiated in
\cite{Hasenfratz:1990zz,Hasenfratz:1990ab}
and developed further in
\cite{Volin:2009wr,Volin:2010cq,Marino:2019eym,Marino:2021dzn,Marino:2022ykm,
Bajnok:2022rtu,Bajnok:2022xgx}.
According to our
philosophy the main result of \cite{Hasenfratz:1990zz,Hasenfratz:1990ab}
is not the free energy itself, but that by comparing the results
of the bootstrap and perturbative calculations,
it became possible to determine exactly the
ratio of the mass parameters of the two approaches,
$m/\Lambda$, where $m$ is the mass of the particles
and $\Lambda$ is the usual Lambda parameter of
asymptotically free perturbation theory.

We will use the Wiener-Hopf (WH) method to \lq\lq
solve" the TBA integral equation in a sense explained
below (see section~\ref{sect3}). The WH method replaces
the original TBA integral equation by an apparently
much more complicated WH integral equation. The latter,
however, has the advantage that it can be systematically
expanded in two small parameters: the bootstrap
running coupling $v$ and an exponentially small
non-perturbative (NP) expansion parameter $\nu$.

The next step is to find the
perturbative solution. In principle, this can be done
by iteratively solving (as in \cite{Hasenfratz:1990zz,Hasenfratz:1990ab}
to second order) the WH integral equation but it is much more
efficient to use Volin's method \cite{Volin:2009wr,Volin:2010cq},
who invented an algorithm to calculate the perturbative  part of
the energy density to arbitrary order algebraically.

Finally we write the complete solution in a form of
a trans-series: a series in powers of the NP expansion
parameter $\nu$ where the coefficient of each power of
$\nu$ is a  perturbative series in the running coupling
$v$. We derived differential equations (see section~\ref{sect1})
which allow us to find all perturbative coefficient
series if we know any one of them. Using Volin's result
all other series are calculable algebraically \cite{Bajnok:2022xgx}.

At this point we found the complete solution of our problem
in the sense that it is given as a trans-series, which can
be calculated to any order in $\nu$ and the coefficient
functions can be given as PT (perturbative or perturbation theory/theoretical) series to any desired order
in the running coupling. Throughout this paper by perturbative we mean a power series in integer powers of the (running) coupling. In this paper we do not discuss
the convergence of the trans-series \cite{Marino:2023epd}
but note that
each power of $\nu$ is multiplied by a coefficient function,
which is not a convergent series but only asymptotic in PT.

To give definite meaning to our result, we assume (as is
customary in this field) that the physical result is
obtained by Borel resummation of the coefficient
functions (see section~\ref{sect7}). We cannot prove the correctness
of this fundamental assumption but, first of all, it can be
checked numerically to very high precision (since the
original TBA equation itself can be solved very precisely
and we can compare). Secondly, it can be shown (see section~\ref{fej4})
that the ambiguities (related to the fact that our coefficient
functions are not Borel summable due to singularities on the
positive real axis of the Borel plane) are cancelled by
similar ambiguities encountered in the derivation of the
WH integral equations.
Moreover, we can study the consequences
of our fundamental assumption and we find that physical
quantities are given by the median resummation
\cite{Marino:2008ya,Aniceto:2013fka,Dorigoni:2014hea}
of the purely PT part of the given quantity. More precisely,
this is the case for all \mO{$N$} models with $N\geq4$. In these
models the purely perturbative part $f_0$ \lq\lq knows" about
all higher terms of the trans-series and the Borel resummation
of all terms of the trans-series is just the median sum of
$f_0$. In this paper, although it is not a universally accepted term, we will call this property \emph{strong resurgence}.
For \mO{3}, which differs from  all higher $N$ models
basically because it is the only model with stable instantons, the
situation is somewhat different (see section~\ref{sect8}). Here we have to
consider three basic functions $f_0,f_1,f_2$, where $f_0$ is
the purely perturbative part while $f_1$ and $f_2$ can be
interpreted as quantum fluctuations around the 1 and 2 instanton
solutions, respectively. In this case the knowledge of $f_0$
alone is not sufficient to reconstruct the full trans-series
because physical quantities are obtained as the median sum of
$f_0+\nu f_1+\nu^2 f_2$. The instanton interpretation of
$f_1$ and $f_2$
is consistent with the results of \cite{Marino:2022ykm}
where it was found
that $f_1$ changes sign (but $f_2$ does not) if one includes
the $\theta$ term in the \mO{3} action and puts $\theta=\pi$.

At the end of the paper (sections~\ref{sect9},\ref{sect10}) we introduce a
physical running coupling (which can also be defined in a field theory,
Lagrangian based study of the same model) and show that the
trans-series in the language of the physical coupling have
similar properties as above except that in the \mO{3}
case now we have $i$-instanton sectors for any $i$ (not just
$i=0,1,2$ as was the case with the bootstrap coupling $v$).

\section{Integral equations, differential relations}
\label{sect1}

In this paper we discuss the solution of the TBA integral equations for
the free energy problem in the two-dimensional \mO{$N$} sigma
models \cite{Bajnok:2022rtu}. However, our methods are also applicable for
similar asymptotically free 2-dimensional integrable models
\cite{Marino:2019eym,Marino:2021dzn,Marino:2022ykm,Bajnok:2022xgx}. We generalized the problem to the case when the Hamiltonian (without the magnetic field) is given by the conserved spin $n$ charge - see \cite{Bajnok:2022xgx}. Although we are not interested in these situations physically, later on it will be clear that this extension is helpful in solving the WH problem.

The equations are of the form  
\begin{equation}\label{TBA}
\chi_n(\theta,B)-\int_{-B}^B{\rm d}\theta^\prime K(\theta-\theta^\prime)
\chi_n(\theta^\prime,B)=\cosh n\theta\qquad\quad \vert\theta\vert\leq B.
\end{equation}
Here the integral kernel is the logarithmic derivative of the S-matrix,
\begin{equation}\label{Ktheta}
K(\theta)=\frac{1}{2\pi i}\partial_{\theta}\log S(\theta).
\end{equation}
In the \mO{$N$} model 
\begin{equation}
  S(\theta)=-\frac{\Gamma(\frac{1}{2}-\frac{i\theta}{2\pi})
    \Gamma(\Delta-\frac{i\theta}{2\pi})\Gamma(1+\frac{i\theta}{2\pi})
    \Gamma(\frac{1}{2}+\Delta+\frac{i\theta}{2\pi})}
  {\Gamma(\frac{1}{2}+\frac{i\theta}{2\pi})
    \Gamma(\Delta+\frac{i\theta}{2\pi})\Gamma(1-\frac{i\theta}{2\pi})
    \Gamma(\frac{1}{2}+\Delta-\frac{i\theta}{2\pi})}.
\end{equation}
The parameter $\Delta$ is related to $N$ as $1/\Delta=N-2$. 

Important quantities are the boundary values
\begin{equation}\label{boundaryvalues}
\rho_n(B)=\chi_n(B,B)  
\end{equation}
and the densities
\begin{equation}\label{densities}
O_{n,m}=\frac{1}{2\pi}\int_{-B}^B{\rm d}\theta\chi_n(\theta,B)\cosh m\theta,
\qquad\qquad O_{n,m}(B)=O_{m,n}(B).  
\end{equation}

From now on we suppress the $B$-dependence for all variables, and use the notation
\begin{equation}
\dot {X}=\frac{{\rm d}{X}}{{\rm d}B},
\end{equation}
for the $B$-derivative of a generic quantity ${X}$.
The variables in (\ref{boundaryvalues}-\ref{densities}) satisfy the following differential relations (see appendix
\ref{appA}).
\begin{equation}
{\dot O}_{n,m}=\frac{1}{\pi}\rho_n \rho_m,
\label{diff1}  
\end{equation}
\begin{equation}
(n^2-m^2)O_{n,m}=\frac{1}{\pi}(\rho_m\dot\rho_n-\rho_n\dot\rho_m),  
\label{diff2}
\end{equation}
\begin{equation}
{\ddot\rho}_n-n^2\rho_n=F\rho_n,
\label{diff3}  
\end{equation}
where $F$ is $n$-independent:
\begin{equation}\label{universalF}
F=A^2+\dot A,\qquad A=\frac{\dot\rho_0}{\rho_0}.  
\end{equation}
Combining the above relations we can write the alternative form
\begin{equation}
{\ddot O}_{n,m}-\frac{2\dot\rho_n}{\rho_n}\dot O_{n,m}=(m^2-n^2)O_{n,m}.  
\end{equation}
The derivation of the differential relations (\ref{diff1})-(\ref{diff3}) are
based on the results of \cite{Ristivojevic:2022vky,Ristivojevic:2022gxr}.
At many points throughout the paper we will take advantage of these basic relations.

The two physical quantities we need are the
particle density ($\rho$) and the energy density ($\varepsilon(\rho)$) given by
\begin{equation}\label{physicaldensities}
\rho=m O_{0,1},\qquad\qquad \varepsilon(\rho)=m^2 O_{1,1}.
\end{equation}
Here $m$ is the mass of the particles (there are no bound states in these
models). The magnetic field in the usual free energy problem is defined as 
\begin{equation}
	h = m \frac{\rho_1}{\rho_0} 
\end{equation}	
in terms of the boundary values for $n=1$ and $n=0$. We will discuss the expression of the free energy itself via $\rho,\epsilon(\rho)$ and $h$ at the beginning of Section \ref{sect9}.

\section{Properties of the integral kernels}
\label{sect2}

We will use the Wiener-Hopf (WH) method to solve the TBA integral equations
\cite{Hasenfratz:1990zz,Hasenfratz:1990ab,Marino:2021dzn,Bajnok:2022rtu}. This section is dedicated to collect the important 
(model-dependent) input data for this method that will govern both the NP and PT structure of the trans-series solution. 
That is, we parametrize the WH integral kernels in Subsection \ref{ssectIntegralKernels} and explore their analytic structure along the imaginary axis in Fourier-space. 
We do so, as the discontinuity through this axis (see Subsection \ref{ssectDisc}) will yield the perturbative expansion of the observables (\ref{boundaryvalues}-\ref{densities}) 
whereas the source of the NP corrections will be the poles (see Subsection \ref{ssectPoles}) sitting on it \cite{Marino:2021dzn,Bajnok:2022rtu}. 
The positions of the poles will interplay to determine the order of the exponentially small NP parameter $\nu$ at which a given correction term in the trans-series arises, 
whereas their residues will correspond to its overall strength in a similar way. By introducing the running coupling $v$ via the trick in Subsection \ref{ssectRunning} to parametrize the kernel in the WH integral equation 
instead of using the boundary parameter $B$ directly, it is possible to obtain each term of the trans-series (the PT part and the NP corrections as well) purely as power serieses in $v$.

\subsection{The integral kernels}
\label{ssectIntegralKernels}

For the WH method we need the Fourier transform of the
kernel \eqref{Ktheta} and the corresponding $G_\pm(\omega)$ decomposition: 
\begin{equation}\label{WHdecomp}
\tilde K(\omega)=\int_{-\infty}^\infty{\rm d}\theta\,
{\rm e}^{i\omega\theta}K(\theta),
\qquad \frac{1}{1-\tilde K(\omega)}=G_+(\omega)G_-(\omega),\qquad
G_-(\omega)=G_+(-\omega).
\end{equation}
We decompose the inverse of $1-\tilde K(\omega)$ as a product of two functions,
where $G_+(\omega)$ is analytic in the upper, and $G_-(\omega)$ in the lower
half-plane. In our subsequent analysis we will need $G_-(\omega)$ also in the
upper half-plane. It is actually also analytic there, except for the positive
imaginary axis, where it has a cut and infinitely many poles.

The generic form for bosonic models is
\begin{equation}
G_+(\omega)=\frac{1}{\sqrt{-i\omega}}H(-i\omega){\rm e}^{-\frac{ia}{2}\omega
\ln(-i\omega)-\frac{ib}{2}\omega}.
\end{equation}
For the \mO{$N$} models we have
\begin{equation}\label{ONGplus}
\begin{split}
{\rm O}(N):\quad H(\kappa)&=\frac{1}{\sqrt{\Delta}}\frac{\Gamma(1+\Delta\kappa)}
{\Gamma(1/2+\kappa/2)}\qquad a=1-2\Delta \qquad b=2\Delta(1-\ln\Delta)-(1+\ln2)\\
{\rm O}(4):\quad H(\kappa)&=\sqrt{2}\frac{\Gamma(1+\kappa/2)}
{\Gamma(1/2+\kappa/2)}\qquad a=b=0\\
{\rm O}(3):\quad H(\kappa)&=\frac{1}{\sqrt{\pi}}\Gamma(1+\kappa/2)
\qquad a=-1\qquad b=1+\ln2.
\end{split}
\end{equation}

\subsection{Discontinuities}
\label{ssectDisc}

For the Wiener-Hopf analysis we will need the quantities ($\epsilon>0$ being an infinitesimal shift)
\begin{equation}\label{sigmaplus}
\sigma_\pm(\kappa)=\frac{G_-(i\kappa\pm\epsilon)}{G_+(i\kappa)},\qquad\qquad
\sigma_+(\kappa)-\sigma_-(\kappa)=-2i\beta(\kappa).
\end{equation}
They are of the form
\begin{equation}
\sigma_\pm(\kappa)=\frac{H(-\kappa)}{H(\kappa)}{\rm e}^{-a\kappa\ln\kappa
  -b\kappa}
\left\{\mp i\cos\frac{a\pi\kappa}{2}-\sin\frac{a\pi\kappa}{2}\right\}.
\label{26}
\end{equation}
and
\begin{equation}
\beta(\kappa)=\frac{H(-\kappa)}{H(\kappa)}{\rm e}^{-a\kappa\ln\kappa
-b\kappa} \cos\frac{a\pi\kappa}{2}
\end{equation}

\subsection{Poles and residues}
\label{ssectPoles}

In the WH analysis we will also need the positions and residues of the
poles of $\sigma_\pm(\kappa)$ and $\beta(\kappa)$. We have
\begin{equation}
\sigma_\pm(\kappa)\approx \frac{\mp iS_j}{\kappa-\kappa_j},
\qquad \beta(\kappa)\approx \frac{S_j}{\kappa-\kappa_j}\quad (\kappa\sim
\kappa_j).
\end{equation}
For our models the Stokes constants\footnote{These will appear as parameters in the multi-parameter trans-series determining the strength of a specific NP correction by which it switches on while going through the Stokes-line in the Stokes-phenomenon \cite{Bajnok:2022xgx}.} $S_j$ are real and
the set of poles is $\{\kappa_\ell\}$,
$\ell=1,2,\dots$, where
\begin{equation}
\kappa_\ell=2\xi_o\ell,\qquad\qquad \xi_o=\left\{\begin{split}
\frac{N-2}{2}\quad &N\ {\rm even}\\
N-2\quad &N\ {\rm odd}\end{split}\right.
\end{equation}
($\xi_o=1$ for $N=3$, $4$.)
We note that for all $N$ $\beta(1)=0$, and for all $N$,
except for $N=3$, $\sigma_\pm(1)=0$. For the
\mO{3} model $\sigma_\pm(1)=1/{\rm e}$.

\subsection{Running coupling}
\label{ssectRunning}

Here we will follow closely Subsection 2.3 of \cite{Bajnok:2022rtu}.
The running coupling\footnote{A similar running coupling was introduced in
\cite{Forgacs:1991rs} in the context of the fermionic GN model.} $v$ is defined by
\begin{equation}
2B=\frac{1}{v}-a\ln v-b-\tilde b,
\end{equation}
where the (arbitrary) constant $\tilde b$ is part of the definition of $v$. For convenience we will also use the notation $L=-b-\tilde b$.

In \cite{Bajnok:2022xgx} we made a choice for $L$ that leads to a simple (model dependent, but generic) parametrization of the perturbative part of the observables (\ref{boundaryvalues}-\ref{densities}), and for the $O(N)$ model we took
\begin{equation}\label{ONLconst}
L = 1-2\Delta+2\Delta\ln\Delta+(1-4\Delta)\ln2 , \qquad \tilde b = 4 \Delta \ln 2,	
\end{equation}
which we will use throughout the paper.
We note that in the case of the \mO{3} model the coupling $v$ used here
is different from the $\beta$ coupling of \cite{Bajnok:2022rtu}.
They are related by
\begin{equation}\label{eq:runningswitch}
2B=\frac{1}{v}+\ln v-1-\ln8=\frac{1}{\beta}+\ln\beta+1-\ln8.
\end{equation}

The advantage of using the running coupling is the formula for $\kappa=vx$
\begin{equation} \label{eq:calAexp}
{\rm e}^{-2B\kappa}\beta(\kappa)={\rm e}^{-x}{\cal A}(x,v),
\end{equation} 
where ${\cal A}(x,v)$ is perturbative (a power series in $v$), and has the explicit form 
\begin{equation}
{\cal A}(x,v)={\rm e}^{vx(\tilde b-a\ln x)}\frac{H(-vx)}{H(vx)}
\cos\frac{axv\pi}{2}.
\end{equation}

It has the expansion
\begin{equation}\label{Axvpert}
{\cal A}(x,v)=\sum_{k=0}^\infty L_k(\ln x)(vx)^k,
\end{equation}
where
\begin{equation}\label{Lkpoly}
L_0(z)=1,\qquad L_k(z): \quad{\rm polynomial\ of\ degree}\ k.
\end{equation}
By transforming the kernel of the WH equations according to \eqref{eq:calAexp} and working with $v$ instead of $B$, one can get rid of the $\log B$ terms appearing when naively expanding in $1/B$.

\section{Wiener-Hopf solutions, trans-series representation of densities}
\label{sect3}

The aim of this section is to summarize the trans-series solution to the WH problem and express the densities with the parameters and functions introduced in Section \ref{sect2}. That is, using only $\kappa_j, S_j, \sigma_+(\kappa)$ for constructing the NP structure, and the perturbative parts of the observables (\ref{boundaryvalues}-\ref{densities}) - asymptotic serieses of the running coupling $v$ - as building blocks. The latter are defined almost completely\footnote{Up to an integration constant in the perturbative part of $O_{0,0}$.} via ${\cal A}(x,v)$ and $H(0)$.

	To this end, at first we define some normalized version of the observables in Subsection \ref{reduced} such that their perturbative parts start with a constant or at most with some (negative) power of $v$. In Subsection \ref{ssectWHn} we present the exact WH integral equations for the $n>0$ case, in the form one arrives at when picking up pole contributions of the kernel after a specific contour deformation \cite{Bajnok:2022rtu, Bajnok:2022xgx}. In solving them, at first we separate their perturbative part ($\text{O}(\nu^0)$ where $\nu$ will be the NP parameter) and construct suitable building blocks that in principle solve them (see Subsubsection \ref{sssectPTn}). As stated in Section \ref{sect0}, Volin's method may be used to effectively calculate these PT serieses instead of using direct methods to solve the truncated integral equations defining them. Finally the PT quantities are used to solve the exact equations via a trans-series ansatz in Subsubsection \ref{sssectTSn}, that essentially boils down to generating the trans-series structure via the Neumann-series of a certain infinite matrix. In Subsection \ref{ssectWH0} we apply the same steps for the more complicated case of solving the WH problem for the $n=0$ TBA. This results in the trans-series representations of the $O_{0,n} = O_{n,0}, O_{0,0}$ and $\rho_0$ quantities, that was not yet worked out in full detail in any of the previous publications \cite{Bajnok:2022rtu, Bajnok:2022xgx}. In the meantime, we comment on the compatibility of the trans-series ans\"atze and the differential relations, and work out some of the apparatus to perform this novel consistency check for our WH result.

\subsection{Reduced densities}
\label{reduced}

Let us define $W_{n,m}$ and $w_n$ by
\begin{equation}
\begin{split}
O_{n,m}&=\frac{1}{4\pi}G_+(in)G_+(im){\rm e}^{(n+m)B} W_{n,m}\quad (n,m>0),\\
\rho_n&=\frac{1}{2}G_+(in){\rm e}^{nB} w_n\quad (n>0).
\end{split}  
\end{equation}
In terms of the reduced variables the differential relations
(\ref{diff1}-\ref{diff3}) read
\begin{equation}
(n+m)W_{n,m}+\dot W_{n,m}=w_nw_m,  
\label{reddiff1}
\end{equation}
\begin{equation}
(n^2-m^2)W_{n,m}=(n-m)w_nw_m+w_m\dot w_n-w_n\dot w_m,
\label{reddiff2}
\end{equation}
\begin{equation}
\ddot w_n+2n\dot w_n=Fw_n.
\label{reddiff3}
\end{equation}
Further definitions:
\begin{equation}\label{O0n}
O_{0,n}=\frac{1}{2\pi}G_+(in){\rm e}^{nB} U_n\quad (n>0),
\end{equation}
\begin{equation}
O_{0,0}=\frac{1}{\pi}D.
\end{equation}
The corresponding differential relations read
\begin{equation}
nU_n+\dot U_n=\rho_0 w_n,
\label{reddiff4}  
\end{equation}
\begin{equation}
n^2U_n=n\rho_0 w_n+\rho_0\dot w_n-\dot\rho_0 w_n,
\label{reddiff5}  
\end{equation}
\begin{equation}
\ddot\rho_0=F\rho_0,
\label{reddiff6}  
\end{equation}
and finally
\begin{equation}
\dot D=\rho_0^2.  
\label{reddiff7}  
\end{equation}

In summary, we introduced the following notations for the reduced densities 
\begin{align}
	O_{n,m} \to W_{n,m}, \qquad\qquad O_{0,n} \to U_n,\qquad\qquad  O_{0,0}\to D,  \nonumber\\
	\rho_n \to w_n, \qquad\qquad \rho_0 \to \rho_0, \qquad\qquad\quad
\end{align}
where $n,m > 0$.

\subsection{WH solution of the $\chi_n$ problem}
\label{ssectWHn}

The exact WH solution of the $\chi_n$ TBA integral equation (see the detailed derivation for $n=1$ in \cite{Bajnok:2022rtu} and a brief description of the generic case in \cite{Bajnok:2022xgx}) leads to the
integral equation\footnote{From now on we suppress the explicit $v$-dependence
of ${\cal A}(x,v)$. Similarly, the $v$-dependence of all asymptotic perturbative
series will not be explicitly indicated.}  
\begin{equation}
Q_n(y)+\frac{1}{\pi}\int_{{\cal C}_+}\frac{{\rm e}^{-x}{\cal A}(x)Q_n(x)}{x+y}
{\rm d}x+\sum_j\frac{d_j q_{n,j}}{\kappa_j+vy}+\frac{\nu^n \sigma_+(n)}{n+vy}=
\frac{1}{n-vy},
\label{WH1}
\end{equation}
where $Q_n(y)$ is related to the Fourier-transform of the auxiliary function of the WH method.
Here
\begin{equation}\label{Qnkappaj}
\nu={\rm e}^{-2B},\qquad d_j=iS_j \nu^{\kappa_j},\qquad
q_{n,j}=Q_n\left(\frac{\kappa_j}{v}\right)
\end{equation}
and the contour ${\cal C}_+$ runs slightly above the real line. 

The first step of the derivation is the extension of the TBA integral equation \eqref{TBA} to the whole $\theta$ line by introducing an auxiliary function. Next we go to Fourier space and using the decomposition \eqref{WHdecomp} we separate the equation into lower and upper analytic pieces. The Fourier transform of the auxiliary function satisfies a linear integral equation. It can also be used to express the densities, including the physical quantities, by an integral over the Fourier variable. The next step is to deform the integration contour, for both the linear integral equation and the expression giving the densities in terms of the auxiliary function, from the real line to a contour that goes around the positive half of the imaginary axis. Using the analytic properties of the kernel this contour integral can be given as an integral running slightly to the left of the positive imaginary axis (where the integrand is proportional to the discontinuity across the cut) plus an infinite series taking into account the contribution of the poles sitting on the positive part of the imaginary axis. After rescaling the argument of the (Fourier space) auxiliary function by an imaginary constant, it becomes the new unknown $Q_n(x)$ and the integral equation reduces to \eqref{WH1}. For details, see \cite{Bajnok:2022rtu} and \cite{Bajnok:2022xgx}.

There is an ambiguity in \eqref{WH1}: we could have used ${\cal C}_-$, which runs below the real
line. This
ambiguity is related to (and cancelled by) a similar ambiguity: the choice of the
Borel resummations $S^\pm$ (see section \ref{fej4}).
The above WH solution is valid for $n>0$, $n\not=\kappa_j$. In terms of this
solution the physical quantities are given by
\begin{equation}\label{wnexact}
w_n=1+\frac{v}{\pi}\int_{{\cal C}_+}{\rm e}^{-x}{\cal A}(x)Q_n(x){\rm d}x
+\sum_j d_j q_{n,j}+\nu^n\sigma_+(n)  
\end{equation}
and
\begin{equation}\label{Wnmexact}
W_{n,m}=\frac{1}{n+m}+\frac{v}{\pi}\int_{{\cal C}_+}
\frac{{\rm e}^{-x}{\cal A}(x)Q_n(x)}{m-vx}{\rm d}x
+\sum_j \frac{d_j q_{n,j}}{m-\kappa_j}+\frac{\nu^n\sigma_+(n)}{m-n}+
\nu^m\sigma_+(m)Q_n\left(\frac{m}{v}\right).
\end{equation}
This last formula is valid for $m,n>0$, $m,n\not=\kappa_j$, $m\not=n$.

\subsubsection{Perturbative solution}
\label{sssectPTn}

Let us denote the purely perturbative part of $Q_n(x)$ by $P_n(x)$, that is, $Q_n(x) = P_n(x)$ up to all NP correction terms proportional to $\nu^\alpha$ with any $\alpha>0$. It satisfies
\begin{equation}\label{Pny}
P_n(y)+\frac{1}{\pi}\int_{{\cal C}_+}\frac{{\rm e}^{-x}{\cal A}(x)P_n(x)}{x+y}
{\rm d}x=\frac{1}{n-vy}.
\end{equation}
$P_n(x)$, as an asymptotic series in $v$, is well-defined for all $n\not=0$. 
We define the asymptotic series
\begin{equation}
P_{n,m}=\frac{v}{\pi}\int_{{\cal C}_+}
\frac{{\rm e}^{-x}{\cal A}(x)P_n(x)}{m-vx}{\rm d}x,
\end{equation}
which is perturbatively well-defined for all $n,m\not=0$. The quantity
\begin{equation}\label{Anmdef}
A_{n,m}=\frac{1}{n+m}+P_{n,m}\qquad (n,m,n+m\not=0)  
\end{equation}
will play an important role in our considerations. First of all, 
\begin{equation}
W_{n,m}\longmapsto A_{n,m}  
\end{equation}
meaning that it is the purely perturbative part of $W_{n,m}$, in the above sense. Secondly, the analytic continuation of this object is perturbatively equivalent to the rescaled solution itself:
\begin{equation}\label{PnyAnm}
	P_n\left(\frac{\tau}{v}\right) \sim A_{n,-\tau},
\end{equation}
as can be seen by comparing (\ref{Pny}-\ref{Anmdef}).

Similarly
\begin{equation}
w_n\longmapsto a_n,  
\end{equation}
where
\begin{equation}
a_n=1+\frac{v}{\pi}\int_{{\cal C}_+}{\rm e}^{-x}{\cal A}(x)P_n(x){\rm d}x.
\end{equation}

The perturbative series satisfy the differential relations (inherited from Subsection \ref{reduced})
\begin{equation}
\begin{split}
(n+m)A_{n,m}+\dot A_{n,m}&=a_na_m,\\  
(n^2-m^2)A_{n,m}&=(n-m)a_na_m+a_m\dot a_n-a_n\dot a_m,\\
\ddot a_n+2n\dot a_n&=fa_n,\\
1+\dot P_{n,-n}&=a_n a_{-n}.
\end{split}
\label{A1234}
\end{equation}
Here $f$ is the perturbative part of the $n$-independent function $F$.

Further perturbative limits are
\begin{equation}
U_n\longmapsto V_n \ (n\not=0),\qquad  
\rho_0\longmapsto s,\qquad  
D\longmapsto D_0  
\end{equation}
satisfying
\begin{equation}
\begin{split}
nV_n+\dot V_n&=sa_n,\\  
n^2V_n&=nsa_n+s\dot a_n-\dot sa_n,\\
\ddot s&=fs,\\
\dot D_0&=s^2.
\end{split}  
\label{V1234}
\end{equation}

Let us calculate the leading (constant) term of $P_n(x)$:
\begin{equation}
P_n(x)=P_n^{(0)}(x)+{\rm O}(v).  
\end{equation}
This is proportional to the function $\psi_0(x)$, calculated explicitly in
\cite{Bajnok:2022rtu}: 
\begin{equation}
P_n^{(0)}(x)=\frac{1}{n}\psi_0(x).  
\end{equation}
This was defined as
\begin{equation}\label{psi0LO}
\psi_0(y) + \frac{1}{\pi}\int_0^\infty \frac{{\rm e}^{-x} \psi_0(x)  }{x+y}\; {\rm d} x = 1,
\end{equation} 
as can be deduced from ${\cal A}(x) = 1 + \text{O}(v)$ due to (\ref{Axvpert}-\ref{Lkpoly}).
Here we need only the integral of this function,
\begin{equation}
\frac{1}{\pi}\int_0^\infty
{\rm e}^{-x}\psi_0(x){\rm d}x=\frac{1}{4}
\end{equation}
to calculate
\begin{equation}
P_{n,m}=\frac{v}{4nm}+{\rm O}(v^2),\qquad
A_{n,m}=\frac{1}{n+m}+\frac{v}{4nm}+{\rm O}(v^2),
\end{equation}
\begin{equation}
a_n=1+\frac{v}{4n}+{\rm O}(v^2),\qquad f=-v^2+{\rm O}(v^3),
\end{equation}
\begin{equation}
s=\frac{\lambda}{\sqrt{v}}\left[1+{\rm O}(v)\right],\qquad
D_0=\frac{\lambda^2}{4v^2}\left[1+{\rm O}(v)\right].
\end{equation}

We see that (nearly) all perturbative building blocks are completely determined
if we know just one of them. For example, using Volin's
method \cite{Volin:2009wr,Volin:2010cq},
we can start from $A_{1,1}$ and calculate successively:
\begin{equation}\label{pertinstructions}
A_{1,1}\longrightarrow a_1\longrightarrow f\longrightarrow\left\{\begin{split}
&a_n\longrightarrow A_{n,m},\ P_{n,-n}\\
&s\ ({\rm up\ to\ overall\ }\lambda)\end{split}\right\}\longrightarrow V_n, D_0
\end{equation}
This way only the overall constant $\lambda$ in $s$ and the integration
constant in $D_0$ is not determined. ($V_n$ is also proportional to $\lambda$
and $D_0$ is proportional to $\lambda^2$.)

\subsubsection{Trans-series representation}\label{transrep}
\label{sssectTSn}

Inspecting the form of the WH integral equation (\ref{WH1}), one may recognize that the last two, 
exponentially suppressed terms on the l.h.s. have similar form as the r.h.s. - i.e. as the source term of the integral equation \eqref{Pny}
that defines the PT solution. Due to the linearity of \eqref{WH1} it is thus possible to linearly combine the $P_n(x)$-s, such that the result solves the exact equation (see below). Thus, we can write the
solution in the form of a trans-series: an expansion in powers of the
non-perturbative parameter $\nu$, where the expansion coefficients are
asymptotic power series in the running coupling $v$. We indicate the
trans-series nature of the solution by a tilde:
\begin{equation}
\tilde Q_n(x)=P_n(x)+\sum_j d_j\tilde q_{n,j} P_{-\kappa_j}(x)+\nu^n\sigma_+(n)
P_{-n}(x).  
\end{equation}
Using the equivalence of the analytically continued perturbative serieses in \eqref{PnyAnm} we have
\begin{equation}\label{TSsol}
\tilde Q_n\left(\frac{\tau}{v}\right)=A_{n,-\tau}+\sum_j d_j\tilde q_{n,j}
A_{-\kappa_j,-\tau}+\nu^n\sigma_+(n)A_{-n,-\tau}  
\end{equation}
and thus in principle we can calculate the previously unknown constants that were defined only implicitely sofar, and will be determined exactly by the following self-consistency conditon
\begin{equation}
\tilde q_{n,j}=\tilde Q_n\left(\frac{\kappa_j}{v}\right).
\label{qeq}
\end{equation}
We may rewrite \eqref{wnexact} following the logic above as
\begin{equation}
\tilde w_n=a_n+\sum_j d_j\tilde q_{n,j}
a_{-\kappa_j}+\nu^n\sigma_+(n)a_{-n}  
\label{wntil}
\end{equation}
and finally express \eqref{Wnmexact} as
\begin{equation}
\tilde W_{n,m}=A_{n,m}+\sum_j d_j\tilde q_{n,j}
A_{-\kappa_j,m}+\nu^n\sigma_+(n)A_{-n,m}+\nu^m\sigma_+(m)
\tilde Q_n\left(\frac{m}{v}\right).  
\label{Wnmtil}
\end{equation}

{\it Matrix-vector notation}

Our formulas become more transparent if we introduce the following compact
notation. We introduce the (infinite) vectors $q^{(n)}$, $p$, $R$, $Y^{(n)}$
with components
\begin{equation}
q^{(n)}_j=\tilde q_{n,j},\qquad p_j=a_{-\kappa_j},\qquad R_j=V_{-\kappa_j},\qquad
Y^{(n)}_j=A_{n,-\kappa_j}  
\end{equation}
and the (infinite) matrices ${\cal N}$, ${\cal D}$, $X$ with entries
\begin{equation}\label{NDX}
{\cal N}_{js}=A_{-\kappa_j,-\kappa_s},\qquad {\cal D}_{js}=d_j\delta_{js},\qquad
X_{js}=\kappa_j\delta_{js}.  
\end{equation}
Moreover, we define
\begin{equation}
{\cal M}={\cal N}{\cal D},\qquad \bar{\cal M}=(1-{\cal M})^{-1}=1+{\cal M}+
{\cal M}^2+\dots  
\end{equation}
and finally the symmetric matrix ${\cal L}$:
\begin{equation}
{\cal L}={\cal D}\bar{\cal M}={\cal D}+{\cal D}{\cal N}{\cal D}+
{\cal D}{\cal N}{\cal D}{\cal N}{\cal D}+\dots  
\end{equation}
With this notation (\ref{qeq}) can be compactly written as
\begin{equation}
q^{(n)}=Y^{(n)}+\nu^n\sigma_+(n)Y^{(-n)}+{\cal M}q^{(n)}
\end{equation}
and it has solution in terms of the Neumann-series $\bar{\cal M}$ of the ${\cal M}$ operator:
\begin{equation}
q^{(n)}=\bar{\cal M}Y^{(n)}+\nu^n\sigma_+(n)\bar{\cal M}Y^{(-n)}.
\end{equation}
The compact form of (\ref{wntil}) is
\begin{equation}
\tilde w_n=a_n+\nu^n\sigma_+(n)a_{-n}+p^{\rm T}{\cal L}Y^{(n)}+\nu^n\sigma_+(n)
p^{\rm T}{\cal L}Y^{(-n)}    
\end{equation}
and finally (\ref{Wnmtil}) becomes
\begin{equation}
\begin{split}
\tilde W_{n,m}&=A_{n,m}+\nu^n\sigma_+(n)A_{-n,m}+\nu^m\sigma_+(m)A_{n,-m}\\
&+\nu^{n+m}\sigma_+(n)\sigma_+(m)A_{-n,-m}+Y^{(m){\rm T}}{\cal L}Y^{(n)}
+\nu^n\sigma_+(n)Y^{(m){\rm T}}{\cal L}Y^{(-n)}\\
&+\nu^m\sigma_+(m)Y^{(-m){\rm T}}{\cal L}Y^{(n)}
+\nu^{n+m}\sigma_+(n)\sigma_+(m)Y^{(-m){\rm T}}{\cal L}Y^{(-n)}.
\end{split}  
\end{equation}
This form is manifestly symmetric in $n,m$, which was not obvious in some
intermediate forms of this quantity.

The most important physical quantities are $\tilde w_1$ and the energy
density $\tilde W_{1,1}$. The latter must be treated as a limiting case because
of the $n\not=m$ restriction for $\tilde W_{n,m}$. To calculate the limit we
introduce
\begin{equation}
\sigma_+(1)=\hatk,\qquad \frac{{\rm d}}{{\rm d}\varepsilon}
\sigma_+(1+\varepsilon)\Big\vert_{\varepsilon=0}=-\frac{M}{2},\qquad
Y^{(1)}=Y,\qquad Y^{(-1)}=Z.
\label{sigmap1}
\end{equation}
We find
\begin{equation}\label{w1til}
\tilde w_1=a_1+\nu \hatk a_{-1}+p^{\rm T}{\cal L}Y+\nu \hatk p^{\rm T}{\cal L}Z
\end{equation}
and
\begin{equation}\label{W11til}
\begin{split}
\tilde W_{1,1}=\lim_{\varepsilon\to0}\tilde W_{1,1+\varepsilon}&=\frac{M}{2}\nu
+2B\nu \hatk+2\nu \hatk P_{1,-1}+A_{1,1}+\nu^2 \hatk^2 A_{-1,-1}\\
&+Y^{\rm T}{\cal L}Y+2\nu \hatk Y^{\rm T}{\cal L}Z
+\nu^2 \hatk^2 Z^{\rm T}{\cal L}Z.
\end{split}  
\end{equation}
(We recall that for all models except \mO{3} $\hatk=0$ and
$\hatk=1/{\rm e}$ for \mO{3}.)

{\it Derivatives}

The differential relations satisfied by the perturbative objects can be
compactly written in matrix form as follows.
\begin{equation}
\begin{split}
\dot{\cal N}&={\cal N}X+X{\cal N}+pp^{\rm T},\\
\dot{\cal D}&=-2X{\cal D},\\
\dot{\cal M}&=X{\cal M}-{\cal M}X+pp^{\rm T}{\cal D},\\
\dot{\bar{\cal M}}&=X\bar{\cal M}-\bar{\cal M}X+\bar{\cal M}pp^{\rm T}{\cal L},\\
\dot{\cal L}&=-X{\cal L}-{\cal L}X+{\cal L}pp^{\rm T}{\cal L},\\
\dot Y&=XY-Y+a_1p,\\
\dot Z&=XZ+Z+a_{-1}p,\\
\dot R&=XR+sp.
\end{split}
\label{matdiff}
\end{equation}
With the help of the above formulas one can verify the
\begin{equation}
{\dot{\tilde W}}_{1,1}+2\tilde W_{1,1}=\tilde w_1^2  
\end{equation}
relation.

\subsection{WH solution of the $\chi_0$ problem}
\label{ssectWH0}

We need to handle the $n=0$ case separately for the following reasons. The WH projection and the $n\to 0$ limit of the Fourier-transformed and extended integral equation are not interchangeable steps. Furthermore, for bosonic models such as the \mO{$N$} models, the $\vert\omega\vert^{-1/2}$-like behaviour of the decomposition $G_+(\omega)$ at the origin superimposed on the $\omega^{-1}$-type source term gives a non-integrable singularity, and therefore one needs to regularize the integrals.

At the end of the calculation the regulator can be removed and we find that the exact WH equations are thus more complicated in this case:
\begin{equation}
\ell(y)+\frac{1}{\pi}\int_{{\cal C}_+}
\frac{{\rm e}^{-x}{\cal A}(x)\ell(x)}{x+y}{\rm d}x
=\frac{g_o}{\pi v\sqrt{v}}\int_{{\cal C}_+}
\frac{{\rm e}^{-x}{\cal A}(x)-1}{(x+y)x\sqrt{x}}{\rm d}x
-\sum_j\frac{d_jy_j}{\kappa_j+vy},  
\end{equation}
\begin{equation}
\begin{split}  
L(\tau)&=-\frac{v}{\pi}\int_{{\cal C}_+}
\frac{{\rm e}^{-x}{\cal A}(x)\ell(x)}{\tau+vx}{\rm d}x
-\frac{g_o\sqrt{v}}{\pi \tau}\int_{{\cal C}_+}
\frac{{\rm e}^{-x}{\cal A}(x)}{(\tau+vx)\sqrt{x}}{\rm d}x\\
&+\frac{g_o}{\pi \tau\sqrt{v}}\int_{{\cal C}_+}
\frac{{\rm e}^{-x}{\cal A}(x)-1}{x\sqrt{x}}{\rm d}x
-\sum_j\frac{d_jy_j}{\tau+\kappa_j},  
\end{split}
\end{equation}
\begin{equation}\label{Lkappaj}
y_j=L(\kappa_j),
\end{equation}
where $g_o=H(0)$. Here $\ell(y)$ is analogous to $Q_n(y)$ in \eqref{WH1}, however, the latter incorporates an explicit pole at $y=n/v$, whereas the first is regular for $y>0$. Note however that \eqref{Lkappaj} is not completely analogous to \eqref{Qnkappaj} since $L(\tau)$ is slightly different from $\ell(\tau/v)$.
The physical quantities, namely the densities $O_{0,n}$ are given by \eqref{O0n} and
\begin{equation}
\begin{split}  
U_n&=\frac{v}{\pi}\int_{{\cal C}_+}
\frac{{\rm e}^{-x}{\cal A}(x)\ell(x)}{n-vx}{\rm d}x
-\frac{g_o\sqrt{v}}{\pi n}\int_{{\cal C}_+}
\frac{{\rm e}^{-x}{\cal A}(x)}{(n-vx)\sqrt{x}}{\rm d}x\\
&-\frac{g_o}{\pi n\sqrt{v}}\int_{{\cal C}_+}
\frac{{\rm e}^{-x}{\cal A}(x)-1}{x\sqrt{x}}{\rm d}x
+\sum_j\frac{d_jy_j}{n-\kappa_j}+\nu^n\sigma_+(n)L(n)  
\end{split}
\end{equation}
and the boundary value as
\begin{equation}
\rho_0=\frac{v}{\pi}\int_{{\cal C}_+}
{\rm e}^{-x}{\cal A}(x)\ell(x){\rm d}x
-\frac{g_o}{\pi \sqrt{v}}\int_{{\cal C}_+}
\frac{{\rm e}^{-x}{\cal A}(x)-1}{x\sqrt{x}}{\rm d}x
+\sum_j d_jy_j.  
\end{equation}

\subsubsection{Perturbative solution}
\label{sssectPT0}

The perturbative part of the problem requires to solve
\begin{equation}
q(y)+\frac{1}{\pi}\int_{{\cal C}_+}
\frac{{\rm e}^{-x}{\cal A}(x)q(x)}{x+y}{\rm d}x
=\frac{1}{\pi}\int_{{\cal C}_+}
\frac{{\rm e}^{-x}{\cal A}(x)-1}{(x+y)x\sqrt{x}}{\rm d}x
\end{equation}
and calculate
\begin{equation}
\begin{split}  
V_n&=\frac{g_o}{\sqrt{v}}\Bigg\{\frac{1}{\pi}\int_{{\cal C}_+}
\frac{{\rm e}^{-x}{\cal A}(x)q(x)}{n-vx}{\rm d}x
-\frac{v}{\pi n}\int_{{\cal C}_+}
\frac{{\rm e}^{-x}{\cal A}(x)}{(n-vx)\sqrt{x}}{\rm d}x\\
&-\frac{1}{\pi n}\int_{{\cal C}_+}
\frac{{\rm e}^{-x}{\cal A}(x)-1}{x\sqrt{x}}{\rm d}x\Bigg\},
\end{split}
\label{Vnrep}
\end{equation}
\begin{equation}\label{sint}
s=\frac{g_o}{\sqrt{v}}\Bigg\{\frac{1}{\pi}\int_{{\cal C}_+}
{\rm e}^{-x}{\cal A}(x)q(x){\rm d}x
-\frac{1}{\pi}\int_{{\cal C}_+}
\frac{{\rm e}^{-x}{\cal A}(x)-1}{x\sqrt{x}}{\rm d}x\Bigg\}.  
\end{equation}

{\it The leading term}

The leading (constant) term of the perturbative solution satisfies
\begin{equation}\label{q0LO}
q_0(y)+\frac{1}{\pi}\int_0^\infty
\frac{{\rm e}^{-x}q_0(x)}{x+y}{\rm d}x
=\frac{1}{\pi}\int_0^\infty
\frac{{\rm e}^{-x}-1}{(x+y)x\sqrt{x}}{\rm d}x.
\end{equation}
In \cite{Bajnok:2022rtu} we calculated the constant - the leading order inside the braces of \eqref{sint} - and obtained 
\begin{equation}\label{s0const}
s_0=\frac{1}{\pi}\int_0^\infty{\rm e}^{-x} q_0(x){\rm d}x
-\frac{1}{\pi}\int_0^\infty\frac{{\rm e}^{-x}-1}{x\sqrt{x}}{\rm d}x  
=\frac{\sqrt{\pi}}{2}.
\end{equation}
This determines the overall constant in
\begin{equation}
s=\frac{\lambda}{\sqrt{v}}\left(1+{\rm O}(v)\right),\qquad  
V_n=\frac{\lambda}{n\sqrt{v}}\left(1+{\rm O}(v)\right),\qquad
\lambda=\frac{g_o\sqrt{\pi}}{2}.
\end{equation}
In \cite{Bajnok:2022xgx} the first few terms of $s$ are given:
\begin{equation}
s=\frac{\lambda}{\sqrt{v}}\left(1+\frac{av}{2}-\frac{5a^2v^2}{8}+\dots\right).  
\label{spert}
\end{equation}
Using
\begin{equation}
{\dot D}_0=-\frac{2v^2}{1+av}D_0^\prime=s^2
\end{equation}
(with the notation $f^\prime={\rm d}f/{\rm d}v$) we find  
\begin{equation}\label{D0prime}
D_0^\prime=-\frac{\lambda^2}{2v^3}-\frac{\lambda^2a}{v^2}+{\rm O}(1)
\end{equation}
and  
\begin{equation}
D_0=\frac{\lambda^2}{4v^2}+\frac{a\lambda^2}{v}+{\rm const.}+{\rm O}(v).
\end{equation}
In our family of models we have
\begin{equation}
a=1-2\Delta,\qquad g_o=\frac{1}{\sqrt{\pi\Delta}},\qquad
\lambda=\frac{1}{2\sqrt{\Delta}}.  
\end{equation}

We see from the absence of the $1/v$ monomial in the derivative \eqref{D0prime} of $D_0$, that there is no log term here, $D_0$ itself is fully perturbative.  
To complete the calculation of all perturbative quantities it remains to find
the integration constant in $D_0$.\footnote{This constant was also discussed in \cite{Bajnok:2022rtu} for the $\mO{3}$ case, where it has been fixed from direct calculations for $O_{0,0}$, a quantity, that is also related to the capacitance of a circular parallel plate capacitor (see e.g. \cite{Reichert:2020ymc}). We omit its derivation here (see also the comment in the footnote above \eqref{DTS}) both because it is lengthy, and also due to the fact that $O_{0,0}$ is not necessary to express the free-energy in terms of the physical running coupling, the focus of the present paper.}

\subsubsection{Trans-series representation}
\label{sssectTS0}

The trans-series solution is of the form
\begin{equation}
\tilde\ell(x)=\frac{g_o}{v\sqrt{v}}q(x)+\sum_j d_j\tilde y_jP_{-\kappa_j}(x),  
\end{equation}
\begin{equation}
\tilde L(\tau)=\sum_j d_j\tilde y_j A_{-\kappa_j,-\tau}+V_{-\tau},
\end{equation}
\begin{equation}
\tilde y_j=\tilde L(\kappa_j).
\label{yeq}
\end{equation}
The physical quantities are given by
\begin{equation}
\tilde U_n=V_n+\sum_j Y^{(n)}_jd_j\tilde y_j+\nu^n\sigma_+(n)\left[V_{-n}+
\sum_j Y^{(-n)}_j d_j\tilde y_j\right],    
\end{equation}
\begin{equation}
\tilde\rho_0=s+\sum_j p_j d_j\tilde y_j.
\end{equation}

It is useful to use the compact matrix-vector notation here too. Introducing
the vector $y$ with components
\begin{equation}
y_j=\tilde y_j  
\end{equation}
(\ref{yeq}) can be compactly written and solved as
\begin{equation}
y=R+{\cal M}y,\qquad y=\bar{\cal M}R.  
\end{equation}
Then
\begin{equation}
\tilde U_n=V_n+R^{\rm T}{\cal L}Y^{(n)}+\nu^n\sigma_+(n)\left[V_{-n}+
R^{\rm T}{\cal L}Y^{(-n)}\right],
\label{Until}
\end{equation}
\begin{equation}
\tilde\rho_0=s+R^{\rm T}{\cal L}p
\label{rho0til}
\end{equation}
and the physical particle density is
\begin{equation}
\tilde U_1=V_1+\nu \hatk V_{-1}+R^{\rm T}{\cal L}Y
+\nu \hatk R^{\rm T}{\cal L}Z.  
\label{U1til}
\end{equation}
Using the differential relations (\ref{matdiff}) the relation
\begin{equation}
\tilde U_1+{\dot{\tilde U}}_1=\tilde\rho_0\tilde w_1  
\label{REL1}
\end{equation}
can be verified.

The last density $D$ could be calculated from the $\varepsilon \to 0$ limit of $O_{0,\varepsilon}$ in principle\footnote{This limiting procedure applied for the perturbative part only would fix the missing constant in $D_0$ as well.}, but it is easier to verify that the trans-series ansatz
\begin{equation}\label{DTS}
	\tilde{D} = D_0 + R^T {\cal L} R
\end{equation}	
satisfies
\begin{equation}
	\dot{\tilde{D}} =  \tilde{\rho}_0^2,
\end{equation}
if the last line of \eqref{V1234} holds for $D_0$.

\subsection{Summary of the trans-series solution}
\label{ssectSumm}
Let us here recollect the notations we used throughout the previous sections in a compact way. We present the trans-series representation of the quantities (\ref{boundaryvalues}-\ref{densities}) for the $\mO{N}$ models, in terms of the bootstrap running coupling $v$ introduced in Subsection \ref{ssectRunning} 
\begin{equation}\label{rcON}
	2B =\frac{1}{v}-(1-2\Delta)\ln v + 1-2\Delta+2\Delta\ln\Delta+(1-4\Delta)\ln2
\end{equation}	
and the small NP trans-series parameter $\nu = {\rm e}^{-2 B}$, where $B$ is the boundary parameter of the TBA equation \eqref{TBA} and $\Delta = 1/(N-2)$.

We have defined reduced versions of the abovementioned quantities in Subsection \ref{reduced}, where we removed ${\rm e}^B$ powers and factors related to the WH decomposition \eqref{WHdecomp} of the kernel
\begin{equation}
	G_+(\omega) = \frac{1}{\sqrt{-i \Delta\omega }} \frac{\Gamma(1-i\Delta\omega)}{\Gamma(1/2-i\omega/2)} {\rm e}^{-i (1/2-\Delta)\omega \ln(-i\omega) - i \left(\Delta(1-\ln\Delta)-(1+\ln 2)/2\right)\omega}.
\end{equation}	
The reduced densities, together with their perturbative limits are\footnote{In this subsection we write $X=X$ inside enumerations to indicate that we would not like to resolve the quantity into lower level terms at that point.} (the PT limit is indicated by $\longmapsto$, these are power serieses of $v$)
\begin{align}
	O_{n,m}=O_{m,n} &=\frac{1}{4\pi} G_+(i n)G_+(i m) {\rm e}^{(n+m)B} W_{n,m} &n,m&>0, & W_{n,m}&\longmapsto A_{n,m} \\
	O_{0,n}=O_{n,0} &= \frac{1}{2\pi} G_+(i n) {\rm e}^{n B} U_n &n&>0, & U_n&\longmapsto V_n  \\
	O_{0,0} &= \frac{1}{\pi} D &&& D&\longmapsto D_0 \\
	\rho_n &=\frac{1}{2} G_+(i n) {\rm e}^{n B} w_{n} &n&>0, & w_n&\longmapsto a_n \\
	\rho_0 &= \rho_0 &&& \rho_0&\longmapsto s  .
\end{align}
The PT limits (see Subsubsections \ref{sssectPTn} and  \ref{sssectPT0}) start out as
\begin{align}
	A_{n,m} &= \frac{1}{n+m} + \frac{v}{4 n m } + \text{O}(v^2) \qquad n+m\neq 0 & n,m&\gtrless0\\
	V_n &= \frac{1}{2 n\sqrt{\Delta v}} \left(1 + \text{O}(v) \right) & n &\gtrless0  \\
	D_0 &=\frac{1}{4 \Delta} \left( \frac{1}{4 v^2} +\frac{1-2\Delta}{v} + \text{const}. + \text{O}(v) \right) & & \\
	a_n &= 1+\frac{v}{4n}+\text{O}(v^2) & n&\gtrless 0 \\
	s &= \frac{1}{2\sqrt{\Delta v}}\left(1+\frac{\left(1-2\Delta\right) v}{2} - \frac{5 \left(1-2\Delta\right)^2 v^2}{8} + \text{O}(v^3) \right) & & 
\end{align}
and may be calculated effectively via Volin's method and their differential relations \eqref{A1234} and \eqref{V1234} as described in \eqref{pertinstructions}. These perturbative building blocks (analytically extended to $n,m<0$) help us to construct the trans-series representations.
The trans-serieses for the reduced quantities are 
\begin{align}
	\tilde{W}_{n,m} &= \tilde{A}_{n,m} + \sigma_+(n)\nu^n \tilde{A}_{-n,m} + \sigma_+(m)\nu^m \tilde{A}_{n,-m} +\sigma_+(n)\sigma_+(m) \nu^{n+m} \tilde{A}_{-n,-m} \label{WnmTS}\\
	\tilde{U}_n &= \tilde{V}_{n} + \sigma_+(n) \nu^n \tilde{V}_{-n} \\
	\tilde{D} &= \tilde{D} \\
	\tilde{w}_n &= \tilde{a}_n + \sigma_+(n)\nu^n \tilde{a}_{-n} \\
	\tilde{\rho_0} &= \tilde{\rho_0}
\end{align}
where $n\neq m$ in the first line as we need to calculate $\tilde{W}_{n,n}$ as a $m\to n$ limit like we did for \eqref{W11TS}. Here $\sigma_+(n) = G_-(i n + \epsilon)/G_+(i n), \; n>0$ was defined and analyzed in Subsection \ref{ssectDisc}, where $\epsilon > 0 $ is an infinitesimal shift. Furthermore, by an abuse of notation, here we defined the intermediate trans-series objects $\tilde{A}_{n,m}, \tilde{a}_n$ and $\tilde{V}_{n}$ that can be evaluated for negative indices as well $n,m \gtrless 0$, similarily to the perturbative limits $A_{n,m}, a_n$ and $V_n$. Their structure is similar to that of  $\tilde{D}$ and $\tilde{\rho}_0$:
\begin{align}
	\tilde{A}_{n,m} &=  A_{n,m}  + Y^{(m) T} {\cal L} Y^{(n)} \qquad n + m \neq 0 & n,m &\gtrless 0\\
	\tilde{V}_{n} &= V_n + R^T {\cal L} Y^{(n)}  &  n &\gtrless 0 \\
	\tilde{D} &= D_0 + R^T {\cal L} R & &\\
	\tilde{a}_n &= a_n + p^T {\cal L} Y^{(n)} & n&\gtrless0\\
	\tilde{\rho}_0 &= s + R^T {\cal L} p,
\end{align}
in terms of our matrix/vector notation introduced in Subsubsections \ref{sssectTSn} and \ref{sssectTS0}. The components of the infinite vectors on the r.h.s. of the above equations are the perturbative building blocks
\begin{equation}
	Y_j^{(n)} = A_{n,-\kappa_j} \quad (n\gtrless0) \qquad R_j = V_{-\kappa_j} \qquad p_j = a_{-\kappa_j}  
\end{equation}
where the infinite matrix ${\cal L}$ is defined as an infinite series via the infinite matrices ${\cal N},{\cal D}$:
\begin{equation}
	{\cal L }= {\cal D}+{\cal D}{\cal N}{\cal D} + 	{\cal D}{\cal N}{\cal D}{\cal N}{\cal D} + \ldots, \qquad {\cal N}_{j s} = A_{-\kappa_j, -\kappa_s} \qquad {\cal D}_{j s} = d_j \delta_{j s}.
\end{equation}
Here $\kappa_j$-s are the positions of the poles in $\sigma_+(\kappa)$ and  $d_j = i S_j \nu^{\kappa_j}$ are related to their residues through $S_j$ (see Subsection \ref{ssectPoles}):
\begin{equation}\label{sigmapoles}
	\sigma_+(\kappa) \approx \frac{-i S_j}{\kappa-\kappa_j}, \qquad \kappa_j = 2^{\left(N \bmod 2\right)}(N-2)j ,\qquad j = 1,2,3,\ldots .
\end{equation}	
Due to the $\nu^{\kappa_j}$ factor in $d_j$ the trans-series solution can be ordered in powers of $\nu$. Throughout the calculations we assumed that the indices $n,m$ will not coincide with any of the $\kappa_j$-s.

Finally the special quantity related to $O_{1,1}$, i.e. the energy density in \eqref{physicaldensities} is 
\begin{equation}\label{W11TS}
	\tilde W_{1,1}=\tilde{A}_{1,1}+ \nu\left[\frac{M}{2} + 2\hatk \left(B + P_{1,-1} +  Y^{(1)\rm T}{\cal L}Y^{(-1)} \right)\right]+\nu^2 \hatk^2 \tilde{A}_{-1,-1},  \quad  \hatk \equiv \sigma_+(1),  
\end{equation}
where the term in the middle of the r.h.s. came from the limit of the second and third term in the expression \eqref{WnmTS} of $\tilde{W}_{n,m}$, while taking  $\tilde W_{1,1} =\lim_{\varepsilon\to0}\tilde W_{1,1+\varepsilon} $. Here $M= -2 \left({\rm d}\sigma_+(1+\varepsilon)/{\rm d}\varepsilon \right)\vert_{\varepsilon=0}$ is a constant, while $P_{n,-m}=A_{n,-m}-\frac{1}{n-m}$ is the non-singular part of $A_{n,-m}$ in the $m\to n$ limit.
The other special quantities, that is $O_{0,1}$, related to the particle density in \eqref{physicaldensities} and $\rho_1,\rho_0$ whose ratio is related to the external magnetic field in the free-energy problem \eqref{eq:hperm} have the respective trans-series representations
\begin{align}
	\tilde{U}_1 &= \tilde{V}_{1} + \nu \hatk \tilde{V}_{-1} \label{U1TS}\\
	\tilde{w}_1 &= \tilde{a}_1 + \nu \hatk \tilde{a}_{-1} \label{w1TS}\\
	\tilde{\rho_0} &= \tilde{\rho_0} \label{rho0TS}.
\end{align}	
Since $\hatk=0$ except for \mO{3}, the other models have
\begin{equation}
	\mO{$N\neq 3$}: \qquad 
	\tilde W_{1,1}=\tilde{A}_{1,1} +  \nu \frac{M}{2}, \qquad
	\tilde{U}_1 = \tilde{V}_{1}, \qquad
	\tilde{w}_1 = \tilde{a}_1, \qquad
	\tilde{\rho_0} = \tilde{\rho_0}. 
\end{equation}
For both $N=4$ and $N=3$, the poles are at even integers $\kappa_j = 2 j$. Thus $\tilde{\rho}_0$ has only powers of $\nu^2$ in both cases, and the same is true for all the other quantities in the \mO{4} case (with the exception of the $\nu M/2$ term in $\tilde{W}_{1,1}$). The odd powers of $\nu$ get switched on for  $\tilde{W}_{1,1}, \tilde{U}_1, \tilde{w}_1$ in the \mO{3} case where $\hatk=1/{\rm e}$. Due to the $2 B \nu \hatk$ term  in \eqref{W11TS}, the leading NP sector of the energy density will contain an $\ln v$ term in the perturbative expansion that comes from the r.h.s. of \eqref{rcON}.

\section{Resummation and the leading NP corrections}
\label{fej4}

In the present and the following section (Sections \ref{fej4}-\ref{sect7}) we will build up our techniques gradually to finally prove our main result for the simpler case of the \mO{$N\geq 4$} models. Namely, that the physical values of the energy and particle densities can be obtained from the so-called \emph{median resummations} of the perturbative parts of the same quantities (up to the bulk-energy constant term $M\nu$ in the energy density \cite{Bajnok:2022rtu}). This elevates the conjectures of \cite{Marino:2021dzn, Bajnok:2022rtu} regarding \emph{strong resurgence} in these models to the level of the proof and generalizes the result of Section 5. in \cite{Bajnok:2022xgx} to the case of the particle density. 

To understand these statements, let us remind ourselves that the end result of the previous section was the trans-series representation of the special densities. These contain imaginary terms, and their physical values must be obtained via lateral Borel-resummation (see below). The latter must cancel the mentioned ambiguous\footnote{Ambiguous, as their sign depended on the choice of ${\cal C}_\pm$ in the WH equations in accord with the choice $S^{\pm}$ for the lateral resummation.} imaginary parts and this means that the singularities of the Borel-plane along the real line (those who render the theory non-Borel-resummable) are related to the NP terms in the trans-series. We will show this cancellation of ambiguities at the leading order first, for the imaginary part of the $M\nu$ term in the energy density. The present section is dedicated to this special case.

Then, in Section \ref{sect7} we will start from the abovementioned fundamental assumption (our main conjecture) that the lateral Borel-resummation of the trans-series solution presented in Section \ref{sect3} must give the real, physical value of these quantities. By the machinery of \emph{resurgence theory} we show in Subsection \ref{ssectStokes} that this condition is equivalent to simple relations for the \emph{alien derivatives} of the PT building blocks, as discussed in Appendix \ref{appB}. Finally, using the latter relations we show in Subsection \ref{subsect72} that (the inverse square-root of) the so-called \emph{Stokes-automorphism} built from the \emph{alien derivatives} generates the whole trans-series of the reduced densities (except the special case of Section \ref{fej4}). In the procedure called \emph{median resummation} this gets laterally resummed and gives the physical values. By \emph{strong resurgence} we mean the fact that it is enough to apply this procedure for the PT parts of the densities only. This has to be contrasted to the \mO{3} case in Section \ref{sect8}, where it will turn out that it is not enough to apply \emph{median resummation} to the perturbative sector only to get the physical results, but one needs to resum extra, NP-suppressed (instanton) sectors, too.

\subsection{Energy and particle densities for the \mO{$N\geq4$} models}
\label{ssectDensON}

In this section we intend to show that the lateral Borel resummation (see Subsection \ref{AppResurgence}) of the perturbative
series naturally cancels the imaginary part of the leading NP corrections.
We will restrict our attention to the $N\geq4$ cases.

For these models the parameter $a$, appearing in the relation
\begin{equation}
\nu=v^a{\rm e}^{-\frac{1}{v}}{\rm e}^{-L}
\end{equation}
is restricted to the interval  
\begin{equation}
0\leq a<1.  
\end{equation}


We recall that the parameters are  
\begin{equation}
a=1-2\Delta,\qquad b+L=-4\Delta\ln2,   
\end{equation}
\begin{equation}
M\nu=-v^a{\rm e}^{-\frac{1}{v}}\frac{\Gamma(1-\Delta)}{\Gamma(1+\Delta)}\,
2^{4\Delta}{\rm e}^{i\pi\Delta}.
\end{equation}
The spectrum of poles is equidistant:  
\begin{equation}
\kappa_j=k j,\qquad{\rm where}\qquad  
k=\left\{\begin{split}N-2\ \ \ \quad&{\rm
(even\ }N)\\ 2(N-2)\ \quad&{\rm (odd\ }N)\end{split} \right.   
\end{equation}
We see that $a\kappa_j$ is always an even integer, hence for these models
all cuts on the Borel-plane of the Borel-transformed perturbative serieses are logarithmic, except the one starting at $t=1$ and inspecting
(\ref{26}) we see that all Stokes constants are indeed real.

For all models the leading
NP corrections to the densities are of the form
\begin{equation}
\hat\epsilon=2\tilde W_{1,1}=2A_{1,1}+M\nu+2iS_1 A_{1,-k}^2\nu^k+{\rm O}(\nu^{2k}),
\quad
\hat\rho=\tilde U_1=V_1+iS_1 V_{-k}A_{1,-k}\nu^k+{\rm O}(\nu^{2k}).
\end{equation}

\subsection{Cancellation of the leading NP term}

The (reduced) energy density in these models (for $N\geq4$) is
\begin{equation}
S^+(2\tilde W_{1,1})=S^+(\Psi)+M\nu+2S^+(Y^{\rm T}{\cal L}Y).  
\label{411}
\end{equation}
Note that we are using $S^+$ here, which corresponds to the choice of the
contour ${\cal C}_+$.
In (\ref{411}) $\Psi=2A_{1,1}$, $M=M_0+iM_1$, and the leading NP contributions
coming from the last term are
\begin{equation}
2Y^{\rm T}{\cal L}Y=2iS_1\nu^k A_{1,-k}^2+{\rm O}(\nu^{2k}),  
\end{equation}
where
\begin{equation}
k=\kappa_1\geq2
\end{equation}
is the position of the nearest pole.\footnote{Note that we introduced the notation $\Psi$ above such that one could compare the following analysis to the definitions in Section \ref{AppResurgence} more easily.}
Since the energy density is real, $S^+(\Psi)$ must compensate the imaginary
part of the leading NP correction,  
\begin{equation}
iM_1v^a{\rm e}^{-\frac{1}{v}}{\rm e}^{-L}.  
\end{equation}
Using (\ref{impart}) we can write
\begin{equation}
\Psi=F+K\Psi^{(-a)},  
\label{decomp}
\end{equation}
where the $\Psi^{(-a)}$ is an elementary asymptotic series defined in \eqref{exam}, which corresponds to a simple branch-cut structure on the Borel-plane, the type of which depends on the parameter $a$. Furthermore,
\begin{equation}
K=-\frac{M_1}{\pi}{\rm e}^{-L}
\end{equation}
and the Borel transform $\hat F(t)$ has no singularity at $t=1$ and must be
regular between $1$ and $k$. $\hat\Psi(t)$ itself was analytic between $-1$
and $1$ and has cuts starting there. $\hat\Psi^{(-a)}(t)$ is analytic for
$t<1$ and has a cut starting there. From (\ref{decomp}) we see that $\hat F(t)$
is analytic between $-1$ and $k$. We have numerically verified this behaviour
of $\hat F(t)$ for $N=6$ and $N=8$ to many digits.

We emphasize once more that our choice of ${\cal C}_+$ and the lateral Borel
resummation $S^+$ are related. Had we chosen ${\cal C}_-$ instead, $M_1$ would
change sign and we would need $S^-$. We will stick to ${\cal C}_+$ and $S^+$
throughout.



\section{The main conjecture}
\label{sect7}

In this section we again restrict our attention to the $N\geq4$ cases not
including the \mO{3} model, where the reduced energy density is of the form
\begin{equation}
\hat\epsilon=2\tilde W_{1,1}=\Psi+M\nu+2Y^{\rm T}{\cal L}Y.  
\end{equation}
Here $\Psi=2A_{1,1}$ is the perturbative part.

The main conjecture is about reconstructing the physical value from the
trans-series~$\hat\epsilon$. In the rest of the paper we will assume that
the physical value (given by the solution of the original TBA equation) can
be obtained by the lateral Borel sum of the trans-series:
\begin{center}
\framebox{$\epsilon_{\rm ph}=\epsilon_{\rm TBA}=S^+(\hat\epsilon)$}
\end{center}
We will analyse and study the consequences of this fundamental assumption.

First of all, we take care of the leading NP correction, $M\nu$. As explained
in section \ref{fej4}
\begin{equation}
{\rm Im}\,S^+\left(K\Psi^{(-a)}\right)=-M_1\nu
\label{M1}
\end{equation}
so we write
\begin{equation}
\epsilon_{\rm ph}=M_0\nu+{\rm Re}\,S^+(\Psi)+i{\rm Im}\,S^+(\Psi_{\rm sub})+
2S^+(Y^{\rm T}{\cal L}Y),  
\end{equation}
where
\begin{equation}
\Psi_{\rm sub}=\Psi-K\Psi^{(-a)}.
\label{psisub}
\end{equation}

All remaining singularities are of logarithmic type and we can use the theory
of alien  derivatives and use the explicit form of the Stokes automorphism, see also Appendix \ref{AppResurgence}.

\subsection{Stokes automorphism}
\label{ssectStokes}

The Stokes automorphism is related to the alien derivative sum 
\begin{equation}
{\cal R}=\sum_{\omega}{\rm e}^{-\omega z}\Delta^{({\rm st})}_\omega=
\sum_\omega \nu^\omega\Delta_\omega,  
\end{equation}
where $z=1/v$, and we replaced the standard definition of the alien derivative ($\Delta^{(\text{st})}_\omega$) via a modified one, that we denote simply as $\Delta_\omega$ (see Subsection \ref{sect6}). The reason for this was to handle the asymptotic serieses in terms of the running coupling $v$, while keeping the NP parameter $\nu={\rm e}^{-2B}$ instead of the standard ${\rm e}^{-z}$, as it is the one that frequently appeared in the WH solution. In the above definition 
\begin{equation}
\omega=jk,\qquad j=1,2,\dots  
\end{equation}

The Stokes automorphism, its inverse and its square root (for later use) are
given respectively as
\begin{equation}
\mathfrak{S}={\rm e}^{-{\cal R}},\qquad
{\mathfrak{S}}^{-1}={\rm e}^{{\cal R}},\qquad
{\mathfrak{S}}^{-1/2}={\rm e}^{{\cal R}/2}.
\end{equation}
The meaning of the Stokes automorphism is that it is a map between the two
lateral Borel resummations of an arbitrary asymptotic series $X$\footnote{Here and throughout this and the following sections $X$ stands for an arbitrary asymptotic series. More precisely, we will usually mean the perturbative part of any physical quantities that appear in Subsection \ref{reduced}. Not to be confused with the infinite matrix $X$ in \eqref{NDX}.}:
\begin{equation}\label{medianX}
S^-(X)=S^+({\mathfrak{S}}^{-1}X).
\end{equation}
For a real asymptotic series we have the relation 
\begin{equation}
2i{\rm Im}\,S^+(X)=S^+\left[(1-{\mathfrak{S}}^{-1})X\right].  
\end{equation}
Using this relation, we write
\begin{equation}
i{\rm Im}\,S^+(\Psi_{\rm sub})
=\frac{1}{2}S^+\left[(1-{\mathfrak{S}}^{-1})\Psi_{\rm sub}\right]    
=\frac{1}{2}S^+\left[(1-{\mathfrak{S}}^{-1})\Psi\right]    
=S^+\left[(1-{\mathfrak{S}}^{-1})A_{1,1}\right].    
\end{equation}
This relation holds because as explained in Section \ref{fej4} the Borel-transform $\hat\Psi^{(-a)}$
has no singularity at $jk$. This observation leads to the relation
\begin{equation}
\epsilon_{\rm ph}=M_0\nu+{\rm Re}\,S^+(2A_{1,1})+
S^+\left[(1-{\mathfrak{S}}^{-1})A_{1,1}\right]+2S^+(Y^{\rm T}{\cal L}Y).    
\end{equation}
Let us spell out the above relation in the first few NP orders. Starting from
\begin{equation}
1-{\mathfrak{S}}^{-1}=-{\cal R}-\frac{1}{2}{\cal R}^2
-\frac{1}{6}{\cal R}^3+\dots 
\end{equation}
and
\begin{equation}
Y^{\rm T}{\cal L}Y=Y^{\rm T}{\cal D}Y+Y^{\rm T}{\cal D}{\cal N}{\cal D}Y
+Y^{\rm T}{\cal D}{\cal N}{\cal D}{\cal N}{\cal D}Y+\dots
\end{equation}
we write the first few NP orders as
\begin{equation}
\epsilon_{\rm ph}=M_0\nu+{\rm Re}\,S^+(2A_{1,1})+S^+(p{\cal H}_1+p^2{\cal H}_2+
p^3{\cal H}_3+\dots),  
\end{equation}
where
\begin{equation}
\begin{split}
{\cal H}_1&=-\Delta_k A_{1,1}+2iS_1 A_{1,-k}^2,\\
{\cal H}_2&=-\Delta_{2k} A_{1,1}+2iS_2 A_{1,-2k}^2-\frac{1}{2}\Delta_k^2 A_{1,1}
-2S_1^2 A_{1,-k}^2 A_{-k,-k},\\
{\cal H}_3&=-\Delta_{3k} A_{1,1}+2iS_3 A_{1,-3k}^2-\frac{1}{2}
(\Delta_k\Delta_{2k}+\Delta_{2k}\Delta_k) A_{1,1}-\frac{1}{6}\Delta_k^3 A_{1,1}\\
&-4S_1S_2 A_{1,-k} A_{1,-2k}A_{-k,-2k}-2iS_1^3A_{1,-k}^2 A_{-k,-k}^2
\end{split}    
\end{equation}
and we introduced the shorthand
\begin{equation}
p=\nu^k.
\end{equation}

Let us start our considerations at O$(p)$. Since, according to our conjecture,
$\epsilon_{\rm ph}$ is real, the purely imaginary ${\cal H}_1$ must vanish and
the following relation must hold:
\begin{equation}
\Delta_k A_{1,1}=2iS_1 A_{1,-k}^2.  
\label{Delta1}
\end{equation}
This relation is of the form of the alien derivative lemma (\ref{lemma0}) and
using (\ref{lemma1}) ${\cal H}_2$ simplifies to
\begin{equation}
{\cal H}_2=-\Delta_{2k}A_{1,1}+2iS_2A_{1,-2k}^2+2S_1^2 A_{1,-k}^2 A_{-k,-k}.  
\end{equation}
Here the third term is real so we find the relation
\begin{equation}
\Delta_{2k} A_{1,1}=2iS_2 A_{1,-2k}^2.  
\label{Delta2}
\end{equation}
If we go up to third NP order, the lateral Borel sum of the remaining real
asymptotic series has also an imaginary part:
\begin{equation}
S^+(p^2{\cal H}_2)=2S_1^2p^2{\rm Re}\,S^+(A_{1,-k}^2 A_{-k,-k})-p^3 S_1^2
S^+(\Delta_k A_{1,-k}^2 A_{-k,-k}).  
\end{equation}
Using also this relation and the alien derivative lemma with (\ref{Delta1})
and (\ref{Delta2}), the O($p^3$) contribution simplifies to:
\begin{equation}\label{calHbar3}
{\cal H}_3=-\Delta_{3k}A_{1,1}+2iS_3 A_{1,-3k}^2+4S_1S_2A_{1,-k}A_{1,-2k}
A_{-k,-2k}.  
\end{equation}
Cancellation of the imaginary part gives
\begin{equation}
\Delta_{3k} A_{1,1}=2iS_3 A_{1,-3k}^2
\label{Delta3}
\end{equation}
and finally, up to O($p^3$) order, the physical density is
\begin{equation}
\epsilon_{\rm ph}=M_0\nu+{\rm Re}\,S^+(2A_{1,1}+2p^2 S_1^2 A_{1,-k}^2 A_{-k,-k}
+4p^3 S_1S_2 A_{1,-k}A_{1,-2k}A_{-k,-2k}+\dots)  
\end{equation}
It can be shown order by order (see appendix \ref{appC}) that for all
higher alien derivatives the
reality condition is equivalent to the relation
\begin{equation}
\Delta_{jk} A_{1,1}=2iS_j A_{1,-jk}^2.
\label{Deltaj}
\end{equation}

\subsection{Median resummation}
\label{subsect72}

The relations (\ref{Deltaj}) and the alien derivative lemma together allow to
write the physical densities in an elegant compact form with the help of the (inverse) Stokes-automorphism.
The square root of the latter generates a trans-series from an asymptotic series. When resummed via $S^+$, this provides the "middle ground" between the two lateral resummations $S^\pm$ of the original asymptotic series, and the result is automatically real. The fact that the trans-series we get this way also coincides with our WH result indicates that the PT part of the analyzed densities contains every information to reconstruct their exact physical values in case of the \mO{$N\geq 4$} models (except the real part of the $M\nu$ term).

We start from the identity
\begin{equation}
\begin{split}
({\cal R}Y)_s&={\cal R}A_{1,-sk}=\sum_{j=1}^\infty p^j\Delta_{jk} A_{1,-sk}=
\sum_{j=1}^\infty 2iS_j p^j A_{1,-jk}A_{-sk,-jk}\\
&=2\sum_{j=1}^\infty {\cal N}_{sj} {\cal D}_{jj} Y_j=2({\cal M}Y)_s. 
\end{split}    
\end{equation}
This can be written in matrix form as
\begin{equation}
{\cal R}Y=2{\cal M}Y  
\end{equation}
and for the transpose we have
\begin{equation}
{\cal R}Y^{\rm T}{\cal D}=2Y^{\rm T}{\cal M}^{\rm T}{\cal D}=  
2Y^{\rm T}{\cal D}{\cal N}{\cal D}=2Y^{\rm T}{\cal D}{\cal M}.
\end{equation}
Similarly,
\begin{equation}
{\cal R}{\cal M}=2{\cal M}^2,\qquad {\cal R}R=2{\cal M}R,\qquad
{\cal R}R^{\rm T}{\cal D}=2R^{\rm T}{\cal D}{\cal M}.  
\end{equation}
We now calculate
\begin{equation}
\begin{split}
{\cal R}&A_{1,1}=\sum_{j=1}^\infty p^j\Delta_{jk}A_{1,1}=\sum_{j=1}^\infty
2iS_j p^j A_{1,-jk}^2=2Y^{\rm T}{\cal D}Y,\\  
{\cal R}^2&A_{1,1}=8Y^{\rm T}{\cal D}{\cal M}Y,\\
&\vdots\\
{\cal R}^k&A_{1,1}=2^k k! Y^{\rm T}{\cal D}{\cal M}^{k-1}Y
\end{split}    
\end{equation}
and find
\begin{equation}
{\mathfrak{S}}^{-1/2} A_{1,1}=A_{1,1}+\sum_{k=1}^\infty \frac{1}{k!}
\left(\frac{\cal R}{2}\right)^k A_{1,1}=A_{1,1}+\sum_{k=1}^\infty Y^{\rm T}
{\cal D}{\cal M}^{k-1}Y=A_{1,1}+Y^{\rm T}{\cal L}Y.
\end{equation}
This way we have proved the relation
\begin{equation}
\hat\epsilon=M\nu+{\mathfrak{S}}^{-1/2}\Psi.  
\end{equation}

Using the definition of the median resummation,
\begin{equation}
S^{\rm med}(X)=S^+({\mathfrak{S}}^{-1/2}X)  
\end{equation}
we see that the physical density can compactly be written as
\begin{equation}
\epsilon_{\rm ph}=M\nu+S^{\rm med}(\Psi).  
\end{equation}
We see that in this family of models the perturbative series $\Psi$ \lq\lq
knows'' about all NP corrections (apart from the bulk energy constant term
$M\nu$).

The case of the particle density is completely analogous:
\begin{equation}
{\cal R}V_1=2Y^{\rm T}{\cal D}R,\qquad {\cal R}^kV_1=2^k k! Y^{\rm T}{\cal D}
{\cal M}^{k-1} R,\qquad
{\mathfrak{S}}^{-1/2}V_1=V_1+Y^{\rm T}{\cal L}R
\end{equation}
and finally we have
\begin{equation}
\hat\rho=\tilde U_1=V_1+R^{\rm T}{\cal L}Y={\mathfrak{S}}^{-1/2}V_1  
\end{equation}
and
\begin{equation}
\rho_{\rm ph}=S^+(\hat\rho)=S^{\rm med}(V_1).
\end{equation}

Finally for the last density we have\footnote{We included the $O_{0,0}$ density into our analysis as well, as for the \mO{3} case it will be related to a completely unrelated, yet historically interesting physical problem. Namely, to the determination of the circular parallel plate capacitor's capacitance \cite{Bajnok:2022rtu}.}
\begin{equation}
{\cal R}D_0=2R^{\rm T}{\cal D}R,\qquad {\mathfrak{S}}^{-1/2}D_0=D_0
+R^{\rm T}{\cal L}R=\tilde D,  
\label{cap1}
\end{equation}
\begin{equation}
D_{\rm ph}=S^+(\tilde D)=S^{\rm med}(D_0).
\label{cap2}
\end{equation}

\section{The \mO{3} model}
\label{sect8}

The \mO{3} model is more complicated than the $N\geq4$ \mO{$N$} models but it is
also the most interesting one. In this section we repeat the steps of Sections \ref{fej4}-\ref{sect7} for the $N=3$ case, and in Subsection \ref{ssectMainO3} we derive very similar alien derivative relations as in Subsection \ref{ssectStokes}, starting from the main conjecture again. Later on we dedicate a separate section (Section \ref{sectCheck}) to verify this result directly from the complex analytic structure of the Borel-transformed asymptotic series $A_{1,1}$ for the \mO{3} case. Following then Subsection \ref{subsect72} we prove that the full trans-series of the energy density may be generated similarily via the median resummation of asymptotic serieses in Subsection \ref{ssectMedianO3}. However, to arrive at the physical value in this case, we need to resum not only the PT part, but also two extra NP terms, suppressed by $\nu$ and $\nu^2$ respectively. For the particle density we will also need an extra, ${\rm O}(\nu)$ sector.  We associate these extra sectors to the effect of instantons following \cite{Marino:2022ykm} (see the Introduction \ref{sectIntro}). Although the fact that these extra sectors exist (contradicting \emph{strong resurgence} and implying a \emph{weak} one) were recognized in a series of papers \cite{Marino:2021dzn, Bajnok:2021zjm, Marino:2022ykm, Bajnok:2022rtu, Bajnok:2022xgx}, now we present also their exact form and their precise role in obtaining the physical values.

In the \mO{3} model
\begin{equation}
\kappa_j=2j,\qquad M=M_0+iM_1=-\frac{i\pi}{{\rm e}}+2\hatk (1-\gamma_E
-\ln2).  
\end{equation}
Here $\hatk=1/{\rm e}$ is defined by (\ref{sigmap1}). The
running coupling $v$ is introduced by
\begin{equation}
2B=\frac{1}{v}+\ln v-1-3\ln2.
\end{equation}
This is the \mO{3} case of the standard choice of \cite{Bajnok:2022xgx}.
The trans-series corresponding to the densities are found in section
\ref{sect3}:
\begin{equation}
\begin{split}
\hat\epsilon&=2{\tilde W}_{1,1}=M\nu+4B\nu\hatk+2A_{1,1}+4\nu\hatk P_{1,-1}
+2\nu^2 {\hatk}^2 A_{-1,-1}\\
&+2Y^{\rm T}{\cal L}Y
+4\nu \hatk Y^{\rm T}{\cal L}Z+2\nu^2 {\hatk}^2 Z^{\rm T}{\cal L}Z,\\
\hat\rho&={\tilde U}_1=V_1+\nu\hatk V_{-1}+
R^{\rm T}{\cal L}Y+\nu \hatk R^{\rm T}{\cal L}Z,\\
\tilde D&=D_0+R^{\rm T}{\cal L}R.
\end{split}
\label{trans123}
\end{equation}
Here
\begin{equation}
Y_j=A_{1,-2j},\qquad Z_j=A_{-1,-2j}, \qquad R_j=V_{-2j}.
\end{equation}
We also introduce
\begin{equation}
\tilde M=M+4B\hatk=iM_1+{\tilde M}_0,\quad
{\tilde M}_0=2\hatk(2B+1-\gamma_E-\ln2)=2\hatk\Big(
\frac{1}{v}+\ln v-\gamma_E-4\ln2\Big).       
\end{equation}
Later we will use the definitions
\begin{equation}
S^\pm(\ln v)=\ln v,\qquad \Delta_\omega \ln v=0.  
\end{equation}
The trans-series $\hat\epsilon$ given in (\ref{trans123}) contains all integer
powers of $\nu$:
\begin{equation}
\hat\epsilon=\sum_{m=0}^\infty \epsilon_m\nu^m.  
\end{equation}
Let us expand this up to quartic order explicitly:
\begin{equation}
\begin{split}
\epsilon_0&=2A_{1,1},\\
\epsilon_1&=iM_1+{\tilde M}_0+4\hatk P_{1,-1},\\
\epsilon_2&=2{\hatk}^2 A_{-1,-1}+2iS_1A_{1,-2}^2,\\
\epsilon_3&=4i\hatk S_1A_{1,-2}A_{-1,-2},\\
\epsilon_4&=2iS_2A_{1,-4}^2-2S_1^2 A_{1,-2}^2A_{-2,-2}
+2i{\hatk}^2S_1A_{-1,-2}^2.
\end{split}    
\end{equation}

\subsection{The main conjecture}
\label{ssectMainO3}

In this subsection we derive the consequences of our main conjecture, which is
the same as for the $N\geq4$ models:
\begin{equation}
\epsilon_{\rm TBA}=\epsilon_{\rm ph}=S^+(\hat\epsilon).  
\end{equation}
We will use the alien derivative operators
\begin{equation}
{\cal R}=\sum_{j=1}^\infty \nu^j\Delta_j,\qquad {\mathfrak{S}}^{-1}
={\rm e}^{\cal R},  
\qquad {\mathfrak{S}}^{-1/2}={\rm e}^{{\cal R}/2}  
\end{equation}
and the relation (valid for real $X$)
\begin{equation}
S^+(X)={\rm Re}\,S^+(X)+\frac{1}{2}S^+\left\{(1-{\mathfrak{S}}^{-1})X\right\}.  
\end{equation}
With the help of the above relation we write (working up to quartic order) 
\begin{equation}
\begin{split}
S^+(\epsilon_0)&={\rm Re}\,S^+(2A_{1,1})-S^+\Big\{\big[\nu\Delta_1    
+\nu^2\Delta_2+\nu^3\Delta_3+\nu^4\Delta_4+\frac{\nu^2}{2}\Delta_1^2\\
&+\frac{\nu^3}{2}(\Delta_1\Delta_2+\Delta_2\Delta_1)+\frac{\nu^4}{2}
(\Delta_1\Delta_3+\Delta_3\Delta_1)+\frac{\nu^4}{2}\Delta_2^2+\frac{\nu^3}{6}
\Delta_1^3\\
&+\frac{\nu^4}{6}(\Delta_1^2\Delta_2+\Delta_1\Delta_2\Delta_1+\Delta_2\Delta_1^2)
+\frac{\nu^4}{24}\Delta_1^4\big]A_{1,1}\Big\}+{\rm O}(\nu^5),\\
\nu S^+(\epsilon_1)&=iM_1\nu+{\tilde M}_0\nu
+4\nu\hatk{\rm Re}\,S^+(P_{1,-1})-2\nu\hatk S^+\Big\{\big[\nu\Delta_1    
+\nu^2\Delta_2+\nu^3\Delta_3+\frac{\nu^2}{2}\Delta_1^2\\
&+\frac{\nu^3}{2}(\Delta_1\Delta_2+\Delta_2\Delta_1)+\frac{\nu^3}{6}
\Delta_1^3\big]P_{1,-1}\Big\}+{\rm O}(\nu^5),\\
\nu^2S^+(\epsilon_2)&=2\nu^2S^+\big\{{\hatk}^2A_{-1,-1}+iS_1A_{1,-2}^2\big\}.
\end{split}    
\end{equation}
At lowest order, O$(\nu)$, the imaginary contribution vanishes if
\begin{equation}
S^+\{-\Delta_1 A_{1,1}\}+iM_1={\rm O}(\nu).
\end{equation}
This means that for these purely imaginary quantities
\begin{equation}
\Delta_1 A_{1,1}=iM_1
\end{equation}
must hold and since $M_1$ is constant, from the considerations in appendix
\ref{appB} we conclude that 
\begin{equation}
\Delta_1 X=0\quad (X\not=A_{1,1}),
\label{Delt1}  
\end{equation}
where $X$ may be any other perturbative quantity like $A_{n,m}, P_{n,-n}, V_n, D_0, a_n, s$, except $A_{1,1}$.

Similarly, at O$(\nu^2)$, we have to require that
\begin{equation}
S^+\big\{-\Delta_2 A_{1,1}+2iS_1A_{1,-2}^2\big\}={\rm O}(\nu),
\end{equation}
but since $S^+$ acts here on a purely imaginary quantity, the relation 
\begin{equation}
\Delta_2 A_{1,1}=2iS_1A_{1,-2}^2
\end{equation}
follows. Now from the alien derivative relations derived in appendix \ref{appB}
we have
\begin{equation}
\Delta_2 A_{n,m}=2iS_1A_{n,-2}A_{m,-2}\quad (n,m\not=2)
\label{Delt2}  
\end{equation}
and analogous relations for other quantities. Using cancellations and
simplifications following from (\ref{Delt1}) and (\ref{Delt2}) we find
\begin{equation}
\begin{split}
S^+(\epsilon_0)&\Rightarrow{\rm Re}\,S^+(2A_{1,1})-S^+\Big\{\big[    
\nu^3\Delta_3+\nu^4\Delta_4+\frac{\nu^4}{2}
\Delta_1\Delta_3\big]A_{1,1}\Big\}\\
&+4S_1^2\nu^4S^+(A_{1,-2}^2A_{-2,-2})+{\rm O}(\nu^5),\\
\nu S^+(\epsilon_1)&\Rightarrow \nu{\tilde M}_0
+4\nu\hatk{\rm Re}\,S^+(P_{1,-1})-2\nu^4\hatk S^+\big(    
\Delta_3P_{1,-1}\big)\\
&-4iS_1\hatk\nu^3S^+(A_{1,-2}A_{-1,-2})+{\rm O}(\nu^5),\\
\nu^2S^+(\epsilon_2)&\Rightarrow
2\nu^2{\hatk}^2{\rm Re}\,S^+(A_{-1,-1})-2i{\hatk}^2S_1\nu^4S^+(A_{-1,-2}^2)
+{\rm O}(\nu^5).
\end{split}    
\end{equation}
After these simplifications we see that at O$(\nu^3)$ we have to require
\begin{equation}
S^+\{-\Delta_3 A_{1,1}\}={\rm O}(\nu),
\end{equation}
which is satisfied only if
\begin{equation}
\Delta_3 A_{1,1}=0
\end{equation}
and consequently for any other quantity $X$ (in the sense of Appendix \ref{appB} again)
\begin{equation}
\Delta_3 X=0\quad (X\not=A_{3,3}).
\label{Delt3}  
\end{equation}
Finally, at O$(\nu^4)$, we have
\begin{equation}
S^+\big\{-\Delta_4 A_{1,1}+2iS_2A_{1,-4}^2\big\}={\rm O}(\nu).
\end{equation}
Again, this means that the purely imaginary quantity itself vanishes: 
\begin{equation}
\Delta_4 A_{1,1}=2iS_2A_{1,-4}^2
\end{equation}
and consequently
\begin{equation}
\Delta_4 A_{n,m}=2iS_2A_{n,-4}A_{m,-4}\quad (n,m\not=2)
\label{Delt4}  
\end{equation}
and $\Delta_4$ acts similarly on all other quantities.

We have thus demonstrated that our main conjecture implies\footnote{So far
we have used only the fact that from the main conjecture the reality of
$\epsilon_{\rm ph}=S^+(\hat\epsilon)$ follows.} the relations (\ref{Delt1}),
(\ref{Delt2}), (\ref{Delt3}), and (\ref{Delt4}). We see that up to quartic
order $\epsilon_{\rm ph}$ is manifestly real:
\begin{equation}
S^+(\hat\epsilon)=\nu{\tilde M}_0+{\rm Re}\, S^+\big\{2A_{1,1}+4\nu\hatk P_{1,-1}
+2\nu^2{\hatk}^2 A_{-1,-1}+2\nu^4S_1^2A_{1,-2}^2A_{-2,-2}\big\}+{\rm O}(\nu^5).
\end{equation}
Going to higher orders we can show, step by step, that for all $j=1,2,\dots$
\begin{equation}
\Delta_{2j}A_{1,1}=2iS_jA_{1,-2j}^2\quad\Rightarrow\quad
\Delta_{2j}A_{n,m}=2iS_jA_{n,-2j}A_{m,-2j}\quad(n,m\not=2j)
\label{alien2j}
\end{equation}
and
\begin{equation}
\Delta_{2j-1}A_{1,1}=iM_1\delta_{j,1} \quad\Rightarrow\quad
\Delta_{2j-1}X=0\quad (X\not=A_{2j-1,2j-1}).
\label{alien2j1}
\end{equation}
($\Delta_{2j}$ acts on other quantities similarly to (\ref{alien2j}).)


\subsection{Median resummation}
\label{ssectMedianO3}

In this subsection  we will use and generalize the results of
subsection \ref{subsect72}. The
even part of the alien derivative operator ${\cal R}$ will be denoted
by $\bar{\cal R}$ in this subsection:
\begin{equation}
\bar{\cal R}=\sum_{j=1}^\infty \nu^{2j}\Delta_{2j}.
\end{equation}
The action of the total ${\cal R}$ on $A_{1,1}$ is
\begin{equation}
{\cal R}A_{1,1}=iM_1\nu+\bar{\cal R}A_{1,1}=iM_1\nu+2Y^{\rm T}{\cal D}Y
\end{equation}
and for higher powers we have
\begin{equation}
{\cal R}^n A_{1,1}={\bar{\cal R}}^nA_{1,1} \quad n\geq2.
\end{equation}
Thus we find
\begin{equation}
{\mathfrak{S}}^{-1/2}A_{1,1}=\frac{iM_1\nu}{2}+A_{1,1}+Y^{\rm T}{\cal L}Y.
\end{equation}
For $P_{1,-1}$ we have
\begin{equation}
{\cal R}^n P_{1,-1}={\bar{\cal R}}^nP_{1,-1} \quad n=1,2,\dots,\quad
\bar{\cal R}P_{1,-1}=2Y^{\rm T}{\cal D}Z
\end{equation}
and using the relation
\begin{equation}
\bar{\cal R}Z=2{\cal M}Z
\end{equation}
we can show that
\begin{equation}
{\mathfrak{S}}^{-1/2}P_{1,-1}=P_{1,-1}+Y^{\rm T}{\cal L}Z.
\end{equation}
Similarly for $A_{-1,-1}$ we find
\begin{equation}
\bar{\cal R}A_{-1,-1}=2Z^{\rm T}{\cal D}Z,\qquad
{\mathfrak{S}}^{-1/2}A_{-1,-1}=A_{-1,-1}+Z^{\rm T}{\cal L}Z.
\end{equation}
Let us introduce
\begin{equation}
f^{(0)}=2A_{1,1},\qquad
f^{(1)}={\tilde M}_0+4\hatk P_{1,-1},\qquad
f^{(2)}=2{\hatk}^2A_{-1,-1}.
\end{equation}
With this notation
\begin{equation}
\hat\epsilon={\mathfrak{S}}^{-1/2}\left(f^{(0)}+\nu f^{(1)}+\nu^2 f^{(2)}\right).
\end{equation}
This relation can be used to prove (\ref{alien2j}) and (\ref{alien2j1})
recursively. We can copy the proof given in appendix \ref{appC} after
noting that starting from order $m-1$ only $f^{(0)}$ contains new quantities
at the next order because $\nu f^{(1)}$ and $\nu^2 f^{(2)}$ are already of the
required form up to order $m$ and $m+1$, respectively.

The case of the particle density is analogous. Here we have
\begin{equation}
\bar{\cal R}V_{-1}=2R^{\rm T}{\cal D}Z,\qquad
{\mathfrak{S}}^{-1/2}V_{-1}=V_{-1}+R^{\rm T}{\cal L}Z.
\end{equation}
Introducing
\begin{equation}
\rho^{(0)}=V_1,\qquad
\rho^{(1)}=\hatk V_{-1}
\end{equation}
we find
\begin{equation}
\hat\rho={\mathfrak{S}}^{-1/2}\left(\rho^{(0)}+\nu \rho^{(1)}\right).
\end{equation}
Finally
\begin{equation}
\epsilon_{\rm ph}=S^+(\hat\epsilon)=
S^{\rm med}\left(f^{(0)}+\nu f^{(1)}+\nu^2 f^{(2)}\right)
\end{equation}
and
\begin{equation}
\rho_{\rm ph}=S^+(\hat\rho)=
S^{\rm med}\left(\rho^{(0)}+\nu \rho^{(1)}\right).
\end{equation}
In this model the perturbative solution, $f^{(0)}$, in itself is not
sufficient to reproduce the full physical result because also $f^{(1)}$ and
$f^{(2)}$ are needed.
Since the \mO{3} model is exceptional also because it is the only one that has
stable instanton solutions, it is tempting to regard $f^{(i)}$ as the
$i$-instanton contribution to the density. If this is indeed the case then
we can say that the perturbative solution plus the perturbative expansions
around the one and two instanton solutions together determine the physical
energy density completely. We do not understand why only these two instanton
sectors matter and also do not understand why only the one instanton sector
contributes for the particle density $\rho_{\rm ph}$.
See, however, (\ref{XX}), which shows that infinitely many instanton sectors
are present in the free energy.

Interestingly, for the capacity\footnote{Here we refer to the reduced version $D$ of the density $O_{0,0}$ related to the capacitance of the circular parallel plate capacitor (see the introduction of \cite{Bajnok:2022rtu} for historical summary). There the parameter $B$ is directly related to the geometry of the setting (to the ratio of the distance and the radii of the plates), and thus the running coupling $v$ and the NP parameter $\nu$ may be regarded directly as a physical parametrization of the trans-series.} the formulas (\ref{cap1}) and (\ref{cap2})
are also valid here (with ${\cal R}\to\bar{\cal R}$) so there is no separate
instanton contribution for the capacity.

\section{Checking the trans-series for the energy density in the \mO{3} model}
\label{sectCheck}

The derivation of our results for the \mO{3} alien derivatives (\ref{alien2j}) was based on the main conjecture. Although there is little doubt
about the correctness of these formulas: they lead to elegant and compact formulas for the full trans-series and in addition (see section~\ref{sect13})
we have numerically checked them to a lot of decimal places. Nevertheless, since the original definition of alien derivatives is based on the study of 
the singularity structure of the Borel transform of the original perturbative series, it is desirable to verify (\ref{alien2j}) directly by studying
the asymptotic behaviour of the coefficients of the perturbative series. In the previous studies \cite{Abbott:2020qnl, Bajnok:2021zjm, Bajnok:2022rtu} the authors typically used repeated asymptotic analysis of the asymptotic coefficients to calculate alien derivatives corresponding to the cuts on the Borel-plane further away from the closest singularity to the origin. This method required either a large number of perturbative coefficients (for the \mO{4} case) or - if obtaining that much data was not an option as in case of the \mO{3} model - it was not performing well enough in reconstructing these alien derivatives. With the procedure presented in the current section however, we were able to verify the alien derivative relations convincingly up to the second closest cut, and did so solely from the limited amount of perturbative data we had for the \mO{3} model.

To this end, in this section we investigate the analytical structure of the energy
density 
\begin{equation}
{O}_{1,1}=\frac{1}{4\pi}G_{+}^{2}(i){\rm e}^{2B}2W_{1,1}
\end{equation}
on the Borel plane based on the perturbative expansion of $W_{1,1}(v)$:
\begin{equation}
A_{1,1}(v)=\sum_{n=0}^{\infty}\alpha_{n}v^{n}
\end{equation}
 in the running coupling $2B=\frac{1}{v}+\ln v-1-3\ln2$. 

These perturbative coefficients can be obtained by implementing Volin's
algorithm \cite{Volin:2009wr,Volin:2010cq} to the \mO{3} model. In the paper \cite{Bajnok:2021zjm} we obtained
336 perturbative coefficients numerically with high precisions and
now we use these data to check the following part of the trans-series
solution
\begin{equation}
\mathfrak{S}^{-1/2}(A_{1,1})=A_{1,1}+i\frac{M_1}{2}\nu+iS_{1}A_{1,-2}^{2}\nu^{2}+(iS_{2}A_{1,-4}^{2}-S_{1}^{2}A_{1,-2}^{2}A_{-2,-2})\nu^{4}+\dots
\end{equation}
 where $\nu=\frac{8e}{v}{\rm e}^{-\frac{1}{v}}$ and $S_{n}=2{\rm e}^{-2n}n^{2n-1}\Gamma_{n}^{-2}$.
We thus check the alien derivatives\footnote{Note that the coefficients of a series might grow faster than $\Gamma(n)$ as assumed in the definition \eqref{eq:aliender} of the standard alien derivative, that is, as $\Gamma(n+\gamma)$, where $\gamma > 0$ is an integer. One has to use \eqref{eq:alienderSplus} and \eqref{impart} to extend the formula \eqref{eq:aliender} to these cases. Then the alien derivative will contain negative powers of $v$, which is exactly the case here. } 
\begin{align}
\Delta^{(\text{st})}_{1}A_{1,1} & = i M_1 (8e/v)\quad;\qquad\Delta^{(\text{st})}_{2}A_{1,1}=2 i S_{1}(8e/v)^{2}A_{1,-2}^{2}\\
\Delta^{(\text{st})}_{4}A_{1,1} & =2 i S_{2}(8e/v)^{4}A_{1,-4}^{2}\quad;\quad\Delta^{(\text{st})}_{2}\Delta^{(\text{st})}_{2}A_{1,1}= -8(8e/v)^{4}S_{1}^{2}A_{1,-2}^{2}A_{-2,-2}
\end{align}
This can be done by investigating the asymptotic behaviour of the
perturbative coefficients
\begin{align}
\alpha_{n}=\alpha\Gamma_{n+1} & +2^{-n}(b_{0}\Gamma_{n+2}+2b_{1}\Gamma_{n+1}+\dots+2^{k}b_{k}\Gamma_{n+2-k}+\dots) \nonumber \\
 & +4^{-n}(c_{0}\Gamma_{n+4}+4c_{1}\Gamma_{n+3}+\dots+4^{k}c_{k}\Gamma_{n+4-k}+\dots)
\end{align}
The leading coefficient $\alpha$ is related to $\Delta^{(\text{st})}_{1}A_{1,1}$ as
$\Delta^{(\text{st})}_{1}A_{1,1}=-2 \pi i \alpha/v$, while the coefficients $-8\pi i b_{n}$
and $-512 \pi i c_{n}$ are the perturbative expansions of $\Delta^{(\text{st})}_{2}A_{1,1}$
and $\Delta^{(\text{st})}_{4}A_{1,1}$, respectively. Typically the coefficients
$b_{n}$ also grow as 
\begin{equation}
\frac{b_{n}}{b_{0}}=2^{-n}(d_{0}\Gamma_{n+2}+2d_{1}\Gamma_{n+1}+\dots+2^{k}d_{k}\Gamma_{n+2-k}+\dots)+\dots
\end{equation}
 and the coefficients $-64\pi^{2}b_{0}d_{n}$ correspond to the expansion
of $\Delta_{2}^{2\,(\text{st})}A_{1,1}$. Since the asymptotic contribution of
the $b_{n}$s are mixing with the $c_{n}$s it is highly non-trivial
to separate $\Delta^{(\text{st})}_{4}A_{1,1}$ from $\Delta_{2}^{2\,(\text{st})}A_{1,1}$. We
are going to do it by manipulating various Borel transformed function. 

In the following we first introduce the first few terms of the building
blocks, $A_{n,m}$. We then perform an asymptotic analysis and extract
the $\alpha$ and $b_{n}$ coefficients using high order Richardson transforms.
Finally, we construct various Borel transformed functions and extract
the $b_{n},c_{n}$ and $d_{n}$ coefficients from their analytical
properties. We find complete agreement in all the cases. 

\subsection{The basis of perturbative functions}

In this subsection we recall how the perturbative basis $A_{n,m}$
can be constructed explicitly. For demonstrations we present the result
for the first four orders in $v$ only, but it can be easily extended
to higher orders.

We start from the perturbative part of the energy density $A_{1,1}(v)$,
which has the expansion
\begin{equation}
2A_{1,1}=1+\frac{v}{2}+\frac{19v^{2}}{8}+\frac{649v^{3}}{48}+\left(\frac{33277}{384}-\frac{27\zeta_3}{16}\right)v^{4}+O\left(v^{5}\right)
\end{equation}
We then use the differential equation 
\begin{equation}
(n+m)A_{n,m}+\dot{A}_{n,m}=(n+m)A_{n,m}+\frac{2v^{2}}{1-v}\frac{dA_{n,m}}{dv}=a_{n}a_{m}
\end{equation}
for $n=m=1$ to obtain the perturbative part of $w_{1}$: 
\begin{equation}
a_{1}=1+\frac{v}{4}+\frac{29v^{2}}{32}+\frac{1501v^{3}}{384}+\left(\frac{116953}{6144}-\frac{27\zeta_3}{32}\right)v^{4}+O\left(v^{5}\right)
\end{equation}
The perturbative part of the universal function $f$ follows from
\begin{equation}
f=2n\frac{\dot{a}_{n}}{a_{n}}+\frac{\ddot{a}_{n}}{a_{n}}\label{eq:diffeqf}
\end{equation}
once used it for $n=1$: 
\begin{equation}
f=-v^{2}-6v^{3}-26v^{4}+\frac{1}{2}(27\zeta_3-193)v^{5}+O\left(v^{6}\right)
\end{equation}
By using again eq. (\ref{eq:diffeqf}) for generic $n$ we can determine
the perturbative part of $w_{n}$:
\begin{align}
a_{n}=1 & +\frac{v}{4n}+\frac{(20n+9)v^{2}}{32n^{2}}+\frac{\left(640n^{2}+636n+225\right)v^{3}}{384n^{3}} \nonumber \\
 & +\frac{v^{4}\left(-576(9\zeta_3-47)n^{3}+43696n^{2}+35160n+11025\right)}{6144n^{4}}+O\left(v^{5}\right)
\end{align}
Finally, using the differential equation for $A_{n,m}$ we can obtain
\begin{align}
A_{n,m} & =\frac{1}{m+n}+\frac{v}{4mn}+\frac{v^{2}(20mn+9m+9n)}{32m^{2}n^{2}}+\nonumber\\
 & \frac{v^{3}\left(m^{2}\left(640n^{2}+636n+225\right)+6mn(106n+39)+225n^{2}\right)}{384m^{3}n^{3}}\\
 & +\frac{v^{4}\left(m^{3}\left(8n\left(72(47-9\zeta_3)n^{2}+5462n+4395\right)+11025\right)\right)}{6144m^{4}n^{4}}\nonumber\\
 & +\frac{v^{4}\left(m^{2}n(16n(2731n+2277)+11475)+15mn^{2}(2344n+765)+11025n^{3}\right)}{6144m^{4}n^{4}}+O\left(v^{5}\right)\nonumber
\end{align}
In checking the first few alien derivatives we need to compare their numerically measured coefficients to the expansions below:
\begin{equation}
A_{1,-2}^{2}=1+\frac{v}{4}+\frac{25v^{2}}{32}+\frac{1159v^{3}}{384}+\left(\frac{84577}{6144}-\frac{27\zeta_3}{32}\right)v^{4}+O\left(v^{5}\right)
\end{equation}

\begin{equation}
A_{1,-4}^{2}=\frac{1}{9}+\frac{v}{24}+\frac{55v^{2}}{384}+\frac{5431v^{3}}{9216}+\left(\frac{90931}{32768}-\frac{9\zeta_3}{64}\right)v^{4}+O\left(v^{5}\right)
\end{equation}

\begin{equation}
A_{1,-2}^{2}A_{-2,-2}=-\frac{1}{4}-\frac{3v^{2}}{32}-\frac{73v^{3}}{128}-\frac{381v^{4}}{128}+O\left(v^{5}\right)
\end{equation}
together with the structure constants: 
\begin{equation}
M_1=-\frac{\pi}{e}\quad;\quad S_{1}=\frac{2}{{\rm e}^{2}}\quad;\quad S_{2}=\frac{16}{{\rm e}^{4}}
\end{equation}

\subsection{Asymptotic analysis}

Using high order Richardson transformation one can investigate the
large $n$ behaviour of the $\alpha_{n}/\Gamma_{n+1}$ series and extract
that 
\begin{equation}
\alpha=4
\end{equation}
with 95 digits precision. Taking into account the value of $M$ with
$\nu$ we can see complete agreement, i.e. $\Delta^{(\text{st})}_{1}A_{1,1}=- 2\pi i \alpha$. 

By subtracting the leading term and inspecting the asymptotics of
$(\alpha_{n}-\alpha\Gamma_{n+1})2^{n}/\Gamma_{n+2}$ we can again use high order
Richardson transform to extract 
\begin{equation}
b_{0}=-\frac{32}{\pi}\quad;\quad\frac{b_{1}}{b_{0}}=\frac{1}{4}\quad;\quad\frac{b_{2}}{b_{0}}=\frac{25}{32}\quad;\quad\frac{b_{3}}{b_{0}}=\frac{1159}{384}\quad;\quad\frac{b_{4}}{b_{0}}=\frac{84577}{6144}-\frac{27\zeta_{3}}{32}
\end{equation}
with precisions of ($39-29$) digits. We could go to higher orders,
but for demonstrative reasons it is enough to stop here. Already at
this level we can see the match with the alien derivative $\Delta^{(\text{st})}_{2}A_{1,1}= 256 i A_{1,-2}^{2}/v^2$. 

Technically, it is very difficult to extract enough $b_{n}$ coefficients
to investigate their asymptotic behaviour. Nevertheless, by performing
an asymptotic analysis on $A_{1,-2}^{2}$ we can determine the coefficients
$d_{n}$ as
\begin{equation}
d_{0}=\frac{16}{\pi}\quad;\quad\frac{d_{1}}{d_{0}}=0\quad;\quad\frac{d_{2}}{d_{0}}=\frac{3}{8}\quad;\quad\frac{d_{3}}{d_{0}}=\frac{73}{32}\quad;\quad\frac{d_{4}}{d_{0}}=\frac{381}{32}
\end{equation}
 Their contribution is actually mixing with the subleading $4^{-n}$
behaviour preventing to extract the $c_{n}$ coefficients at all.
In order to avoid this complication we switch to work directly on
the Borel plane where we can disentangle the contributions of the
various sub-divergent behaviours. 

\subsection{The Borel plane with square root cuts}

Our calculation will be based on \cite{Aniceto:2018uik,Costin:2020pcj}. From the asymptotic coefficients
$\alpha_{n}$ (after subtracting the trivial leading behaviour) we
can define the generalized Borel transforms
\begin{equation}
B_{k}(s)=\sum_{n=0}^{\infty}\frac{(\alpha_{n}-4\Gamma_{n+1})}{\Gamma_{n+k+1}}s^{n+k}=s^{k}\hat{B}_{k}(s)
\end{equation}
 This function has the closest cut to the origin starting at $s=2$
and the next closest at $s=4$. The nature of the cut depends on $k$.
For $k$ integer the cuts are logarithmic and the functions multiplying
the logarithms are directly related to the alien derivatives, see \cite{Dorigoni:2014hea}
for details. For numerical investigations, however it is advantageous
to take $k$ half integer as in this case the cuts are of square root
types. In the following we analyse the $s=2$ cut with $k=\frac{3}{2}$
and the $s=4$ cut with $k=\frac{7}{2}$. 

\subsubsection{The cut at $s=2$}

Let us subtract the leading contribution and switch to the Borel plane
with $k=\frac{3}{2}$. The closest singularity of $B_{\frac{3}{2}}(s)$
to the origin is a square root type cut with singular behaviour 
\begin{equation}
B_{\frac{3}{2}}(s)=\frac{4\pi}{\sqrt{2-s}}\left(\frac{b_{0}}{\Gamma_{\frac{1}{2}}}+\frac{b_{1}(s-2)}{\Gamma_{\frac{3}{2}}}+\dots+\frac{b_{k}(s-2)^{k}}{\Gamma_{k+\frac{1}{2}}}+\dots\right)+\mathrm{regular}
\end{equation}
 where the regular piece has a Taylor expansion in $(s-2)$, i.e.
it does not have any square roots. 

Following the idea of \cite{Costin:2020pcj} we open up the cut by introducing
$s(z)=2-z^{2}$ and expand around $z=0$, which leads to 
\begin{equation}
z B_{\frac{3}{2}}(2-z^{2})=4\pi\left(\frac{b_{0}}{\Gamma_{\frac{1}{2}}}-\frac{b_{1}z^{2}}{\Gamma_{\frac{3}{2}}}+\dots+\frac{(-1)^{k}b_{k}z^{2k}}{\Gamma_{k+\frac{1}{2}}}+\dots\right)+\textrm{odd powers in }z
\end{equation}
Thus the singular and regular parts in $s$ can be separated by focusing
on the odd and even powers in $z$, respectively. 

This is particularly useful in the practical implementation as we
know only the first $336$ perturbative coefficients numerically.
We thus calculate the Borel transform upto order $336$ and prepare
the Pade approximant of the corresponding $\hat{B}_{\frac{3}{2}}(s)$,
which we denote by $P\!\hat{B}_{\frac{3}{2}}(s)$. We then open up
the cut by introducing $s(w)=2-(1-w)^{2}$, such that $w=0$ corresponds
to $s=1$, where the original perturbative series is convergent and
the Pade approximant is reliable.\footnote{This intermediate expansion point $s=1$ was introduced for the following reason. The Borel-transform has a cut starting at $s=2$, and this limits the convergence radius of the Pade approximant, as the cut manifests itself as a condensation of the poles of the approximant starting around $s=2$. Thus, when re-expanding directly around $s=2$, even after the mapping $s \to w$, these singularities would render the expansion unreliable (i.e. they will not disappear in the same way the branchcut of the exact Borel transform itself would disappear as a result of the mapping). The additional Pade approximant in $w$ around a point where the original complex function was convergent captures the analytic structure of the $w$-plane (where the cut is opened) better.} After this we expand 
\begin{equation}\label{Padesw}
(1-w)s(w)^{\frac{3}{2}}P\negthinspace\hat{B}_{\frac{3}{2}}(s(w))
\end{equation}
 around $w=0$ in $w$ up to $336$ order and prepare its Pade approximant,
which we expand around $w=1$. As a result we could extract the same
$b_{n}$ coefficients as before, but with much higher ($60-50$ digits)
precisions than in the asymptotic analysis. 

\subsubsection{The cut at $s=4$}

In investigating the cut at $s=4$ it is more natural to analyse the
$B_{\frac{7}{2}}(s)$ Borel function. The two functions are related
to each other by differentiation $d^{2}B_{\frac{7}{2}}(s)/ds^{2}=B_{\frac{3}{2}}(s)$.
The closest singularity of $B_{\frac{7}{2}}(s)$ to the origin is
still a square root type cut at $s=2$ containing the contribtutions
of the $b_{n}$ coefficients as
\begin{equation}
B_{\frac{7}{2}}(s)= - 4\pi \sqrt{2-s}\left(\frac{b_{0}(s-2)}{\Gamma_{\frac{5}{2}}}+\frac{b_{1}(s-2)^{2}}{\Gamma_{\frac{7}{2}}}+\dots+\frac{b_{k}(s-2)^{k+1}}{\Gamma_{k+\frac{5}{2}}}+\dots\right)+\mathrm{regular}
\end{equation}
where we assumed that $s$ is larger than $2$ as we analyze the contribution
of this function near $s=4$. In doing so we need the asymptotical
behaviour of the $b_{n}$ coefficient and use them in
\begin{equation}
B_{\frac{7}{2}}(s)= -4\pi  \frac{\sqrt{2-s}}{\sqrt{s-2}} \sum_{n=0}^{\infty}\frac{b_{n}(s-2)^{n+\frac{3}{2}}}{\Gamma_{n+\frac{5}{2}}}+\dots
\end{equation}
But this is exactly the $k=\frac{3}{2}$ Borel transformed function
with coefficients $b_{n}$ leading to the their contributions to the
$s\sim4$ behaviour as 
\begin{equation}
B_{\frac{7}{2}}(s)= - \frac{\sqrt{2-s}}{\sqrt{s-2}}\frac{16\pi^{2}b_{0}}{\sqrt{4-s}}\left(\frac{d_{0}}{\Gamma_{\frac{1}{2}}}+\frac{d_{1}(s-4)}{\Gamma_{\frac{3}{2}}}+\dots+\frac{d_{k}(s-4)^{k}}{\Gamma_{k+\frac{1}{2}}}+\dots\right)+\dots
\end{equation}
This is not all the square root type behaviour at $s\sim4$ as
we also have the direct contribution of the $c_{n}$ coefficients:
\begin{equation}
B_{\frac{7}{2}}(s)=\frac{256\pi}{\sqrt{4-s}}\left(\frac{c_{0}}{\Gamma_{\frac{1}{2}}}+\frac{c_{1}(s-4)}{\Gamma_{\frac{3}{2}}}+\dots+\frac{c_{k}(s-4)^{k}}{\Gamma_{k+\frac{1}{2}}}+\dots\right)+\dots
\end{equation}
The stark difference between these contributions are in their reality
properties, which enables us to disentangle the $c$ and $d$ coefficients. 

In investigating the square root cut at $s=4$ we would like to do
the same trick as we did before to open up that cut. Unfortunately
the cut starting at $s=2$ overlaps with this cut, thus we should
disentangle the two cuts first. This can be done by introducing a
conformal map as 
\begin{equation}
s(z)=2\frac{4z}{(1+z)^{2}}
\end{equation}
This map resolves the cut at $s=2$ and maps the cut at $s=4$ to
$z=\pm i$. We can rotate this cut back to the real line using 
\begin{equation}
z(w)=-iw
\end{equation}
in order to be able to see clearly the real and imaginary parts. (The prefactor $\frac{\sqrt{2-s}}{\sqrt{s-2}}$ after this map for real $w$ will give $i$).
We then resolve the cut at $w=1$ by introducing $w(t)=1-t^{2}$, expanding
around $t=0$ and focusing on the odd powers. 

Technically, we prepare the Pade approximant of $\hat{B}_{\frac{7}{2}}(s(z))$,
which we denote as \newline $P\!\hat{B}_{\frac{7}{2}}(s(z(w)))$ and desingularize
the cut at $w=1$ by introducing 
\begin{equation}
w(t)=1-(t-2^{-1/2})^{2}
\end{equation}
We then expand
\begin{equation}
(t-2^{-1/2})s(z(w(t)))^{\frac{7}{2}}P\!\hat{B}(s(z(w(t)))
\end{equation}
around $t=0$ and prepare its Pade approximant, which we expand around
$t=2^{-\frac{1}{2}}$. From the even terms in the expansion we can
recalculate the corresponding $d_{n}$ and $c_{n}$ coefficients as
\begin{equation}
c_{0}=-\frac{256}{9\pi}\quad;\quad\frac{c_{1}}{c_{0}}=\frac{3}{8}\quad;\quad\frac{c_{2}}{c_{0}}=\frac{165}{128}\quad;\quad\frac{c_{3}}{c_{0}}=\frac{5431}{1024}\quad;\quad\frac{c_{4}}{c_{0}}=\frac{818379}{32768}-\frac{81\zeta_{3}}{64}
\end{equation}
 
\begin{equation}
d_{0}=\frac{16}{\pi}\quad;\quad\frac{d_{1}}{d_{0}}=0\quad;\quad\frac{d_{2}}{d_{0}}=\frac{3}{8}\quad;\quad\frac{d_{3}}{d_{0}}=\frac{73}{32}\quad;\quad\frac{d_{4}}{d_{0}}=\frac{381}{32}
\end{equation}
with ($16,13,12,9,8$) digits precision. 

Clearly they completely agree with the various alien derivatives.

\section{Free-energy problem in terms of the physical coupling}
\label{sect9}

In the present section we turn to the usual free-energy problem of the \mO{$N$} models in general. At first we establish notations in Subsection \ref{ssectPhysical} for the trans-series of the (dimensionless) free-energy and the
corresponding physical coupling. The latter depends only on the ratio of the $\Lambda$ scale (or equivalently, the dynamically generated mass $m$) and the magnetic field $h$. Both may be expressed via $v$ and $\nu$. In Subsection \ref{ssectFT} we introduce the field theoretical version of the same free-energy. From perturbative 1-loop calculations one may fix the mass gap $m/\Lambda$ and the 2-loop result is then already consistent with the WH one. Although a direct field-theoretical calculation was not attempted for the NP effects in these models\footnote{For the \mO{3} model in magnetic field the instantons are not known, and the renormalon singularities on the Borel-plane for the \mO{$N$} models were only related to standard PT calculations in the large $N$ limit \cite{Marino:2021dzn}. We also do not know whether the latter have a semiclassical interpretation at all.}, we may still assume that it is possible to write a trans-series ansatz for the free-energy in terms of the physical coupling itself. We can then take a practical approach, and calculate the coefficients in this ansatz simply by requiring that substituting the $(v,\nu)$ trans-series of the physical coupling into it reproduces our WH result for the free-energy density.

In Subsection \ref{ssectExampleO4} we show that in the simplest case of the \mO{4} model we find a similar resurgent structure in terms of the physical running coupling as we had in terms of the bootstrap one ($v$), at least up to the leading order. Based on this observation we state conjecture A in Subsection \ref{ssectConjA}, namely that the Stokes-automorphism can be used to generate the trans-series of the free-energy in terms of the physical running coupling almost exclusively from its perturbative part (just like it was able to do so in terms of $v$). Moreover, we also formulate conjecture B in Subsection \ref{ssectConjB}, stating that the lateral resummation of this trans-series gives the physical result (analogously to our main conjecture in Section \ref{sect7}). These conjectures mean essentially that the composition of strongly resurgent asymptotic serieses are also compatible with the median resummation, and thus it is also strongly resurgent. We prove these statements partially in Appendix \ref{appD}, up to second order for the \mO{4} case. 
Although some resurgent analysis was performed and the trans-series was obtained (partially) before in terms of the physical coupling \cite{Marino:2021dzn,Marino:2023epd}, with this approach we would like to emphasize that most probably the above, conjectured natural properties of the composition\footnote{These may be even provable as mathematical theorems more generally.} are sufficient to prove strong resurgence for the \mO{$N\geq 4$} models in the physical coupling too (where according to Section \ref{sect7} it already holds in the bootstrap running coupling $v$).

\subsection{Physical quantities}
\label{ssectPhysical}

The main players in our considerations are the particle density $\rho$ and
the energy density~$\varepsilon$. Eliminating $B$ (or equivalently $v$) we can
establish a relation $\varepsilon=E(\rho)$ between them. They can also be used
to calculate the free energy ${\cal F}$ and a related dimensionless function
$\Omega_1$:
\begin{equation}\label{eq:freeenergydef}
{\cal F}=\varepsilon-h\rho,\qquad
\Omega_1=-\frac{\cal F}{h^2},  
\end{equation}
where
\begin{equation}
h=E^\prime(\rho).
\end{equation}

We use the TBA integral equation to calculate 
\begin{equation}
\varepsilon=m^2 O_{1,1}(B),\qquad\quad \rho=mO_{1,0}(B)  
\end{equation}
and apply the Wiener-Hopf method to analyse the solution. Here we
recall some important relations.

Using the differential equations derived in section \ref{sect1} we have
\begin{equation}\label{eq:hperm}
h(B)=m\frac{\rho_1(B)}{\rho_0(B)}.  
\end{equation}
In this section the quantities $O_{1,1}(B)$, $O_{1,0}(B)$,
$\rho_1(B)$, and $\rho_0(B)$ will play a role. They are parametrized as\footnote{One should not confuse the normalized version of the energy density $F(v)$ with the universal function $F$ in \eqref{universalF}.}
\begin{equation}\label{physicals}
\begin{split}
O_{1,1}(B)&=\frac{G_+^2(i)}{8\pi}\,{\rm e}^{2B}\,F(v),
\qquad F(v)=2W_{1,1}(v),\\   
O_{1,0}(B)&=\frac{G_+(i)}{2\pi}\,{\rm e}^B\,\frac{\lambda}{\sqrt{v}}u(v),\qquad
u(v)=\frac{\sqrt{v}}{\lambda}U_1,\qquad
\lambda=\frac{g_o\sqrt{\pi}}{2},\qquad g_o=H(0),\\
\rho_1(B)&=\frac{G_+(i)}{2}\,{\rm e}^B\,w_1(v),\\
\rho_0(B)&=\frac{\lambda}{\sqrt{v}}\,\sigma(v).
\end{split}    
\end{equation}
Then the free energy and the dimensionless function in \eqref{eq:freeenergydef} are combinations of these:
\begin{equation}\label{FOmega1}
	{\cal F} = m^2 \left(O_{1,1}-\frac{\rho_1}{\rho_0} O_{1,0}\right), \qquad  \Omega_1 = \frac{O_{1,0} \rho_0  }{\rho_1}  - \frac{ O_{1,1} \rho_0^2}{\rho_1^2}, 
\end{equation}
and then we may substitute the r.h.s. of \eqref{physicals}.
It is also useful to introduce the rescaled quantity $\Omega$:
\begin{equation}
\Omega_1=\frac{g_o^2}{4}\Omega,\qquad\Omega=\frac{1}{2v}
\left(\frac{2u\sigma}{w_1}-\frac{F\sigma^2}{w_1^2}\right).
\end{equation}
From now on we will use the set of variables $\{F,u,w_1,\sigma\}$ and the
corresponding trans-series $\{\tilde F,\tilde u,\tilde w_1,\tilde\sigma\}$. The latter are normalized\footnote{Note that the reason for introducing the new notations $F(v), u(v)$ and $\sigma(v)$ for these trivial rescalings is exactly the simplicity of their normalization.} so that they all are of the form $1+{\rm O}(v)$. More precisely, excuding the \mO{3} case,
\begin{equation}\label{normalizedtoone}
\begin{split}
\tilde F&=M\nu+2A_{1,1}+2Y^{\rm T}{\cal L}Y=1+\frac{v}{2}+\Big(\frac{9}{8}-
\frac{5a}{4}\Big)v^2+\dots,\\  
\tilde u&=\frac{\sqrt{v}}{\lambda}\big(V_1+R^{\rm T}{\cal L}Y\big)
=1+\Big(\frac{a}{2}-\frac{3}{4}\Big)v+\Big(a-\frac{15}{32}-
\frac{5a^2}{8}\Big)v^2+\dots,\\  
\tilde w_1&=a_1+p^{\rm T}{\cal L}Y=1+\frac{v}{4}+\Big(\frac{9}{32}-\frac{5a}{8}
\Big)v^2+\dots,\\
\tilde\sigma&=\frac{\sqrt{v}}{\lambda}\big(s+R^{\rm T}{\cal L}p\big)=
1+\frac{av}{2}-\frac{5a^2v^2}{8}+\dots
\end{split}    
\end{equation}
An explicit formula for $\tilde u$ and $\tilde F$ (up to certain $v$ and $\nu$ orders) for the \mO{4} model can be found in \eqref{eq:O4uTS} and \eqref{eq:O4FTS}, while the same equations for the \mO{3} are spelled out in \eqref{eq:O3uTS} and \eqref{eq:O3FTS} respectively.
Our main conjecture is that physical quantities are obtained by laterally
Borel resumming the corresponding trans-series:
\begin{equation}
F(v)=S^+(\tilde F)(v),\quad  
u(v)=S^+(\tilde u)(v),\quad  
w_1(v)=S^+(\tilde w_1)(v),\quad  
\sigma(v)=S^+(\tilde \sigma)(v).
\label{main}
\end{equation}
For the rescaled $\Omega$ we have
\begin{equation}
\tilde\Omega=\frac{1}{2v}\left(\frac{2\tilde u\tilde\sigma}{\tilde w_1}
-\frac{\tilde F{\tilde\sigma}^2}{\tilde w_1^2}\right)=\frac{1}{2v}
+{\rm O}(1).
\end{equation}
Assuming our main conjecture (\ref{main}) and using the properties of $S^\pm$
discussed above, we arrive at
\begin{equation}
\Omega(v)=S^+(\tilde\Omega)(v).  
\end{equation}

At this point we introduce the physical running coupling $\beta$.
Its definition is
\begin{equation}
\frac{1}{\beta}+\xi\ln\beta=\ln\frac{h}{\Lambda},\qquad\xi=\frac{1-a}{2},  
\end{equation}
where $\Lambda=\Lambda_{\overline{\rm MS}}$ is the non-perturbative scale generated
via dimensional transmutation in the model. Expressed in terms of our
variables, 
\begin{equation}
\frac{1}{\beta}+\xi\ln\beta=P+\frac{1}{2v}+\xi\ln(2v)+\ln(1+S),
\label{physcoup}
\end{equation}
where $P$ is a constant,
\begin{equation}
P=\ln\frac{m}{\Lambda}+\ln\frac{H(1)}{\sqrt{\pi}H(0)}+\frac{b+L}{2}-\xi\ln2 
\end{equation}
and we have written
\begin{equation}
\frac{w_1}{\sigma}=1+S  
\end{equation}
so that the trans-series corresponding to $S$ is \lq\lq small'':
\begin{equation}
\tilde S=\frac{\tilde w_1}{\tilde\sigma}-1=\Big(\frac{1}{4}-\frac{a}{2}\Big)v+
{\rm O}(v^2).  
\end{equation}
We will need the leading terms in the following expansion:
\begin{equation}\label{Fexpand}
2v \tilde \Omega_1 = \frac{2\tilde u\tilde\sigma}{\tilde w_1}-\frac{\tilde F{\tilde\sigma}^2}
{\tilde w_1^2}=1+(a-2)v+\Big(2a-a^2-\frac{3}{2}\Big)v^2+\dots.
\end{equation}

We can somewhat simplify the problem of calculating the physical coupling
by writing
\begin{equation}
\beta=b(v)=\frac{2v}{1+vI(v)},
\end{equation}
where $I(v)$ satisfies
\begin{equation}
\frac{1}{2}I(v)-\xi\ln[1+vI(v)]=P+\ln(1+S(v)).  
\end{equation}
The trans-series corresponding to $I(v)$ satisfies 
\begin{equation}
\frac{1}{2}\tilde I(v)-\xi\ln[1+v\tilde I(v)]=P+\ln(1+\tilde S(v)).  
\label{tildeI}
\end{equation}
The expansion of $\tilde I(v)$ starts as
\begin{equation}
\tilde I(v)=y_1+y_2v+{\rm O}(v^2),\qquad y_1=2P,\qquad
y_2=2\xi y_1+\frac{1}{2}-a.
\end{equation}
Using the properties \eqref{ssectSplusprop} of $S^+$ we have
\begin{equation}\label{Iv}
I(v)=S^+(\tilde I)(v)
\end{equation}
and further
\begin{equation}
b(v)=S^+(\tilde b)(v),\qquad \tilde b(v)=\frac{2v}{1+v\tilde I(v)}.
\end{equation}

\subsection{Field theory}
\label{ssectFT}

The physical quantity $\Omega_1$ is related to the free energy density
in the presence of an external field of strength $h$ and can, in principle,
independently be calculated from field theory by functional integration.
It is difficult to perform functional integration exactly, but, at least,
the corresponding trans-series in terms of the physical coupling $\bar\beta$
is well-defined in field theory:
\begin{equation}
\Omega_1=d_o\tilde G(\bar\beta),
\end{equation}
where $d_o$ is some overall constant introduced for later convenience
and the physical running coupling is defined in field theory by
\begin{equation}
\frac{1}{\bar\beta}+\eta\ln\bar\beta=\ln\frac{h}{\Lambda},
\qquad\eta=\frac{\beta_1}{2\beta_0^2},  
\end{equation}
where $\beta_0$, $\beta_1$ are the two scheme independent coefficients of the
field theory beta function.
Of course, in practice so far only a few leading contributions have been
calculated \cite{Hasenfratz:1990zz,Hasenfratz:1990ab,Bajnok:2008it}:
\begin{equation}
\tilde G(\bar\beta)=\frac{1}{\bar\beta}+f_0+f_1\bar\beta+{\rm O}(\bar\beta^2).  
\label{LEAD}
\end{equation}
Consistency with the Wiener-Hopf calculation requires that the two definitions
of the physical coupling coincide:
\begin{equation}
\bar\beta=\beta,\qquad \eta=\xi,\qquad d_o=\frac{g_o^2}{4}.
\label{consist1}  
\end{equation}
Further consistency conditions follow from the comparison of the first few
expansion coefficients:
\begin{equation}
\begin{split}
{\rm (from\ FT)}\quad \tilde\Omega&=\frac{1}{2v}\Big\{
1+(y_1+2f_0)v+(y_2+4f_1)v^2+\dots\Big\}\\  
{\rm (from\ WH)}\quad \tilde\Omega&=\frac{1}{2v}\Big\{
1+(a-2)v+\Big(2a-a^2-\frac{3}{2}\Big)v^2+\dots\Big\}.
\end{split}
\end{equation}
We obtain
\begin{equation}
y_1=a-2-2f_0,\qquad\quad y_2=2a-a^2-\frac{3}{2}-4f_1.  
\end{equation}
The first one can be used to find the link between bootstrap (TBA) and field
theory data, $m/\Lambda$:
\begin{equation}
\ln\frac{m}{\Lambda}=\frac{a-2-2f_0}{2}-\ln\frac{H(1)}{\sqrt{\pi}H(0)}
-\frac{b+L}{2}+\xi\ln2.  
\end{equation}
The second one gives
\begin{equation}
f_1=\xi f_0.
\label{consist2}  
\end{equation}


We can check that both (\ref{consist1}) and (\ref{consist2}) are satisfied in
the \mO{$N$} case, where
\begin{equation}
\begin{split}
a&=1-2\Delta,\qquad b+L=-4\Delta\ln2,\qquad
H(0)=\frac{1}{\sqrt{\pi\Delta}},\qquad
H(1)=\frac{\Gamma(1+\Delta)}{\sqrt{\Delta}},\\
\beta_0&=\frac{1}{4\pi\Delta},\qquad \beta_1=\frac{1}{8\pi^2\Delta},\qquad
\eta=\xi=\Delta,\qquad d_o=\beta_0,\\
f_0&=-\frac{1}{2},\qquad f_1=-\frac{\Delta}{2},\qquad
\ln\frac{m}{\Lambda}=\Delta(\ln8-1)-\ln\Gamma(1+\Delta).
\end{split}  
\end{equation}
Full consistency means that the two completely different approaches lead
to the same result:
\begin{equation}
\tilde G\circ\tilde b=\tilde\Omega,\qquad\quad
\tilde G(\tilde b(v))=\tilde\Omega(v).  
\label{fullcon}
\end{equation}
In practice, in the absence of more data beyond (\ref{LEAD}), we cannot check
full consistence. However, we can calculate the trans-series $\tilde G(\beta)$
from (\ref{fullcon}). Explicit formulas (up to certain PT and NP orders) for the \mO{3} and \mO{4} models can be found in \eqref{eq:O3F} and \eqref{gO4TS}.

\subsection{Example: \mO{4} model expansion to $1^{\rm st}$ order}
\label{ssectExampleO4}

In this (simplest) case
\begin{equation}
a=b=0,\quad L=-2\ln2,\quad M=-2i,\quad y_1=-1,\quad
\nu={\rm e}^{-2B}=4{\rm e}^{-\frac{1}{v}},
\end{equation}
and (\ref{physcoup}) can be rewritten as
\begin{equation}
\frac{\nu}{v(1+S)^2}=\frac{8}{{\rm e}\beta}{\rm e}^{-\frac{2}{\beta}}.  
\end{equation}

Taking care of the exceptional $M\nu$ term we write 
\begin{equation}
\Omega=\frac{i\nu}{v}\,\frac{1}{(1+S)^2}+\Omega_{\rm reg},\qquad
\tilde\Omega_{\rm reg}(v)=\sum_{s=0}^\infty \Omega_s(v)\nu^{2s}  
\end{equation}
and analogously
\begin{equation}\label{O4Gr}
\tilde G(\beta)=\frac{8i}{{\rm e}\beta}{\rm e}^{-\frac{2}{\beta}}+
\tilde G_{\rm reg}(\beta),\qquad \tilde G_{\rm reg}(\beta)=
\sum_{r=0}^\infty {\cal G}_r(\beta){\rm e}^{-\frac{4r}{\beta}}.
\end{equation}
For later convenience we introduce
\begin{equation}
{\cal G}_r(\beta)=G_r(\beta)\left(\frac{8}{{\rm e}\beta}\right)^{2r}  
\end{equation}
and write
\begin{equation}
\tilde G_{\rm reg}(\tilde b(v))=\sum_{r=0}^\infty G_r(\tilde b(v))\left[
\frac{\nu}{v(1+\tilde S(v))^2}\right]^{2r}=\tilde\Omega_{\rm reg}(v).    
\end{equation}
We will use the trans-series expansions
\begin{equation}
\tilde b(v)=\sum_{r=0}^\infty b_r(v)\nu^{2r},\qquad  
\tilde I(v)=\sum_{r=0}^\infty I_r(v)\nu^{2r},\qquad  
\end{equation}
and (for convenience) two representations of $\tilde S(v)$:
\begin{equation}
\tilde S(v)=\sum_{r=0}^\infty S_r(v)\nu^{2r},\qquad  
[1+\tilde S(v)]^2=\bar S_0(v)
\left\{1+\sum_{r=1}^\infty \bar S_r(v)\nu^{2r}\right\}.  
\end{equation}

The leading (purely perturbative) part of (\ref{tildeI}) is
\begin{equation}
I_0(v)-\ln[1+v I_0(v)]=-1+\ln(1+S_0(v))^2  
\label{tildeI0}
\end{equation}
and after having solved (\ref{tildeI0}) for $I_0(v)$ as an asymptotic series,
we write
\begin{equation}
b_0(v)=\frac{2v}{1+vI_0(v)}  
\end{equation}
and all higher trans-series components can be calculated algebraically from
linear equations like
\begin{equation}
\begin{split}
-\frac{2b_1}{b_0^2}&+\frac{b_1}{b_0}=\bar S_1,\\
-\frac{2b_2}{b_0^2}&+\frac{b_2}{b_0}+\frac{2b_1^2}{b_0^3}
-\frac{1}{2}\frac{b_1^2}{b_0^2}=\bar S_2-\frac{1}{2}\bar S_1^2.
\end{split}    
\end{equation}
Next we can calculate $G_0(\beta)$, $G_1(\beta)$, $G_2(\beta)$ etc. from the
implicit relations
\begin{equation}
\begin{split}  
\Omega_0&=G_0(b_0),\\
\Omega_1&=G_0^\prime(b_0)b_1+\frac{G_1(b_0)}{(v\bar S_0)^2},\\
\Omega_2&=G_0^\prime(b_0)b_2+\frac{1}{2}G_0^{\prime\prime}(b_0)b_1^2+
\frac{G_2(b_0)}{(v\bar S_0)^4}+\frac{G_1^\prime(b_0)b_1}{(v\bar S_0)^2}
-2\bar S_1\frac{G_1(b_0)}{(v\bar S_0)^2},
\end{split}
\label{Omega012}
\end{equation}
and so on.

In section \ref{sect7} we established the relations
\begin{equation}
\tilde W_{1,1}={\mathfrak{S}}^{-1/2}A_{1,1},\qquad\quad  
\tilde U_1={\mathfrak{S}}^{-1/2}V_1,\qquad\quad  
\tilde D={\mathfrak{S}}^{-1/2}D_0.  
\end{equation}
From the second one it follows by \eqref{ssectStokesprop} that
\begin{equation}\label{medianu0}
\tilde u={\mathfrak{S}}^{-1/2}u_0,  
\end{equation}
where
\begin{equation}
\tilde u(v)=\sum_{r=0}^\infty u_r(v)\nu^{2r}.
\end{equation}
Since ${\cal R}$ commutes with $\frac{{\rm d}}{{\rm d}B}$ and this property is
inherited by ${\mathfrak{S}}^{-1/2}$, we can write  
\begin{equation}
\tilde w_1^2={\dot{\tilde W}}_{1,1}+2\tilde W_{1,1}={\mathfrak{S}}^{-1/2}
(\dot A_{1,1}+2A_{1,1})={\mathfrak{S}}^{-1/2}a_1^2
=\left({\mathfrak{S}}^{-1/2}a_1\right)^2  
\end{equation}
and conclude that
\begin{equation}
\tilde w_1={\mathfrak{S}}^{-1/2}a_1.  
\end{equation}
Similarly
\begin{equation}
{\dot{\tilde D}}=\tilde\rho_0^2\quad\Longrightarrow\quad \tilde\rho_0=
{\mathfrak{S}}^{-1/2}s  
\end{equation}
and after rescaling
\begin{equation}
\tilde\sigma={\mathfrak{S}}^{-1/2}\sigma_0,\qquad\quad \tilde\sigma
=\sum_{r=0}^\infty\sigma_r\nu^{2r}.
\end{equation}
This structure is inherited by the composite objects:
\begin{equation}
1+\tilde S={\mathfrak{S}}^{-1/2}(1+S_0),\quad  
\tilde\Omega_{\rm reg}={\mathfrak{S}}^{-1/2}\Omega_0,\quad  
\tilde I={\mathfrak{S}}^{-1/2}I_0,\quad  
\tilde b=\frac{2v}{1+v\tilde I}={\mathfrak{S}}^{-1/2}b_0.  
\label{mathfrakOmega}
\end{equation}
At leading order we have
\begin{equation}
\Omega_1=\frac{1}{2}\Delta_2\Omega_0,\qquad  
b_1=\frac{1}{2}\Delta_2b_0,  
\end{equation}
where
\begin{equation}
\Delta_\omega=4^{-\omega}\Delta_\omega^{\rm (st)}.
\end{equation}
Using this in (\ref{saee}) we first write
\begin{equation}
\Omega_1(v)=\frac{1}{2}\,\left[\frac{{\rm e}b_0(v)}{8v\bar S_0(v)}\right]^2
(D_2G_0)(b_0(v))+G^\prime_0(b_0(v))b_1(v)  
\label{saee2}
\end{equation}
and comparison with (\ref{Omega012}) then gives
\begin{equation}
G_1(\beta)=\frac{1}{2}\left(\frac{{\rm e}\beta}{8}\right)^2(D_2G_0)(\beta),
\qquad {\cal G}_1(\beta)=\frac{1}{2}(D_2G_0)(\beta).
\label{firstG1}
\end{equation}

\subsection{Conjecture A}
\label{ssectConjA}

It seems natural to conjecture that, analogously to the $\tilde\Omega_{\rm reg}$
relation (\ref{mathfrakOmega}), $\tilde G_{\rm reg}$ satisfies 
\begin{equation}
\tilde G_{\rm reg}(\beta)=\left(\Sigma^{-1/2}G_0\right)(\beta),
\label{conA}
\end{equation}
where we introduced the Stokes-automorphism $\Sigma$ in terms of the alien derivative $D_\omega$ defined in Appendix \ref{compAsy}, which acts on asymptotic serieses of the physical running coupling as
\begin{equation}
\Sigma^{-1/2}={\rm e}^{\frac{1}{2}{\cal T}},\qquad {\cal T}=\sum_\omega
{\rm e}^{-\omega x}D_\omega=\sum_{r=1}^\infty{\rm e}^{-2r x}D_{2r},\qquad
x=\frac{2}{\beta}.      
\end{equation}
The relation between $D_\omega$ and $\Delta_\omega$ is given by \eqref{saee}.

Expanding (\ref{conA}) to leading order
\begin{equation}
G_0+\frac{1}{2}{\rm e}^{-2x}D_2G_0+\dots=G_0+{\cal G}_1{\rm e}^{-2x}+\dots  
\end{equation}
we see that our result (\ref{firstG1}) shows that Conjecture A is satisfied, at
least to first order.

\subsection{Conjecture B}
\label{ssectConjB}

From our main conjecture, using the properties of $S^+$ and
${\mathfrak{S}}^{-1/2}$ it follows that
\begin{equation}
\Omega_{\rm reg}=S^+(\tilde\Omega_{\rm reg})(v)=S^{\rm med}(\Omega_0)(v),\qquad  
\beta=S^+(\tilde b)(v)=S^{\rm med}(b_0)(v)  
\end{equation}
and therefore it is natural to conjecture that also
\begin{equation}
G_{\rm reg}=S^+(\tilde G_{\rm reg})(\beta)  
\end{equation}
holds. In a more abstract notation we conjecture that
\begin{equation}
S^+\left(\tilde G_{\rm reg}\circ\tilde b\right)(v)=S^+(\tilde G_{\rm reg})
\left(S^+(\tilde b)(v)\right).  
\end{equation}

In appendix \ref{appD} the validity of conjectures A and B is discussed in the
special case of the \mO{4} model.

\section{Conjecture A: instanton sectors of the \mO{3} model}
\label{sect10}

In this section we discuss the \mO{3} model analogue of conjecture A. 
It is not in our focus to prove conjecture B, as we expect that it can be easily adapted to the \mO{3} case once conjecture A is proven, based on the calculations in Appendix \ref{appD}. 
The main difference to the \mO{$N\geq 4$} case (that was the focus of Section \ref{sect9}) is that here the median resummation of the PT sector is not enough to arrive at the physical result, additional sectors are needed. This resembles the situation of \emph{weak resurgence} in Section \ref{sect8}. However, there is a further complication, namely that here the number of these extra sectors is not finite anymore, an \emph{infinite number} of \emph{instanton sectors} are present. This fact, culminated in \eqref{mainresult} of Subsection \ref{conjAsumm} is the main result of our paper. 

Before we start, here are some \mO{3} specific definitions and relations.

\begin{equation}
a=-1,\qquad L=-1-\ln8,\qquad \nu=\frac{8{\rm e}}{v}{\rm e}^{-1/v},
\end{equation}
\begin{equation}
u=2\sqrt{v}U_1,\qquad
r=2\sqrt{v}\rho_0,\qquad
\bar V_n=2\sqrt{v} V_n,\qquad
s_o=2\sqrt{v}s,\qquad
\end{equation}
\begin{equation}
{\overset{\circ}{\Delta}}_\omega={\rm e}^{-\omega/v}\Delta_\omega^{({\rm st})}=
\nu^\omega\Delta_\omega,
\end{equation}
\begin{equation}
\Omega=-\frac{4\pi}{h^2}(\varepsilon-h\rho)=\frac{1}{2v}\left(
\frac{2ur}{w_1}-\frac{Fr^2}{w_1^2}\right),\qquad F=2W_{1,1},
\end{equation}
\begin{equation}
F=-\frac{i\pi}{{\rm e}}\nu+F_{\rm reg},
\end{equation}
\begin{equation}
{\cal L}=\frac{r^2}{vw_1^2},\qquad \Omega=\frac{i\pi}{2{\rm e}}\nu{\cal L}+
\Omega_{\rm reg}, \qquad\Omega_{\rm reg}=\frac{1}{2v}\left(
\frac{2ur}{w_1}-\frac{F_{\rm reg}r^2}{w_1^2}\right). 
\end{equation}
The regular part\footnote{In this section ${\mathfrak{S}}$ denotes the regular
part of the Stokes automorphism (not including the $\Delta_1$ part).}
of the Stokes automorphism in this model is
\begin{equation}
{\mathfrak{S}}=\exp\left\{-\sum_{r=1}^\infty\nu^{2r}\Delta_{2r}\right\}
\end{equation}
and for the trans-series we have
\begin{equation}
\tilde F_{\rm reg}={\mathfrak{S}}^{-1/2}\left(f_0+\nu f_1+\nu^2 f_2\right),
\end{equation}
where\footnote{Here $\hatk=1/{\rm e}$ again.}
\begin{equation}
\begin{split}
f_0&=2A_{1,1},\\
f_1&=4\hatk P_{1,-1}+2\hatk\left(\frac{1}{v}+\ln v-\gamma_E-4\ln2\right),\\
f_2&=2\hatk^2A_{-1,-1}.
\end{split}
\end{equation}
Further, 
\begin{equation}
\tilde u={\mathfrak{S}}^{-1/2}\left(\bar V_1+\nu \hatk \bar V_{-1}\right),\qquad
\tilde r={\mathfrak{S}}^{-1/2}s_o,\qquad\tilde w_1={\mathfrak{S}}^{-1/2}
\left(a_1+\nu \hatk a_{-1}\right).  
\end{equation}
This structure is inherited by $\tilde{\cal L}$:
\begin{equation}
\tilde{\cal L}={\mathfrak{S}}^{-1/2}{\cal K},  
\end{equation}
where
\begin{equation}
{\cal K}=\frac{s_o^2}{v(a_1+\nu \hatk a_{-1})^2}.
\end{equation}
Note that
\begin{equation}
{\cal K}=\sum_{r=1}^\infty {\cal K}_r\nu^r
\end{equation}
is purely real.

Similarly,
\begin{equation}
\tilde\Omega_{\rm reg}={\mathfrak{S}}^{-1/2}X,  
\end{equation}
where
\begin{equation}
X=\frac{1}{2v}\left[\frac{2s_o(\bar V_1+\nu
\hatk\bar V_{-1})}{(a_1+\nu \hatk a_{-1})}-
\frac{s_o^2(f_0+\nu f_1+\nu^2 f_2)}{(a_1+\nu \hatk a_{-1})^2}\right].
\label{XX}
\end{equation}
\begin{equation}
X=\sum_{r=1}^\infty X_r\nu^r
\end{equation}
is also purely real.

\subsection{Physical running coupling}

In the \mO{3} case the running coupling is defined by
\begin{equation}
\frac{1}{\beta}+\ln\beta=\ln\frac{h}{\Lambda}.
\end{equation}
Equivalently, it is the solution of
\begin{equation}
\frac{2}{\beta}+2\ln\beta=\ln4-2+\frac{1}{v}+\ln v-\ln{\cal L}.  
\end{equation}
This last equation can also be formulated as
\begin{equation}
\frac{1}{\beta^2}{\rm e}^{-2/\beta}=\frac{{\rm e}\nu}{32}{\cal L}.  
\end{equation}
It is also useful to introduce the variable $I$ by
\begin{equation}
\beta=\frac{2v}{1+vI}.  
\end{equation}
This satisfies
\begin{equation}
\frac{1}{2}I-\ln[1+vI]=-1-\frac{1}{2}\ln(v{\cal L})  
\end{equation}
and the corresponding trans-series relation is
\begin{equation}
\frac{1}{2}\tilde I-\ln[1+v\tilde I]=-1-\frac{1}{2}\ln(v\tilde{\cal L}).  
\end{equation}
The above equation is solved by
\begin{equation}
\tilde I={\mathfrak{S}}^{-1/2} Y,
\end{equation}
where $Y$ is the solution of             
\begin{equation}
\frac{1}{2}Y-\ln[1+vY]=-1-\frac{1}{2}\ln(v{\cal K}).  
\end{equation}
Since $Y$ is purely real,
\begin{equation}
b=\frac{2v}{1+vY}=\sum_{r=0}^\infty b_r\nu^r
\end{equation}
is also real and
\begin{equation}
\tilde \beta={\mathfrak{S}}^{-1/2} b.
\end{equation}
The defining equation in terms of $b$ becomes
\begin{equation}
\frac{1}{b}+\ln b=\frac{1}{2}\ln\frac{32}{\rm e}+B-\frac{1}{2}\ln{\cal K}.
\end{equation}
We note, for later purposes, that this implies the relations
\begin{equation}
\left(\frac{1}{b_0}-\frac{1}{b_0^2}\right)b_1=-\frac{1}{2}\frac{{\cal K}_1}
{{\cal K}_0},\qquad  
\left(\frac{1}{b_0}-\frac{1}{b_0^2}\right)\dot b_0=
1-\frac{1}{2}\frac{\dot{\cal K}_0}{{\cal K}_0}.  
\label{lat}
\end{equation}
Also for later purposes we note that the relation (\ref{saee})
in the \mO{3} special case reads
\begin{equation}
\Delta_\omega[q(b_0)]=\left(\frac{{\rm e}b_0^2}{32}{\cal K}_0\right)^\omega
(D_\omega q)(b_0)+q^\prime(b_0)\Delta_\omega b_0.  
\end{equation}

\subsection{Rephrasing and consequences of the conjecture}
\label{conjAsumm}

The physical quantity $\Omega_{\rm reg}$ is represented as a trans-series
in the variable $\beta$:
\begin{equation}\label{O3Gr}
	\tilde G_{\rm reg}(\beta)=\sum_{r=0}^\infty{\cal G}_r(\beta){\rm e}^{-2r/\beta}.
\end{equation}
It is convenient to introduce the rescaled coefficients $G_r(\beta)$ by
\begin{equation}
{\cal G}_r(\beta)=\left(\frac{32}{{\rm e}\beta^2}\right)^r G_r(\beta).
\end{equation}
This is convenient because 
\begin{equation}\label{Om1G}
\tilde\Omega_{\rm reg}=\tilde G_{\rm reg}(\tilde\beta)=\sum_{r=0}^\infty
G_r(\tilde\beta)\left(\frac{32}{{\rm e}\tilde\beta^2}{\rm e}^{-2/\tilde\beta}
\right)^r=\sum_{r=0}^\infty G_r(\tilde\beta)\left(\nu\tilde{\cal L}\right)^r.  
\end{equation}
As shown in Appendix \ref{AppO3proof} by expanding both sides of \eqref{Om1G} in $\nu$, the first two coefficient serieses ${\cal G}_{0}(\beta)$ and ${\cal G}_1(\beta)$ are real, while the following ones are complex
\begin{equation}
{\cal G}_s(\beta)={\cal G}_s^{({\rm r})}(\beta)+{\cal G}_s^{({\rm i})}(\beta),\qquad s=2,3,
\end{equation}
where we denoted their real and imaginary parts via $({\rm r})$ and $({\rm i})$ respectively. It turned out that the imaginary parts are related to the alien derivatives of previous coefficients as
\begin{equation}
{\cal G}_2^{({\rm i})}(\beta)=\frac{1}{2}(D_2{\cal G}_0)(\beta), \qquad {\cal G}_3^{({\rm i})}(\beta)=\frac{1}{2}(D_2{\cal G}_1)(\beta).  
\end{equation}
In this way we have verified the \mO{3} version of conjecture A (up to $3^{\rm rd}$ order):
\begin{equation}\label{mainresult}
\tilde\Omega_{\rm reg}=\Sigma^{-1/2}\left[
I_0(\beta)+I_1(\beta){\rm e}^{-2/\beta}+I_2(\beta){\rm e}^{-4/\beta}+
I_3(\beta){\rm e}^{-6/\beta}+\dots\right]\Bigg\vert_{\beta=\tilde\beta},  
\end{equation}
where
\begin{equation}
I_0(\beta)={\cal G}_0(\beta),\qquad
I_1(\beta)={\cal G}_1(\beta),\qquad
I_2(\beta)={\cal G}_2^{({\rm r})}(\beta),\qquad
I_3(\beta)={\cal G}_3^{({\rm r})}(\beta)
\end{equation}
and $\Sigma$ is the Stokes automorphism in the $\beta$ variable:
\begin{equation}
\Sigma=\exp\left\{-\sum_{r=1}^\infty {\rm e}^{-4r/\beta}D_{2r}\right\}.  
\end{equation}
We can interpret $I_r(\beta)$ as the contribution of the $r$-instanton sector.
We see that there are infinitely many instanton sectors in the field theory
calculation.

\section{Numerical results}
\label{sect13}

To back up our analytical calculations, we present some high-precision numerical evidence that our trans-series indeed reproduces the exact solution of the integral equation, way beyond the leading non-perturbative correction. That is, we demonstrate that if one truncates the trans-series at some given non-perturbative level, and takes its lateral Borel resummation, the result's difference to the (numerically evaluated) TBA is of the order of the next non-perturbative correction coming from the same trans-series. The figures \ref{fig:VO3}-\ref{fig:FO4} below show exactly this: each coloured line corresponds to such a truncation, and each of them is proportional essentially to some integer power of ${\rm e}^{-2 B}$, which would appear as a straight line with negative slope on a logarithmic plot. We analyse the \mO{3} model, and as before, our other example from the \mO{$N \geq 4$} family will be \mO{4} itself. In each case, we compared our TBA and trans-series data at integer values of $B$, in the range $1\leq B \leq 20$, however, in some cases this interval was restricted by the numerical errors. 

To calculate the physical quantities such as $\rho, \varepsilon$ or $\mathcal{F}$ numerically directly from the integral equation \eqref{TBA} one may expand its solution on the basis of even Chebyshev polynomials. As higher-order polynomials encode finer details of the solution, one may truncate the series at a given level to reach a certain precision. Evaluating the integral equation at a finite number of specific rapidities then translates it to a matrix equation (the kernel being represented as a square-matrix), which one can then feed to a linear solver \cite{Abbott:2020qnl}. This method works well for any model, however, for the \mO{3} kernel (due to its specific form) one can even rephrase the problem as a recursion \cite{ristivojevic2019conjectures}, which may be solved significantly faster, practically leading to a higher precision. 
The method described here is essentially a small $B$ expansion, therefore the error of the TBA solution increases with the size of the integration domain - see Figure \ref{fig:WVO4} for the \mO{4} model.

On the other side of the comparison, one needs to calculate the trans-series of the above-mentioned quantities for relatively high perturbative orders (at every non-perturbative order) in the running couplings $v$ or $\beta$.
To achieve this, at first one needs to determine the perturbative building blocks as described in \eqref{pertinstructions} in terms of the bootstrap running coupling $v$. 

Then, the trans-series of the different quantities like $w_1, W_{1,1}, \rho_0$ or $U_1$  - see  \eqref{w1til}, \eqref{W11til}, \eqref{rho0til}
and \eqref{U1til} - may be calculated in a recursive way, order-by-order in $\nu$. 
These contain terms expressed via the infinite matrix-vector notation introduced in Subsection \ref{transrep}. To calculate the truncation of such a term (in other words, that of matrix $\mathcal{L}$) at order $\nu^{\kappa_1 M}$ one may set up a two-dimensional recursion for $v$-dependent (asymptotic) power serieses $q_{s,M}$ (these are truncated at some power $v^L$ as well):
\begin{align}
	q_{s,M}(v) &= \sum_{l=1}^M i S_l \, A_{-\kappa_s, -\kappa_l}(v) q_{l, M-l}(v) + \rmO(v^{L+1}),  & q_{s,0}(v)&=f_{-\kappa_s}(v) + \rmO(v^{L+1}) 
\end{align}
where at each step one has to multiply two (truncated) power series together - namely a previous result $q_{l, M-l}$ and the building block $A_{-\kappa_s, -\kappa_l}$. This can be efficiently performed by a discrete convolution of their coefficients.\footnote{Note that this works only if $\kappa_j =  \kappa_1 j$, i.e. the Wiener-Hopf kernel has equally spaced poles. Also, notice that $\kappa_1=2$ for both the \mO{3} and \mO{4} models.}
The starting point $f_{-\kappa_s}$ of the recursion is another perturbative building block, based on which vector-matrix-vector product (mentioned above) we would like to calculate. In the end, one needs to "close" the recursion by yet another perturbative building block $g_{-\kappa_l}$:
\begin{equation}\label{fginner}
	\langle g, f\rangle_M (v) = \sum_{l=1}^M  i S_l \, g_{-\kappa_l}(v) q_{l,M-l}(v) + \rmO(v^{L+1}).
\end{equation}
For example, to calculate the coefficient of $\nu^{\kappa_1 M}$ in the term $p^T \mathcal{L} Y$  in \eqref{w1til} one needs to choose $f_n = A_{1,n} $ and $g_n = a_n$ as starting- and end-points.\footnote{Due to the building blocks $A_{n,m}$ being symmetric in their indices $n\leftrightarrow m$,  expression \eqref{fginner} is also symmetric for $f \leftrightarrow g$.} 

Next, the lateral resummation is done similarly to the Borel-Pad\'e resummation method explained in \cite{Abbott:2020mba, Abbott:2020qnl}.\footnote{In this section we avoided the conformal mapping method. The reason is, that although it improves the resummations' accuracy significantly (even by several orders of magnitude), yet for our purposes - that is, comparing exponentially small contributions plotted on a logarithmic scale - it proved to have about the same performance as the Borel-Pad\'e method.} After the Borel transformation of our asymptotic series we apply a Padé-approximation, and finally integrate it numerically along a contour tilted slightly above the positive real axis. The lateral resummation needs to be performed on each of the non-perturbative levels of the trans-series individually (themselves being asymptotic serieses in $v$), then combined via the appropriate power of $\nu$. The resulting quantity at a finite truncation of the trans-series is in general complex: the real part approximates the TBA, whereas the imaginary part will decrease and eventually approach zero when including more and more terms from the trans-series (see Figure \ref{fig:VO3im}). 

The analysis in terms of the physical coupling requires the assembly of the trans-series of the free energy density. A naive but working approach is to take an ansatz for $\tilde{G}(\beta)$ where the coefficients of the asymptotic serieses $\mathcal{G}_r(\beta)$ in \eqref{O4Gr} and \eqref{O3Gr} are a priori undetermined. Then, the trans-series $\tilde{b}(v)$ of the running coupling needs to be determined by solving the equation \eqref{tildeI} order-by-order (both in terms of variable $v$ and in  $\nu$, formally treated as independent expansion parameters) for $\tilde{I}(v)$. Finally, the composition of the two serieses, namely our ansatz for $\tilde{G}(\beta)$ and $\tilde{b}(v)$, needs to be compared to $\tilde{\Omega}(v)$ as in \eqref{fullcon}. After a $v,\nu$ expansion of both sides one arrives at realations fixing the coefficients of the ansatz, and one can calculate their value iteratively again. The lateral resummations in $\beta$ were performed in a very similar manner to the analysis in the bootstrap coupling $v$.  

\subsection{Particle and energy density}\label{subsec:energydensity}

Instead of the physical densities $\rho,\varepsilon$ we analyse their normalized variants $u(v), F(v)$ as defined in \eqref{physicals}. For our models, these look as
\begin{align}\label{eq:epsrho}
\frac{\rho}{m}&=\frac{1}{4}{\rm e}^{B}\frac{u(v)}{\sqrt{\pi v}}\cdot\,\begin{cases}
		\sqrt{\frac{e}{2\pi}} & \text{O}(3)\\
		1 & \text{O}(4),
\end{cases}
	\quad & 	
\frac{\varepsilon}{m^{2}}&=\frac{1}{16}{\rm e}^{2B}F(v)\cdot\begin{cases}
	\frac{e}{\pi} &  \text{O}(3)\\
	1 &  \text{O}(4)
\end{cases}
\end{align}
where - as we have seen previously - the bootstrap couplings are defined via \eqref{ONLconst} as
\begin{equation}\label{eq:twoB}	
	2 B = \frac{1}{v} + \begin{cases} \ln v - 3 \ln 2 -1, \quad\quad 0.2171 \gtrsim v > 0   & \text{O}(3) \\ \quad\quad - 2 \ln 2, \quad\quad\quad\;\;\,  0.7213 \gtrsim  v > 0 & \text{O}(4),\end{cases}
\end{equation}
for the two cases. The numeric bounds indicate the largest possible values of these couplings when approaching $B \simeq 0$.

We aim to compare the resummed trans-series of the quantities in \eqref{eq:epsrho} to their exact value. The latter gets very precisely approximated when directly evaluating the integral equation in the Chebyshev basis as explained in the introductory part of the present section, and then taking the moments of the solution to arrive at the quantities \eqref{eq:epsrho}. That is, one has an independent way to calculate the physical result at a certain $B$ value and also full control over its accuracy. We denote the densities obtained like this as $u_{\text{TBA}}(v)$ and $F_{\text{TBA}}(v)$. To get rid of the prefactor $1/\sqrt{v}$ in $\rho$ one needs to invert \eqref{eq:twoB} as well. 

To quantify the discrepancy between the TBA and the trans-series taken into account up to a given order in $\nu$ we define the differences
\begin{align}\label{eq:dudF}
\delta u^{(k)}(v) &\equiv  u_{\text{TBA}}(v) - \text{Re}\,S^+(\tilde{u}^{(k)})(v), & \delta F^{(k)}(v) &\equiv  F_{\text{TBA}}(v) - \text{Re}\,S^+(\tilde{F}^{(k)})(v),
\end{align}	
where the upper index indicates at which NP term we truncate the trans-series: 
\begin{align}
	\tilde{u}^{(k)} &= \sum_{r=0}^k u_r \nu^r, &  \tilde{F}^{(k)} &= \sum_{r=0}^k F_r \nu^r.
\end{align}	
Here $u_r$ and $F_r$ are the formal power serieses at order $\nu^r$ in the full trans-series, and $S^+$ acts on them individually.

\subsubsection{The \mO{3} model}

We spell out these trans-serieses for the \mO{3} model below, giving the first two non-vanishing perturbative orders for each NP order we analyse in this section. For the particle density, this looks as
\begin{align}\label{eq:O3uTS} 
	\tilde{u}(v) & \overset{\text{\tiny \mO{3}}}{=} \Bigg\lbrace 1-\frac{5 v}{4}+\rmO\left(v^2\right) \Bigg\rbrace
-\frac{\nu}{e}  \Bigg\lbrace 1+\frac{v}{4}+\rmO\left(v^2\right) \Bigg\rbrace
+ i\frac{\nu^2}{{\rm e}^2} \Bigg\lbrace 1+\frac{ v^2}{8}+\rmO\left(v^3\right) \Bigg\rbrace \nonumber \\
&+ i \frac{\nu^3}{{\rm e}^3} \Bigg\lbrace \frac{1}{3}-\frac{1}{6}  v+\rmO\left(v^2\right) \Bigg\rbrace 
+\frac{\nu^4}{{\rm e}^4}\Bigg\lbrace \left(\frac{1}{2}+\frac{4 i}{3}\right)-\left(\frac{1}{8}+\frac{i}{6}\right) v+\rmO\left(v^2\right) \Bigg\rbrace \nonumber\\
	&+ \frac{\nu^5}{{\rm e}^5} \Bigg\lbrace\left(\frac{1}{6}+\frac{4 i}{5}\right) - \left(\frac{1}{8}+\frac{i}{2}\right) v+\rmO\left(v^2\right)\Bigg\rbrace    \nonumber\\
	&+ \frac{\nu^6}{{\rm e}^6}\Bigg\lbrace \left(\frac{20}{9}+\frac{19 i}{5}\right)-\left(\frac{11}{18}+\frac{11 i}{20}\right) v+\rmO\left(v^2\right) \Bigg\rbrace + \rmO(\nu^7),  
\end{align}
while for the energy density, it is
\begin{align}\label{eq:O3FTS}
	\tilde{F}(v) & \overset{\text{\tiny \mO{3}}}{=} \Bigg\lbrace 1+\frac{v}{2}+\rmO\left(v^2\right)   \Bigg\rbrace 
	+\frac{\nu}{e}\Bigg\lbrace \frac{1}{v} + \left(\ln v - i \pi -2 \gamma_E -4 \ln 2 + 2\right)   - v+\rmO\left(v^2\right)   \Bigg\rbrace \nonumber \\
	& - \frac{\nu^2}{{\rm e}^2}  \Bigg\lbrace (1-4 i)-\left(\frac{1}{2}+i\right) v+\rmO\left(v^2\right)   \Bigg\rbrace 
	 + i \frac{\nu^3}{{\rm e}^3}\Bigg\lbrace \frac{8 }{3}-\frac{2  v}{3}+\rmO\left(v^2\right)   \Bigg\rbrace \nonumber\\
	&+\frac{\nu^4}{{\rm e}^4}\Bigg\lbrace  (2+4 i)+i v+\rmO\left(v^2\right)  \Bigg\rbrace 
	+\frac{\nu^5}{{\rm e}^5}\Bigg\lbrace  \left(\frac{4}{3}+\frac{64 i}{15}\right)-\left(\frac{2}{3}+\frac{8 i}{15}\right) v+\rmO\left(v^2\right) \Bigg\rbrace   \nonumber \\
	&+\frac{\nu^6}{{\rm e}^6}\Bigg\lbrace  \left(\frac{22}{3}+10 i\right)+\left(\frac{2}{3}+\frac{7 i}{2}\right) v+\rmO\left(v^2\right)  \Bigg\rbrace+\rmO(\nu^7) . 
\end{align}

Clearly, we only know a finite number of coefficients in an asymptotic series like $u_r$ or $F_r$ for a given $r$. From these, we may only approximate the exact value of the lateral sum. A Pad\'e-approximant of the truncated Borel-transform captures the analytic structure of the Borel-plane already from these coefficients, and it allows us to reconstruct the Borel-transform beyond its convergence radius in the presence of singularities. In this way, we approximate the integrand of the inverse transform \eqref{eq:inverseBorel}, which in principle should be the analytically continued Borel-transform if we knew all its coefficients and were able to sum it up.

Our procedure (Borel-Pad\'e resummation) to evaluate $S^+$ of a formal asymptotic series $\Psi(v)$ from the finitely many numerical coefficients then looks
\begin{align}\label{eq:resumproc}
	\Psi(v) \; \overset{\text{Borel-transform}}{\longrightarrow} \;  \mathcal{B}_{\Psi}(t) \; \overset{\text{diagonal Pad\'e}}{\longrightarrow} \; \big[ \tfrac{N}{2} \big/ \tfrac{N}{2} \big]_{\mathcal{B}_{\Psi}}(t) \; \overset{\substack{\text{Laplace-tf. over}\\\text{tilted contour}}}{\longrightarrow} \; S^+_{N}(\Psi)(v) 
\end{align}	
where the series is known up to a certain order $N$: 
\begin{align}
	\Psi(v) &= \sum_{n=0}^{N}\psi_n v^n + \rmO(v^{N+1}), & \mathcal{B}_{\Psi}(t) &= \sum_{n=0}^{N} \frac{\psi_n}{n!} t^n  + \rmO(t^{N+1}).
\end{align}
Note that in this section the Borel transform (and its inverse) are defined slightly differently from, but equivalently to \eqref{eq:Borel} and \eqref{eq:inverseBorel}. Now the diagonal Pad\'e-approximant of $\mathcal{B}_{\Psi}(t)$ is a ratio of two polynomials of order $\frac{N}{2}$, i.e. $P_{\frac{N}{2}}(t)$ and $Q_{\frac{N}{2}}(t)$
\begin{align}\label{eq:resum}
	\big[ \tfrac{N}{2} \big/ \tfrac{N}{2} \big]_{\mathcal{B}_{\Psi}}(t) &=\frac{P_{\frac{N}{2}}(t)}{Q_{\frac{N}{2}}(t)}, & \text{such that} & &  \mathcal{B}_{\Psi}(t) &=\frac{P_{\frac{N}{2}}(t)}{Q_{\frac{N}{2}}(t)} +\rmO(t^{N+1}),
\end{align}
which we integrate along a contour above the positive real line 
\begin{align}
	S^+_{N}(\Psi)(v) &= v^{-1} \int\limits_0^{{\rm e}^{i \varphi} \infty} \mathrm{d}t \, {\rm e}^{-t/v} \big[ \tfrac{N}{2} \big/ \tfrac{N}{2} \big]_{\mathcal{B}_{\Psi}}(t),
\end{align}
where $\varphi$ is some acute angle between the real axis and the contour of integration to perform the numerical integrals.  

The technique described above was used to evaluate the trans-series of the \mO{3} model's densities \eqref{eq:epsrho}.\footnote{All the integrations were performed at $\varphi=3/5$.} There were $N_{\max}=336$ terms at our disposal for the perturbative sector from the data obtained by Volin's method \cite{Volin:2010cq} previously in \cite{Bajnok:2021zjm,Bajnok:2022rtu}, and thus at every NP order we had the same amount of trans-series coefficients.\footnote{See Appendix D of \cite{Bajnok:2022rtu} on how to re-express the $B^{-1}, \ln B$ expansion of the original algorithm \cite{Volin:2010cq} in terms of the running coupling $v$. See also the remark above \eqref{eq:runningswitch}.}

\begin{figure}[t]
	\centering
	\includegraphics[width=0.48\textwidth]{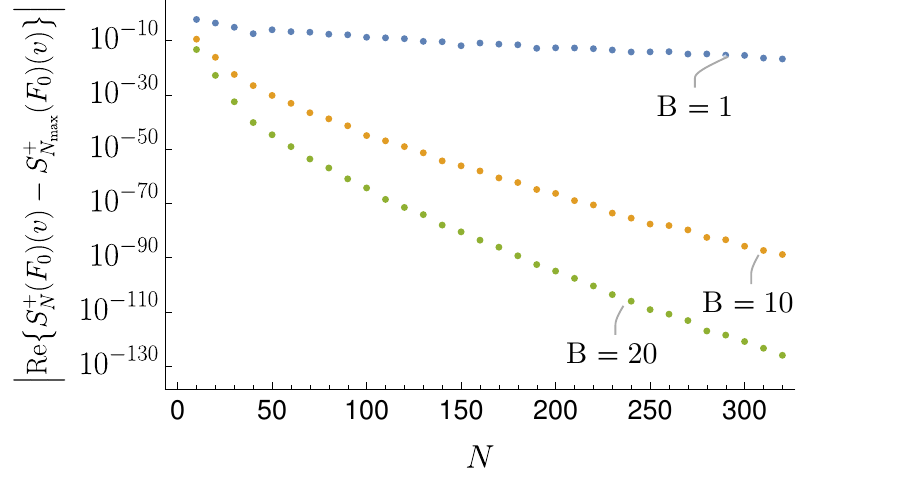}
	\includegraphics[width=0.48\textwidth]{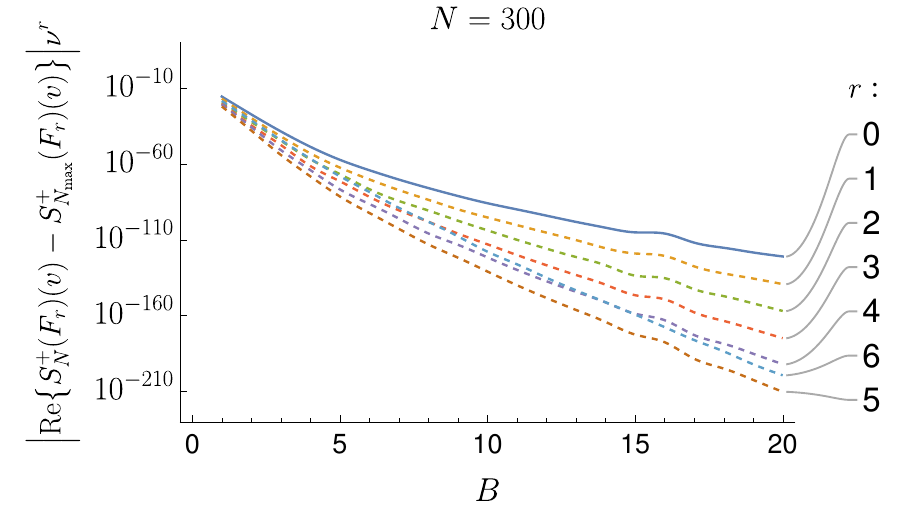}
	\caption{
		Left: Convergence of the $\text{Re}\, S^+_N(F_0)(v)$ sums for increasing $N$ to the most precise resummation available at $N=N_{\max}$; plotted for three different $B$ values. On a logarithmic scale, the difference shows a qualitatively exponential decay for large $N$. 
	Right: Comparison of the relevance of the different $\text{Re}\, S^+_N(F_r) \nu^r$ terms in the resummed trans-series as error sources. The solid line shows that the estimated error from the perturbative sector dominates.}
	\label{fig:NBplot}
\end{figure}

For a rough estimate of the magnitude of the resummation's error, we compared the Borel-Pad\'e sum evaluated for only $\approx 90 \%$ of these coefficients (i.e. $N=300$) to the same resummation performed for all known data.\footnote{{See the discussion on this choice later.}} Based on the left of Figure \ref{fig:NBplot}, this comparison is an upper estimate for the error, as we expect the difference of the $\text{Re}\, S^+_{N_{\max}}$ sum to the exact value orders of magnitude smaller than the difference of $ \text{Re}\, S^+_N$ to $ \text{Re}\, S^+_{N_{\max}}$ for $N < N_{\text{max}}$. Also, when assessing the error, we only have to consider the leading, perturbative sector as our main source of inaccuracies. This is because the error of the resummation is approximately the same for each asymptotic series (e.g. $S^+(F_r)$ for each $r$), yet these uncertainties are suppressed by the exponential factor $\nu^r$ (this is shown on the right of Figure \ref{fig:NBplot}). 

\begin{figure}[t]
	\centering
	\includegraphics[width=0.9\textwidth]{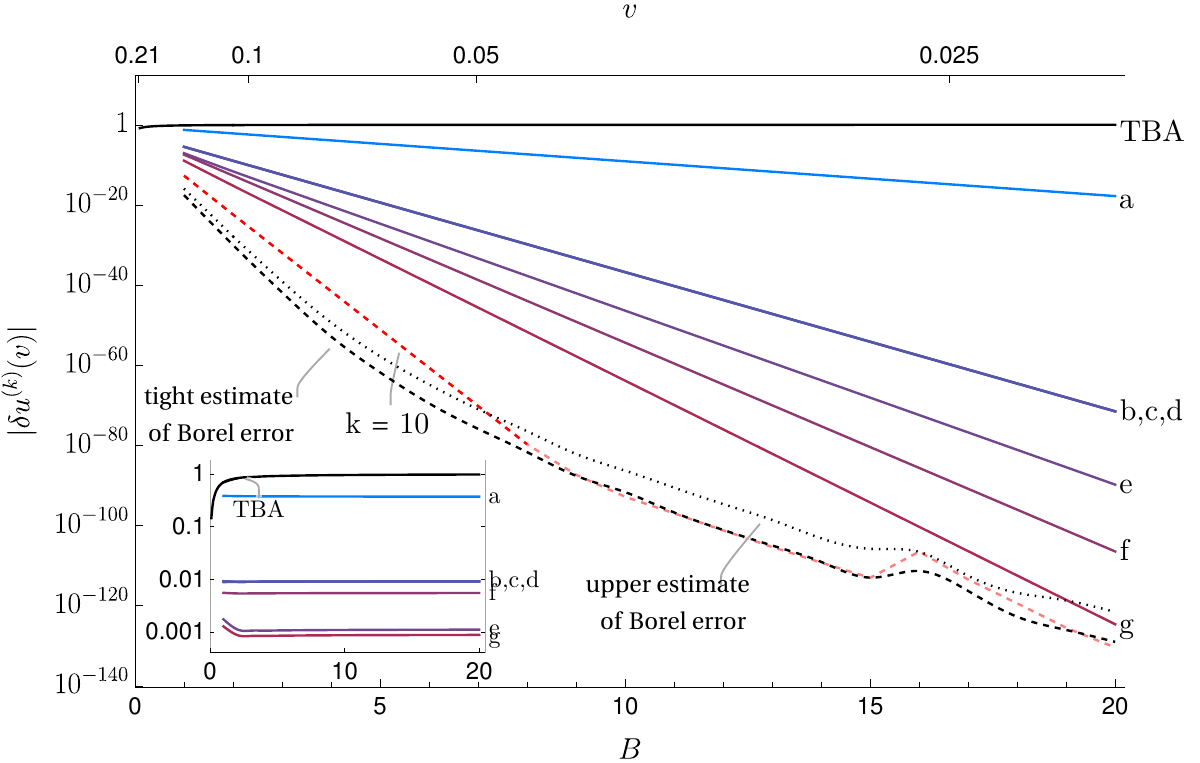}
	\caption{The difference of the TBA and the various truncations of the resummed trans-series of the normalized particle density $u(v)$ for the \mO{3} model as a function of $B$ (see Table \ref{tab:WV} for the notations a-g of the coloured lines). The black line shows the TBA value itself. The same data is plotted on the inset as on the main plot, but the lines on the latter are rescaled by the appropriate inverse power of $\nu$ (see Table \ref{tab:WV}), to have a constant large $B$ asymptotics. The red dashed line for $\delta u^{(k=10)}(v)$ is added to show how the error of the resummation would influence further subtractions. }
	\label{fig:VO3}
\end{figure}

\begin{figure}[t]
	\centering
	\includegraphics[width=0.9\textwidth]{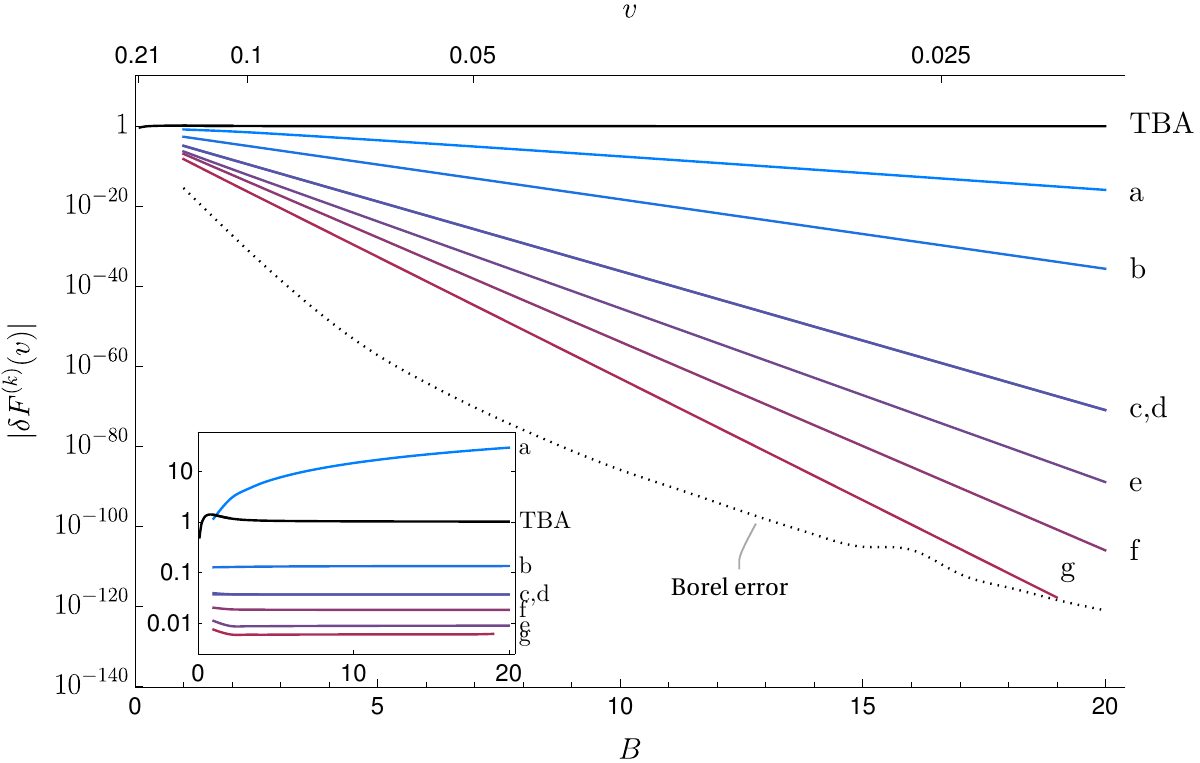}
	\caption{Similar demonstration as in Figure \ref{fig:VO3} for the normalized energy density $F(v)$ of the \mO{3} model. See the explanation for the labeling in Table \ref{tab:WV}. The line $a$ of the inset shows linear growth in $B$ (on a log scale) due to the $1/v+\ln v$ term in $F_1$ - see \eqref{eq:O3FTS}.
	}
	\label{fig:WO3}
\end{figure}

The estimated relative precision of the TBA result for the \mO{3} model was $\simeq 10^{-257}$ even at the largest $B=20$ value. To reach this accuracy we used the recursive method in \cite{ristivojevic2019conjectures}, and a basis of even Chebyshev polynomials up to the order 1400 was chosen. The relative magnitudes of the last and first expansion coefficients of the solution in this finite Chebyshev basis indicated the above-mentioned relative error. As this value is several orders of magnitude smaller than the last subtractions plotted in Figures \ref{fig:VO3} and \ref{fig:WO3} (lines labeled as $g$), it does not affect our analysis over the range $1 \leq B\leq 20$.

\begin{table}[t]
	\centering
	\begin{tabular}{c c c c c c c c c c}
		&	& & a & b & c & d & e & f & g \\ 
		\cline{2-10}
		 & & $k$  & 0 & 1  & 2  & 3 & 4 & 5 & 6 \\  
		\cline{2-10}
		\rule{0pt}{3ex}
		\multirow{8}{*}{\mO{3}} 	& \multirow{5}{*}{$u(v)$} & $\tilde{u}^{(k)}-\tilde{u}^{(k-1)}$ & $\mathbb{R}\nu^0$ & $\mathbb{R}\nu^1$  & $i\mathbb{R}\nu^2$  & $i\mathbb{R}\nu^3$  & $\mathbb{C}\nu^4$ & $\mathbb{C}\nu^5$ & $\mathbb{C}\nu^6$ \\  
								& & $ \text{Re}\, S^+(u_k)\nu^k $  & $+\nu^0$ & $-\nu^1$  &  $-\nu^4$  & $-\nu^5$  & $+\nu^4$ & $+\nu^5$ & $+\nu^6$ \\
								  & &$\delta u^{(k)}$ & $-\nu^1$ & $-\nu^4$  &  $+\nu^4$  & $+\nu^4$  & $+\nu^5$ & $+\nu^6$ & $+\nu^7$ \\
		\cline{4-10}\rule{0pt}{3ex}						  
						&&		$ \text{Im}\, S^+(u_k)$ & $-\nu^2$ & $ -\nu^3$  & $+\nu^2$  & $ +\nu^3$  & $ +\nu^4$ & $ +\nu^5$ & $ +\nu^6$ \\
		 				&&		$ \text{Im}\, S^+(\tilde{u}^{(k)})$ & $-\nu^2$ & $ -\nu^2$  & $-\nu^3$  & $ -\nu^4$  & $ -\nu^5$ & $ -\nu^6$ & $ -\nu^7$\\ 
					\cline{3-10}\rule{0pt}{3ex}
					& \multirow{3}{*}{$F(v)$} & $\tilde{F}^{(k)}-\tilde{F}^{(k-1)}$ & $\mathbb{R}\nu^0$ & $i M_1\nu^1 +\mathbb{R}\nu^1$  & $\mathbb{C}\nu^2$  & $i\mathbb{R}\nu^3$  & $\mathbb{C}\nu^4$ & $\mathbb{C}\nu^5$ & $\mathbb{C}\nu^6$ \\
								  & & $ \text{Re}\, S^+(F_k)\nu^k $   & $+\nu^0$ & $+\nu^1$  &  $-\nu^2$  & $-\nu^5$  & $+\nu^4$ & $+\nu^5$ & $+\nu^6$ \\ 
								  & & $\delta F^{(k)}$ & $+\nu^1$ & $-\nu^2$  & $+\nu^4$  & $+\nu^4$  & $+\nu^5$ & $+\nu^6$ & $+\nu^7$ \\ 

		\cline{2-10} 
		  & & $k$  & 0 & 2  & 4  & 6 & 8 &  &  \\ 
		\cline{2-8} 
		\rule{0pt}{3ex}
		\multirow{6}{*}{\mO{4}} 	&  \multirow{3}{*}{$u(v)$}     		& $\tilde{u}^{(k)}-\tilde{u}^{(k-2)}$	& $\mathbb{R}\nu^0$ 	& $i\mathbb{R}\nu^2$  	& $\mathbb{C}\nu^4$  	& $\mathbb{C}\nu^6$  & $\mathbb{C}\nu^8$ &		& 	  \\	
									&	& $\text{Re}\, S^+(u_k)\nu^k$	& $+\nu^0$ 		& $-\nu^4$		& $+\nu^4$		& $+\nu^6$	     & $+\nu^8$		 & 		&  	  \\
									&	&  $\delta u^{(k)}$	& $-\nu^4$ 		& $+\nu^4$  		& $+\nu^6$  		& $+\nu^8$  	     & $+\nu^{10}$	 &   		&  	  \\ 
					\cline{3-8}\rule{0pt}{3ex}
					& \multirow{3}{*}{$F(v)$}		& $\tilde{F}^{(k)}-\tilde{F}^{(k-2)}$ 	& $\mathbb{R}\nu^0$ 	& $iM_1\nu^1+i\mathbb{R}\nu^2$  & $\mathbb{C}\nu^4$  	& $\mathbb{C}\nu^6$  	& $\mathbb{C}\nu^8$ 	&		&   \\
									&	&  $ \text{Re}\, S^+(F_k)\nu^k $	& $+\nu^0$ 		& $-\nu^4$			& $+\nu^4$		& $+\nu^6$		& $+\nu^8$		&  		&   \\
									&	& $\delta F^{(k)}$			& $-\nu^4$ 		& $+\nu^4$  			& $+\nu^6$  		& $+\nu^8$  		& $+\nu^{10}$     	&  	  	&  \\ 
 
		\cline{2-8}
	\end{tabular}			
	\caption{Explanation for the alphabetical notations in Figures \ref{fig:VO3},\ref{fig:WO3},\ref{fig:WVO4}. For each model \mO{3},\mO{4} and both quantities $u(v),F(v)$ the table presents three different rows (except for $u(v)$ of \mO{3}, where two extra rows were added for the imaginary parts). In every case, the upper one indicates the magnitudes (in powers of $\nu$) and the reality properties ($\mathbb{R}:\text{real}$,  $i\mathbb{R}:\text{imaginary}$, $\mathbb{C}:\text{complex}$) of the last term(s) included in the truncated trans-serieses $\tilde{u}^{(k)},\tilde{F}^{(k)}$  (i.e. $u_k \nu^k, F_k \nu^k$, except for the $i M_1 \nu^1$ terms). The middle one shows the order of the magnitude and the sign for the real part of the lateral resummation of the same terms - that is $\text{Re}\, S^+ (u_k)\nu^k,\; \text{Re}\, S^+ (F_k)\nu^k $.  Finally, the lower row tells us the sign and the order of magnitude for the differences $\delta u^{(k)}, \delta F^{(k)}$ themselves.}
	\label{tab:WV}
\end{table}
Our results for the \mO{3} model i.e. $\delta u^{(k)}(v), \delta F^{(k)}(v)$ for $k=0,1,\ldots,6$ are shown on Figures \ref{fig:VO3},\ref{fig:WO3}. These differences are all exponentially small in $B$ and each tends to a straight line on a logarithmic scale, corresponding to some integer power of $\nu$. This power can be easily determined (the insets of these plots show the same data divided by exactly this power of $\nu$), and it is always larger or equal\footnote{ 
Note that for the asymptotic series where $u_r,F_r$ itself has only purely imaginary coefficients, the scaling of the real part of their resummation is exponentially suppressed even compared to $\nu^r$. This means an additional power of $\nu^p$ where $p>0$ is the position $t=p$ of the closest singularity to the origin on the transformed series' Borel plane along the positive real axis (see Table \ref{tab:WV}).} to the one for the previous subtraction. 
The importance of this is that if an NP correction term from the resummed trans-series would be incorrect, then we were stuck at a given $\nu$ power endlessly when subtracting more and more trans-series terms (being orders of magnitude smaller) from the TBA result. Moreover, it turns out that the exponential order of $\delta u^{(k)}(v), \delta F^{(k)}(v)$ is the same as that of the next term in the trans-series which would have a leading exponential scaling after taking the real part of its Borel-Pad\'e resummation (see Table \ref{tab:WV}).
	
Although it is easy to generate the coefficients $u_r,F_r$ for $r>0$, we stopped at $k=6$ when plotting the subtractions. (See some demonstration of the accuracy for this truncation in Table \ref{tab:prec}.) We opted to do this, since over the chosen range of $B$ values, the $\delta u^{(k)}(v), \delta F^{(k)}(v)$ curves for $k>6$ would reach the magnitude of the resummation's error (see the dotted lines on Figures \ref{fig:VO3},\ref{fig:WO3}) way below the upper end $B=20$. 
\begin{table}[t]
	\centering
	\begin{tabular}{c c c c}
		 	&  $B = 1$	& $B=10$ & $B = 20$   \\
		\cline{2-4}\rule{0pt}{3ex}
		$u_{\text{TBA}}(v)$  				& {\small $0.69095550${\tiny 602} }  & { \small $0.94918\ldots943253${\tiny 265} } &  {\small $0.97235\ldots201968${\tiny 450}  }\\
		$\text{Re}\, S^+_{N=336}(\tilde{u}^{(k=6)})$ 	& {\small $0.69095550${\tiny 496} }  & { \small $0.94918\ldots943253${\tiny 128} } & {\small  $0.97235\ldots201968${\tiny 231}  }\\
		digits in agreement 					& $8$ 			   & $63$		& $124$  	\\										  
	\end{tabular}
	\caption{The accuracy of our best numerics in the case of the particle density of the \mO{3} model. The agreeing number of digits after the decimal point is shown, together with the following three digits in smaller font. }
	\label{tab:prec}
\end{table}

\begin{figure}[t]
	\centering
	\includegraphics[width=0.9\textwidth]{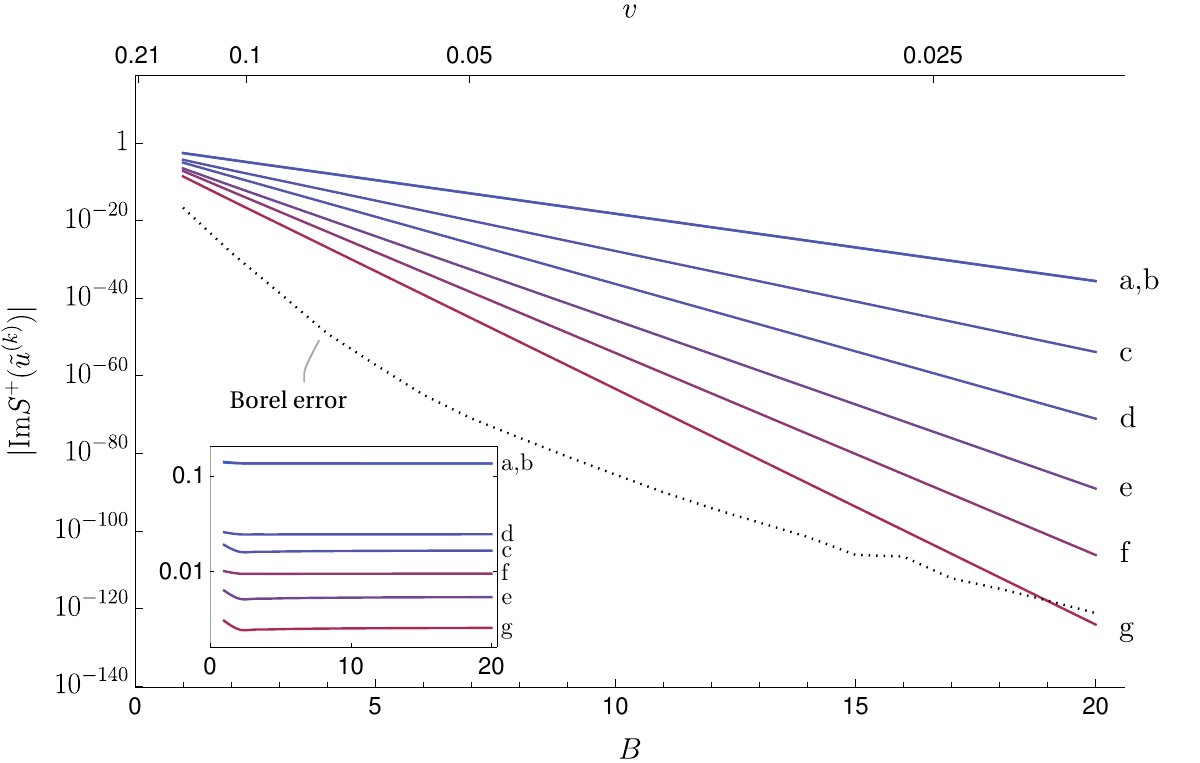}
	\caption{{The imaginary part of the resummed trans-series at different truncations (see Table \ref{tab:WV}) as a function of $B$, in case of the \mO{3} model's particle density. When including more terms from the trans-series $\tilde{u}$, the sum decreases. The inset shows the same data divided by the corresponding $\nu$ power taken from Table \ref{tab:WV}. The error was estimated in a similar way, as for the real part.}}
	\label{fig:VO3im}
\end{figure}

Our experience is that indeed, when reaching this approximate error level, the sofar straight lines start to diverge. To demonstrate this, we added the red dashed line corresponding to $\delta u^{(k=10)}(v)$ to Figure \ref{fig:VO3} which stops behaving in the expected way at around $B=8$. Comparing its $B>8$ part (which is clearly influenced by the error) to the black dotted line it is also clear that our upper estimate for the error was slightly pessimistic. {We could have used a much tighter estimation on the error by comparing the real parts of $S^+_{N_{\max}-2}$ and $S^+_{N_{\max}}$ (see the black dashed line on Figure \ref{fig:VO3}). Although this seems to predict the magnitude at which our subtraction method will fail more precisely in most cases, it also seems to underestimate the error for certain $B$ values. Therefore we opted for the more conservative estimate in case of Figures \ref{fig:WO3}-\ref{fig:FO4}, to be on the safe side.}

In the end, the closest approximation to the TBA we were able to obtain over this domain from the perturbative data we had is then for $B=20$ and $B=19$, and is of the order $ \simeq 10^{-125}$ and $\simeq 10^{-118}$ respectively for $\delta u^{(k=6)}(v)$ and $\delta F^{(k=6)}(v)$, as can be deduced from the plots.

{For a consistency check, we also calculated the imaginary parts $\text{Im}\, S^+(\tilde{u}^{(k)})$ for the same truncations $k=0,1,\ldots,6$. As expected, for increasing $k$, the sums are decreasing in magnitude, i.e. they show a $\nu^p, \; p \in \mathbb{Z}^+ $ behaviour with increasing values of the power $p$, just as the differences $\delta u^{(k)}(v)$ did. See the extra rows for the particle density of the \mO{3} model in Table \ref{tab:WV}, and the corresponding data on Figure \ref{fig:VO3im}.
Throughout this section, similarily to the above case, we followed the cancellation of the imaginary parts of the resummed trans-series terms for every other quantity in Tables \ref{tab:WV}, \ref{tab:F}. In each case, we confirmed the reduction in the sum's magnitude (scaling as some integer power in the corresponding NP parameter of the problem) when truncating the trans-series at higher orders.
}

{At this point we ought to make a remark about the constants at which the lines on the insets of Figures \ref{fig:VO3},\ref{fig:WO3} tend to, for large values of $B$. Each is equal to the real part of the leading perturbative coefficient of the first such NP term in the trans-series, which was not included in the truncation yet - and whose leading coefficient has a real part at all.\footnote{By Table \ref{tab:WV}, after the first few $\nu$ orders, each NP term has complex perturbative coefficents.} Both for $u(v)$ and $F(v)$ the magnitude of these constants tend to decrease, with increasing order of the NP parameter. This is true even for the \mO{4} model (see Figure \ref{fig:WVO4}). For the free energy of both models in terms of the physical running coupling, from the insets of Figures \ref{fig:FO3},\ref{fig:FO4} we see that it is plausible, that the same leading coefficients do not grow with the NP order too fast. That is, for a small enough value of the NP parameter the resummed trans-serieses of these quantities seem to have good convergence properties. This is consistent with the findings of \cite{Marino:2023epd} where numerical evidence for the convergence of the free energy's Borel-resummed trans-series was presented in the case of the \mO{$N\geq 4$} models. The argument in the latter work was based on the convergence properties of the LO and NLO perturbative coefficients of each NP term in the trans-series (trans-asymptotics). By knowing the full trans-series up to arbitrarily high perturbative orders in principle, the convergence can be tested by direct evaluation of the Borel-resummed trans-series. We plan to publish such investigations in our following paper \cite{Bajnok:2022prepi}.}

\subsubsection{The \mO{4} model}

\begin{figure}[t]
	\centering
	\includegraphics[width=0.9\textwidth]{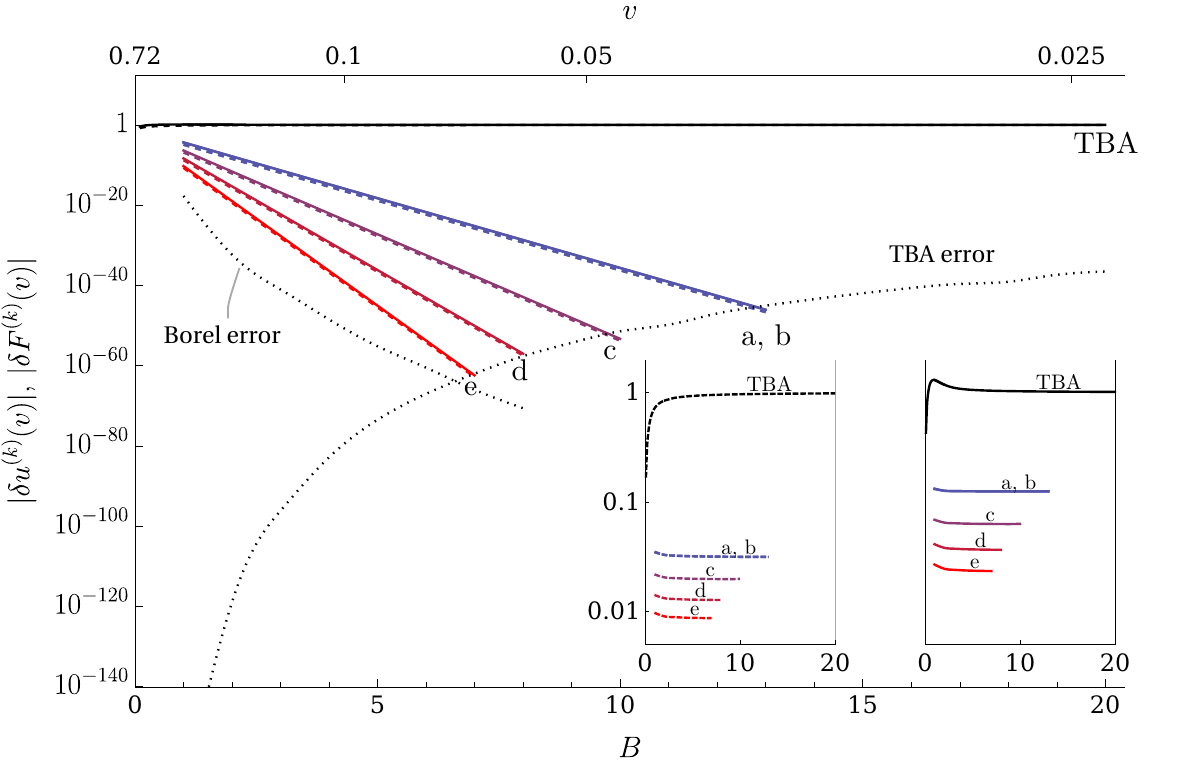}
	\caption{ Comparison of the TBA result and the truncated and resummed trans-series for the normalized particle and energy densities of the \mO{4} model (dashed vs solid lines for $u(v)$ and $F(v)$ respectively).}
	\label{fig:WVO4}
\end{figure}

Now let us turn to the case of the \mO{4} model. The trans-series of the particle density starts as
\begin{align}\label{eq:O4uTS} 
	\tilde{u}(v) & \overset{\text{\tiny \mO{4}}}{=}\Bigg\lbrace 1-\frac{3 v}{4}+\rmO\left(v^2\right)  \Bigg\rbrace 
	+ i \frac{\nu^2}{2^2} \Bigg\lbrace 1+\frac{v}{2}+\rmO\left(v^2\right)  \Bigg\rbrace \nonumber \\
	&+ \frac{\nu^4}{2^4} \Bigg\lbrace \left(\frac{1}{2}+\frac{3 i}{4}\right)+\left(\frac{1}{8}+\frac{9 i}{32}\right) v +\rmO\left(v^2\right)   \Bigg\rbrace \nonumber\\
	&+ \frac{\nu^6}{2^6} \Bigg\lbrace \left(\frac{5}{4}+i\right)+\left(\frac{9}{32}+\frac{5 i}{12}\right) v  +\rmO\left(v^2\right)  \Bigg\rbrace \nonumber \\
	&+ \frac{\nu^8}{2^8} \Bigg\lbrace \left(\frac{103}{32}+\frac{103 i}{64}\right)+\left(\frac{293}{384}+\frac{827 i}{1024}\right) v  +\rmO\left(v^2\right)  \Bigg\rbrace
	+ \rmO(\nu^{10}), 
\end{align}
while that of the energy density as
\begin{align}\label{eq:O4FTS}	
	\tilde{F}(v) & \overset{\text{\tiny \mO{4}}}{=} \Bigg\lbrace  1+\frac{v}{2}  +\rmO\left(v^2\right)  \Bigg\rbrace  -2 i \nu 
	+ i \frac{\nu^2}{2^2} \Bigg\lbrace 4 +v   +\rmO\left(v^2\right)  \Bigg\rbrace \nonumber \\
	&+ \frac{\nu^4}{2^4} \Bigg\lbrace (2+2 i)+\frac{3 i v}{4}  +\rmO\left(v^2\right)   \Bigg\rbrace \nonumber\\
	&+ \frac{\nu^6}{2^6} \Bigg\lbrace (4+2 i)+\left(\frac{1}{2}+\frac{3 i}{2}\right) v  +\rmO\left(v^2\right)  \Bigg\rbrace \nonumber \\
	&+ \frac{\nu^8}{2^8} \Bigg\lbrace \left(\frac{37}{4}+\frac{9 i}{4}\right)+\left(\frac{33}{16}+\frac{207 i}{64}\right) v  +\rmO\left(v^2\right)  \Bigg\rbrace
	+ \rmO(\nu^{10}) .
\end{align}
For previous works \cite{Abbott:2020mba, Abbott:2020qnl} we took advantage of Volin's method to generate 2000 of the coefficients for the perturbative sector. In principle, this would allow us to determine the same amount of coefficients for all the NP corrections. For the present work we restricted ourselves to half of them, that is, we performed the Borel-Pad\'e resummations as described in \eqref{eq:resumproc} for $N_{\max} = 1000$ coefficients at each NP order\footnote{With $\varphi=1/10$ integration angle.}. 
To get a picture of the error of this procedure we made a comparison to $S^+_N$ taken at $N=900$. In this case, however, the error source limiting our comparison's domain of applicability was the accuracy of the numeric TBA solution as can be seen from Figure \ref{fig:WVO4}. This happened since due to the more complicated functional form of the \mO{4} model's TBA kernel, we had to stick to the less efficient method in \cite{Abbott:2020qnl}, and thus the highest Chebyshev polynomial involved in our expansion was only the $200^{\text{th}}$. 

Note that as the NP orders in the trans-series for the \mO{4} case are suppressed by even powers in $\nu$ (with the exception of the $i M_1 \nu $ term), we were able to check fewer terms from the trans-series compared to the \mO{3} case, as the smallest ones (lines labeled $e$ on Figure \ref{fig:WVO4}) have already higher order in $\nu$ than the last ones, i.e. lines $g$ on Figures \ref{fig:VO3},\ref{fig:WO3}. 

The best approximation of the TBA by the trans-series we could reach this way was of the order $\simeq 10^{-63}$ at $B=7$  for both $\delta u^{(k=8)}(v)$ and $\delta F^{(k=8)}(v)$.

\subsection{Free energy density}\label{subsec:freeenergy}

When analysing the physical free energy density, for notational consistency we define and denote its normalized version as $G(\beta)$ in terms of the physical running coupling:\footnote{Here $\Delta = 1 $ for \mO{3} and $\Delta=1/2$ for \mO{4} as usual.}
\begin{align}\label{eq:gbeta}
	\mathcal{F} &= - \frac{h^2}{4 \pi \Delta} G(\beta), & G(\beta) = \frac{g(\beta)}{\beta}.
\end{align}
However, due to its leading $1/\beta$ behaviour - see \eqref{LEAD} - it is more convenient to work with the rescaled quantity $g(\beta)$ defined above, whose trans-series $\tilde{g}(\beta)$ starts with unity - just like those of $\tilde{F}(v)$ and $\tilde{u}(v)$.

The physical couplings are defined as: 
\begin{align}\label{betalimsO3}
	\frac{1}{\beta} \; + \; \ln \beta &= \ln\left(\frac{h}{m}\right) \; + \; \ln \left(\frac{8}{e}\right) &  \begin{matrix}    0.6881 & \gtrsim  \beta > & 0 \quad\;  \\   \quad 1 &<  \frac{h}{m} < & \infty \quad\quad \\   \quad 0 & < B < & \infty \quad\quad \end{matrix}  &&& \text{for }\text{O}(3) \\
		\nonumber	\\ \label{betalimsO4}
		\frac{1}{\beta} + \frac{1}{2}\ln \beta &= \ln\left(\frac{h}{m}\right) + \frac{1}{2} \ln \left(\frac{32}{e\pi} \right) &  \begin{matrix}  \quad\quad\; 2 &> \beta > & 0  \quad\quad \\ 1.2045 &\lesssim \frac{h}{m} < & \infty  \quad\quad  \\ 0.6688 & \lesssim B < &\infty  \quad\quad \end{matrix}  &&& \text{for }\text{O}(4),
\end{align}	
where the approximate bound on $B$ for \mO{4} was determined from the TBA result of $h/m$.

Analogously to the differences introduced in \eqref{eq:dudF}, we compare the TBA vs the real part of the Borel resummation via
\begin{align}\label{eq:dg}
	\delta g^{(k)}(\beta) &= g_{\text{TBA}}(\beta) - \text{Re} \, S_+(\tilde{g}^{(k)})(\beta),   &   \tilde{g}^{(k)}  &= \sum_{r=0}^k g_r \mu^r, 
\end{align}
where for notational simplicity we introduced the NP parameter $\mu$ as $\mu \equiv {\rm e}^{-2/\beta}$ for \mO{4} while $\mu \equiv \frac{2}{\beta} {\rm e}^{-2/\beta}$ for \mO{3}. In the latter case, the extra $\frac{2}{\beta}$ factor was introduced to shift the $g_r$ asymptotic serieses such that each (except $g_1$) starts with a constant instead of some negative power of $\beta$.   

\subsubsection{The \mO{3} model}

The coefficients of the \mO{3} model up to the first $\zeta_3$-containing term in $\beta$ at each analyzed NP order are given below:
\begin{align}\label{eq:O3F}
	\tilde{g}(\beta) &\overset{\text{\tiny \mO{3}}}{=} 1
	-\frac{\beta }{2}
	-\frac{\beta^2}{2}
	- \left(\frac{7}{8} - \frac{3}{32}\zeta_3 \right) \beta^3  + \rmO(\beta^4) \nonumber \\ 
	&+\left(\frac{32}{{\rm e}^2 \beta} \right){\rm e}^{-2/\beta}  \Bigg\lbrace -\frac{2}{\beta } 
	+ \left(\frac{i\pi }{2} + \gamma_E + 5 \ln 2 - 3  - \ln\beta  \right) \nonumber \\
	&\quad\quad\quad\quad\quad\quad\quad\quad\quad \quad\quad\quad\quad\quad -\frac{\beta }{2} +\frac{1}{32}  (4-3 \zeta_3) \beta^3 + \rmO(\beta^4) \Bigg\rbrace   \nonumber\\
	&+\left(\frac{32}{{\rm e}^2 \beta}  \right)^2 {\rm e}^{-4/\beta} \Bigg\lbrace \frac{1}{2}\left(1-i\right) 
	    -\frac{1}{8}\left(1-\frac{3 i}{2}\right) \beta 
	    +\frac{1}{8}\left(1-\frac{i}{32}\right) \beta ^2 \nonumber\\
	    &\quad\quad\quad\quad\quad\quad\quad\quad+ \frac{1}{32} \left[\left(3-\frac{129 i}{64}\right)-\frac{3}{2}\left(1-i\right) \zeta_3\right] \beta ^3 + \rmO(\beta^4)  \Bigg\rbrace   \nonumber\\
	&+\left(\frac{32}{{\rm e}^2 \beta}  \right)^3 {\rm e}^{-6/\beta} \Bigg\lbrace  \frac{1}{4}\left(1-2 i\right)
	- \frac{1}{24}\left(1 - \frac{13 i}{2}\right) \beta  
	+\frac{1}{12}\left(1-\frac{83 i}{64}\right) \beta ^2 \nonumber\\
	&\quad\quad\quad\quad\quad\quad\quad\quad+\frac{1}{32}   \left[\left(5-\frac{317 i}{64}\right)-3\left(\frac{1}{2}- i\right) \zeta_3\right] \beta^3  + \rmO(\beta^4) \Bigg\rbrace + \rmO({\rm e}^{-8/\beta}).
\end{align}	
The Borel-Pad\'e resummation was performed in a similar manner as we defined the Borel-resummation \eqref{eq:inverseBorel} for an asymptotic series starting from some negative power of the perturbative variable. That is, we separated the $\beta^{-p},\; p\geq 1$ terms from the power series $g_r (2/\beta)^r$, and used the method as explained in \eqref{eq:resum} for the rest of the coefficients (now in $\beta$). Finally, we added the separated terms to this sum thus obtaining the total value. We evaluated the $\ln \beta$ term in $g_1$ in the same way. 

To carry out the numerics we used only $N_{\max}=100$ coefficients ($N=90$ to estimate the error), and were calculating $\delta g^{(k)}$ up to $k=3$ only, as obtaining more coefficients proved to be time-consuming with the method explained in the introductory part of the present section (i.e. fixing an ansatz for $\tilde{G}(\beta)$ and obtaining its coefficients from $\tilde{G}(\tilde{b}(v))\overset{!}{=}\Omega(v)$). 

The TBA data was the same that we used for Figures \ref{fig:VO3},\ref{fig:WO3}. The definition \eqref{eq:freeenergydef} and the normalization \eqref{eq:gbeta} was used to determine $g_{\text{TBA}}(\beta)$ at a given $B$ value. The $\beta$ values to $B=1,2,\ldots,20$ were obtained from numerically solving \eqref{betalimsO3} using $h/m$ as an input from the boundary values of the TBA solutions as in \eqref{eq:hperm}.

\begin{figure}[t]
	\centering
	\includegraphics[width=.9\textwidth]{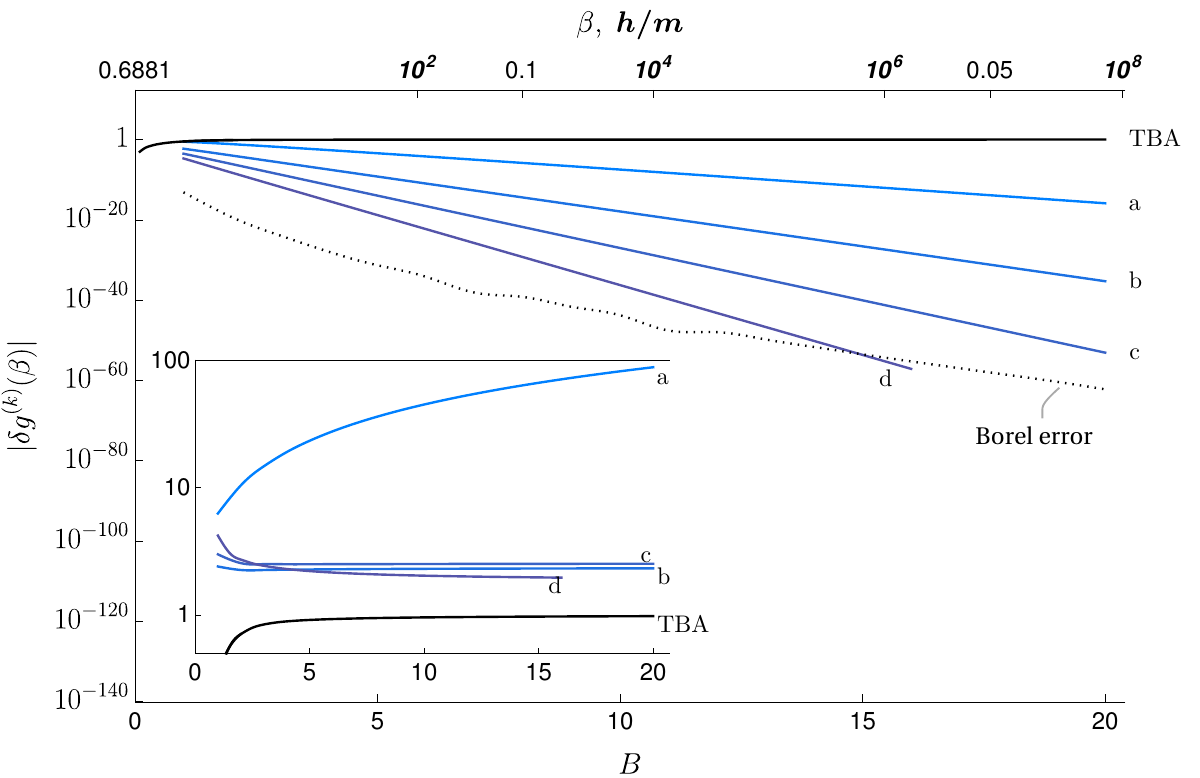}
	\caption{The comparison plot for the normalized free energy density $g(\beta)$ of the \mO{3} model in terms of $B$. The upper scale shows the corresponding $\beta$ and $h/m$ values simultaneously. See Table \ref{tab:F} for explanations of the labeling. The inset shows the same data as the main plot divided by the appropriate power of $\mu=\frac{2}{\beta} {\rm e}^{-2/\beta}$ to have a constant behaviour for large $B$, as given in the table. Line $a$ of the inset grows linearly (on a log-scale) in terms of $B$ due to the $-\frac{2}{\beta} - \ln \beta $ term appearing in the second line of \eqref{eq:O3F}.}
	\label{fig:FO3}
\end{figure}

The results are shown in Figure \ref{fig:FO3}. With this data, we could verify the agreement to the TBA in the physical coupling up to $\simeq 10^{-57}$ at $B=16$ based on line $d$.

\subsubsection{The \mO{4} model}

\begin{figure}[h]
	\centering
	\includegraphics[width=.9\textwidth]{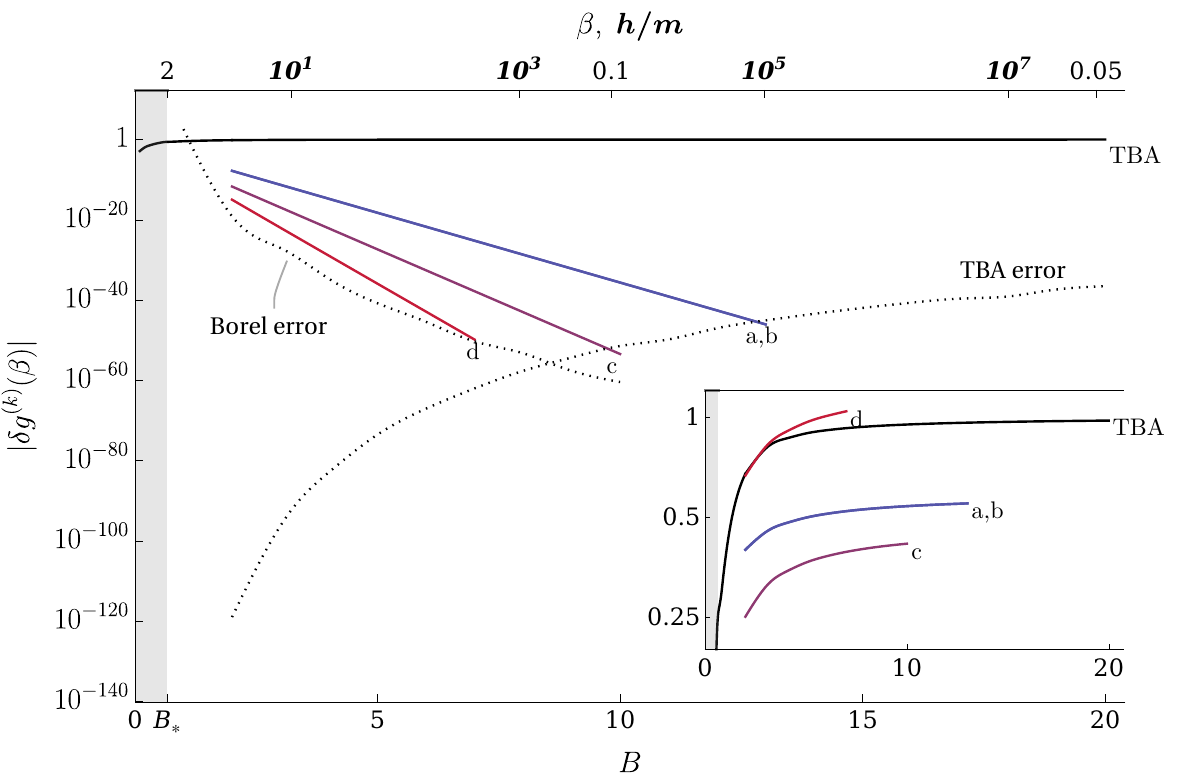}
	\caption{The truncated trans-series vs TBA plot for the normalized free energy density $g(\beta)$ of the \mO{4} model (see also Table \ref{tab:F}). The inset shows the main plot's data divided by the correct power of $\mu={\rm e}^{-2/\beta}$. The shaded region indicates that the $\beta$ coupling is defined - and so the lateral resummation of the trans-series makes sense - above $B_* \simeq 0.6688 $ or below $\beta < 2$ according to \eqref{betalimsO4}. Both the resummation's error and the TBA uncertainty limit the lines shown. The first even restricts the domain from below, that is why the data was plotted only for $B\geq 2$.}
	\label{fig:FO4}
\end{figure}

For our last numerical investigation, we took the free energy density of the \mO{4} model under scrutiny. The coefficients of the trans-series (up to the first terms containing $\zeta_3$) are presented below:
\begin{align}\label{gO4TS}
	\tilde{g}(\beta) & \overset{\text{\tiny \mO{4}}}{=} \Bigg\lbrace 1
		-\frac{\beta }{2}
		-\frac{\beta ^2}{4}
		-\left(\frac{5}{16} - \frac{3 \zeta _3}{32}\right) \beta ^3  + \rmO(\beta^4) \Bigg\rbrace + \frac{8 i}{e}{\rm e}^{-2/\beta} \nonumber \\
		& - i \left( \frac{2}{e}\right)^2 {\rm e}^{-4/\beta}\Bigg\lbrace 2 
	-\frac{3   }{4}\beta
	+\frac{13 }{64} \beta^2
	-\left(\frac{29 }{512}+ \frac{3  \zeta _3}{16}\right) \beta^3  + \rmO(\beta^4) \Bigg\rbrace  \nonumber\\
	&-\left(\frac{2}{e}\right)^4 {\rm e}^{-8/\beta} \Bigg\lbrace \left(2+\frac{i}{2}\right)
	-\left(\frac{7}{4}+\frac{15 i}{32}\right) \beta 
	+\left(\frac{9}{8}+\frac{237 i}{1024}\right) \beta^2 \nonumber \\
	&\quad\quad\quad\quad\quad\quad\quad-\left[\left(\frac{51}{128}-\frac{73 i}{16384}\right) + \left(\frac{9}{16}+\frac{9 i}{64}\right) \zeta_3 \right] \beta^3 + \rmO(\beta^3) \Bigg\rbrace   \nonumber\\
	&-\left(\frac{2}{e}\right)^6 {\rm e}^{-12/\beta} \Bigg\lbrace (3-5 i)
	-\left(\frac{67}{16}-\frac{151 i}{24}\right) \beta 
	+\left(\frac{1703}{512}-\frac{705 i}{128}\right) \beta^2 \nonumber\\
	&\quad\quad\quad\quad\quad\quad\quad-\left[\left(\frac{45}{32}-\frac{75 i}{32}\right) \zeta_3
	+\left(\frac{10159}{8192}-\frac{8711 i}{3072}\right)\right] \beta^3 + \rmO(\beta^4) \Bigg\rbrace + \rmO({\rm e}^{-16/\beta}).
\end{align}
At each NP order, we used $N_{\max}=600$ coefficients for the Borel-Pad\'e resummation ($N=540$ to estimate its error), except for the last term $k=6$ in Table \ref{tab:F} (line $d$ of Figure \ref{fig:FO4}), where we obtained the expansion only up to $N_{\max}=150$ in $\beta$. The same TBA data were used as those for the analysis in the bootstrap coupling $v$. Compared to the latter case (see Figure \ref{fig:WVO4}) here we managed to subtract one less NP term in the trans-series from the TBA, as the error of the resummation was higher due to using fewer (600 instead of 1000) coefficients to resum the perturbative level $k=0$. Based on line $c$ of Figure \ref{fig:FO4} we managed to verify the trans-series $\tilde{g}(\beta)$ up to $\simeq 10^{-53}$ at $B=10$ for the \mO{4} model.

\begin{table}[!h]
	\centering
	\begin{tabular}{c c c c c c c}
			& & a & b & c & d  \\
		\cline{2-6}
					&	$k$			&	0		&	1					&	2		&		3			\\
		\cline{2-6}
		\rule{0pt}{3ex}
		\multirow{3}{*}{\mO{3}} 	& $\tilde{g}^{(k)}-\tilde{g}^{(k-1)}$ 	& $\mathbb{R}\mu^0$ 	& $-i M_1\frac{8}{e}\mu^1 +\mathbb{R}\mu^1$  	& $\mathbb{C}\mu^2$  	& $\mathbb{C}\mu^3$    \\ 
					& $\text{Re}\,S^+(g_k)\mu^k$ 				& $+\mu^0$ 		& $-\mu^1$  	& $+\mu^2$  						& $+\mu^3$    \\ 
					& $\delta g^{(k)}$ 				& $-\mu^1$ 		& $+\mu^2$  	& $+\mu^3$  						& $+\mu^4$    \\ 
		\cline{2-6}
					&	$k$			&	0		&	2					&	4		&		6			\\
		\cline{2-6}\rule{0pt}{3ex}
		\multirow{3}{*}{\mO{4}} 	&  $\tilde{g}^{(k)}-\tilde{g}^{(k-2)}$ 	& $\mathbb{R}\mu^0$ 	& $-i M_1\frac{4}{e}\mu^1+i\mathbb{R}\mu^2$  	& $\mathbb{C}\mu^4$  	& $\mathbb{C}\mu^6$  \\
					&  $\text{Re}\,S^+(g_k)\mu^k$ 		& $+\mu^0$ 		& $+\mu^4$  					& $-\mu^4$  		& $-\mu^6$  \\
					&  $\delta g^{(k)}$				& $+\mu^4$ 		& $-\mu^4$  					& $-\mu^6$  		& $+\mu^8$  \\ 
	\end{tabular}			
	\caption{Explanation for the labels in Figures \ref{fig:FO3},\ref{fig:FO4}. The NP parameter $\mu$ is defined as stated below \eqref{eq:dg}, and the meaning of the rows is similar to what was explained in Table \ref{tab:WV}: 
	the upper row shows the magnitude and reality of the last trans-series term $g_k \mu^k$ subtracted, the middle one tells us the sign and $\mu$-order of its resummation's real part, and lastly, the lower one indicates the sign and order of $\delta g^{(k)}(\beta)$. 
	}
	\label{tab:F}
\end{table}

\section{Conclusions}
\label{sect14}

We used the Wiener-Hopf technique to solve the integral equation governing
the groundstate energy density in the two dimensional non-linear \mO{$N$}
$\sigma$ model. This linear integral equation is the thermodynamic
limit of the Bethe ansatz equation, it has a finite support $[-B,B]$
and its kernel is the logarithmic derivative of the scattering
matrix. We obtained the solution in a trans-series form, which was
a double series in the perturbative running coupling $v$ defined by
$2B=\frac{1}{v}-a\log v+const$,
and in the non-perturbative, exponentially small quantity
$\nu={\rm e}^{-2B}\sim v^{a}{\rm e}^{-\frac{1}{v}}$.
The structure of the non-perturbative corrections was governed
by the pole structure of the Wiener-Hopf kernel. Every power of the
non-perturbative parameter $\nu$ was multiplied by an asymptotic perturbative
series in $v$. The latter were expressed in terms of the basic quantities
$A_{n,m}$. The $A_{n,m}$ are asymptotic series, which are the perturbative
parts of the generalized moments of the generalized energy densities
$O_{n,m}$, such that $O_{1,1}$ is related to the energy density,
while $O_{1,0}$ to the particle density. We derived a system of differential
equations relating these observables to each other, which enabled
us to calculate all of them in terms of just a single one, say from $A_{1,1}$,
which was already known from Volin's method. When deriving our solution, we
paid particular attention to the degenerate cases $O_{n,0}$, which included
the calculation of the particle density itself.

In order to connect our trans-series solution to the physical observables
all the asymptotic series have to be summed up. This is understood
via lateral Borel resummations. Although there were many indications
along the derivation that this is the right resummation prescription
it does not follow directly from our solution. We thus assumed that
this is the case and thoroughly investigated its consequences. In
particular, the ambiguity free nature of the result requires that
the asymptotic behaviour of the asymptotic series in $v$ of a given
power of $\nu$ is related to the perturbative expansion in $v$ of
other powers of $\nu$. We investigated these \emph{resurgence} relations
separately for the $N>3$ and $N=3$ cases, and determined the Stokes
automorphism, which relates the two lateral resummations. We found
that for $N>3$ the perturbative series contains all the information
about the non-perturbative terms and its median resummation reproduces
the result of the numerical solution of the TBA equation. However, in the $N=3$
case we could identify three sectors: in addition to the perturbative sector
the $\nu$ and $\nu^{2}$ sectors. We called them the perturbative,
the one-instanton and the two-instanton sector, respectively.
We found that the physical result is the median resummation of the sum of
these three sectors. By carefully
investigating the asymptotics of the perturbative sector we showed
that the other two sectors are indeed indepent. We further supported
our trans-series solution with very high precision numerical investigations,
in which we compared the numerical solution of the TBA equations with
the resummation of various orders of the trans-series. {In doing so we truncated the trans-series at some given non-perturbative level, and evaluated its lateral Borel resummation. Since the result's difference to the (numerically evaluated) TBA was always of the order of the next non-perturbative correction coming from the same trans-series, we confirmed the correctness of all tested non-perturbative order. Any discrepancy of a given non-perturbative order would prevent us from getting more and more precise results as we were stuck at the wrong non-perturbative order. Considering how the values of the individual resummed asymptotic serieses at each non-perturbative level grow, we also found some indication that the trans-series will be convergent.
}

We then switched to the physical couplings, which appears in the Lagrangean
formulation of the \mO{$N$} models, when one of the global charges
is coupled to a magnetic field. The observable of interest is the
free energy density, which is the Legrendre transform of the energy
density. Since the physical coupling is defined through the particle density,
the resurgence properties of the free energy density in this coupling
is more involved. Indeed, we found infinitely many sectors, meaning
that the physical result is the median resummation of infinitely many
independent asymptotic series, which multiply the various non-perturbative
sectors.

It would be extremely interesting to relate our results to direct
instanton calculations based on the Lagrangean description of the \mO{3}
$\sigma$ model. First one should identify instanton solutions in large
magnetic fields, then one should calculate the small fluctuations
around them. The appearing series, when expressed in the running physicial
coupling should reproduce our results.

\vspace{5ex}

\section*{Acknowledgements}

The authors would like to thank the organizers of the workshops
“Physical Resurgence: On Quantum, Gauge, and Stringy” (Isaac Newton Institute)
and “Workshop on Quantization and Resurgence” (SwissMAP) for the invitation and the workshop participants for many interesting discussions. Many of the ideas and results of this work were actually developed during our participation in these events. We also thank Gerald Dunne for correspondence on the asymptotic analysis of the \mO{3} perturbative coefficients and Ovidiu Costin for explaining their desingularization method.  This work has been supported in part by the NKFIH grant K134946. The work of IV was also partially supported by the NKFIH excellence grant TKP2021-NKTA-64.

\par\bigskip

\appendix

\section{Proof of the differential relations}
\label{appA}

In this appendix\footnote{Parts of the results derived in this section was known before, yet we do not know about any appearance of the second line of \eqref{EVEN} and \eqref{secondordereven} in the literature in their present form and generality. The same is true for the relations involving the odd quantities (\ref{ODD}-\ref{ODDFROMEVEN}).} we will consider the TBA type integral equations
\begin{equation}
g_\alpha(\theta,B)-\int_{-B}^B{\rm d}\theta^\prime K(\theta-\theta^\prime)
g_\alpha(\theta^\prime,B)=R_\alpha(\theta,B),\qquad\qquad
-B\leq\theta\leq B,
\label{TBAtype}
\end{equation}
where the integral kernel $K(\theta)$ is assumed to be symmetric:
\begin{equation}
K(-\theta)=K(\theta).  
\end{equation}
(In most applications the kernel is given by the logarithmic derivative of
the S-matrix. In this case the symmetry of the kernel follows from unitarity.)

The following considerations are based on the crucial assumption that the
solution of the integral equation (\ref{TBAtype}) is unique, in other words
\begin{equation}
\Phi(\theta,B)-\int_{-B}^B{\rm d}\theta^\prime K(\theta-\theta^\prime)
\Phi(\theta^\prime,B)=0,\qquad\qquad
-B\leq\theta\leq B,
\end{equation}
implies 
\begin{equation}
\Phi(\theta,B)\equiv 0.
\end{equation}

The symmetry of the kernel combined with the uniqueness assumption allows us to
study the even and odd cases separately. Given an even or odd right-hand side,
\begin{equation}
R_\alpha(-\theta,B)=\varepsilon_\alpha R_\alpha(\theta,B),\qquad\quad
\varepsilon_\alpha=\pm1,  
\end{equation}
it is easy to see that the solution $g_\alpha(\theta,B)$ inherits the
even/odd property:
\begin{equation}
g_\alpha(-\theta,B)=\varepsilon_\alpha g_\alpha(\theta,B).  
\end{equation}

Let us give here some definitions. We will use the prime and dot notation
for derivatives:
\begin{equation}
\frac{\partial}{\partial\theta}\Phi(\theta,B)=\Phi^\prime(\theta,B),\qquad\quad  
\frac{\partial}{\partial B}\Phi(\theta,B)=\dot\Phi(\theta,B),\qquad\quad
\frac{{\rm d}}{{\rm d}B} f(B)=\dot f(B).
\end{equation}
An important quantity is the boundary value of the solution:
\begin{equation}
h_\alpha(B)=g_\alpha(B,B),\qquad\quad g_\alpha(-B,B)=\varepsilon_\alpha h_\alpha(B).
\end{equation}
In this appendix we will often use the special right-hand side functions
\begin{equation}
R_\pm(\theta,B)=K(\theta-B)\pm K(\theta+B),\qquad\quad \varepsilon_\pm=\pm1,
\end{equation}
and the derivatives of the right-hand side functions:
\begin{equation}
R_{(\alpha 0)}(\theta,R)=\dot R_\alpha(\theta,B),\qquad\quad  
R_{(\alpha 1)}(\theta,R)=R^\prime_\alpha(\theta,B).  
\end{equation}
Finally the central objects are the generalized moments\footnote{Note that the
parity of the two functions must be the same otherwise the integral vanishes.}  
\begin{equation}
\Omega_{\alpha\beta}(B)=\frac{1}{2\pi}\int_{-B}^B{\rm d}\theta\, g_\alpha(\theta,B)
R_\beta(\theta,B).  
\end{equation}
These are symmetric in their indices:
\begin{equation}
\begin{split}
\Omega_{\alpha\beta}(B)&
=\frac{1}{2\pi}\int_{-B}^B{\rm d}\theta\, g_\alpha(\theta,B)
\Big\{g_\beta(\theta,B)-\int_{-B}^B{\rm d}\theta^\prime K(\theta-\theta^\prime)
g_\beta(\theta^\prime,B)\Big\}\\
&=\frac{1}{2\pi}\int_{-B}^B{\rm d}\theta^\prime\, g_\beta(\theta^\prime,B)
\Big\{g_\alpha(\theta^\prime,B)-\int_{-B}^B{\rm d}\theta K(\theta-\theta^\prime)
g_\alpha(\theta,B)\Big\}\\
&=\frac{1}{2\pi}\int_{-B}^B{\rm d}\theta^\prime\,g_\beta(\theta^\prime,B)
R_\alpha(\theta^\prime,B)=\Omega_{\beta\alpha}(B).
\end{split}
\label{relati1a}
\end{equation}
Later we will use the identity (with $\varepsilon_\alpha=\pm1$)
\begin{equation}
\begin{split}
\Omega_{\pm\alpha}(B)&=\frac{1}{2\pi}\int_{-B}^B{\rm d}\theta\,
g_\alpha(\theta,B)\Big\{K(\theta-B)\pm K(\theta+B)\Big\}\\  
&=\frac{1}{2\pi}\Big[h_\alpha(B)-R_\alpha(B,B)\Big]\pm\frac{1}{2\pi}
\Big[\varepsilon_\alpha h_\alpha(B)-\varepsilon_\alpha R_\alpha(B,B)\Big]\\
&=\frac{1}{\pi}\Big[h_\alpha(B)-R_\alpha(B,B)\Big].
\end{split}
\label{urel}
\end{equation}

Let us now take the $B$-derivative of (\ref{TBAtype}):
\begin{equation}
\begin{split}  
\dot g_\alpha(\theta,B)&-\int_{-B}^B{\rm d}\theta^\prime\, K(\theta-\theta^\prime)
\dot g_\alpha(\theta^\prime,B)-K(\theta-B)h_\alpha(B)\\
&-K(\theta+B)\varepsilon_\alpha h_\alpha(B)=\dot R_\alpha(\theta,B).
\end{split}
\end{equation}
By the uniqueness property of the integral equation this implies
\begin{equation}
\dot g_\alpha(\theta,B)=g_{(\alpha 0)}(\theta,B)+h_\alpha(B)
g_{\varepsilon_\alpha}(\theta,B).  
\end{equation}
Using this and the relation (\ref{urel}) we can calculate the $B$ derivative
of the generalized densities:
\begin{equation}
\begin{split}  
\dot\Omega_{\alpha\beta}&=\frac{1}{2\pi}\int_{-B}^B{\rm d}\theta\,
R_\beta(\theta,B)\Big[g_{(\alpha0)}(\theta,B)+h_\alpha(B)
g_{\varepsilon_\alpha}(\theta,B)\Big\}\\
&+\Omega_{\alpha(\beta0)}(B)+\frac{1}{\pi}h_\alpha(B)R_\beta(B,B)\\
&=\Omega_{(\alpha0)\beta}(B)+\Omega_{\alpha(\beta0)}(B)+h_\alpha(B)\Bigg\{
\frac{1}{\pi} R_\beta(B,B)+\Omega_{\varepsilon_\alpha \beta}(B)\Bigg\}\\
&=\Omega_{(\alpha0)\beta}(B)+\Omega_{\alpha(\beta0)}(B)+\frac{1}{\pi}h_\alpha(B)
h_\beta(B).
\end{split}  
\label{relati2a}
\end{equation}

Next we take the $\theta$-derivative of (\ref{TBAtype}) and after partial
integration we obtain
\begin{equation}
\begin{split}
g^\prime_\alpha(\theta,B)&-\int_{-B}^B{\rm d}\theta^\prime\,K(\theta-\theta^\prime)
g^\prime_\alpha(\theta^\prime,B)+K(B-\theta)h_\alpha(B)\\  
&-K(B+\theta)\varepsilon_\alpha h_\alpha(B)=R_{(\alpha1)}(\theta,B),
\end{split}    
\end{equation}
which implies
\begin{equation}
g^\prime(\theta,B)=g_{(\alpha1)}(\theta,B)-h_\alpha(B)
g_{-\varepsilon_\alpha}(\theta,B).  
\end{equation}
Using this we calculate
\begin{equation}
\begin{split}
\frac{1}{2\pi}\int_{-B}^B
{\rm d}\theta\,g^\prime_\alpha(\theta,B)&R_\beta(\theta,B)=
\Omega_{(\alpha1)\beta}(B)-h_\alpha(B)\Omega_{-\varepsilon_\alpha \beta}(B)\\  
&=\Omega_{(\alpha1)\beta}(B)-\frac{1}{\pi}h_\alpha(B)h_\beta(B)+\frac{1}{\pi}
h_\alpha(B)R_\beta(B,B).
\end{split}
\label{calc1}
\end{equation}
Calculating the same integral directly by partial integration we obtain
\begin{equation}
\frac{1}{2\pi}\int_{-B}^B
{\rm d}\theta\,g^\prime_\alpha(\theta,B)R_\beta(\theta,B)=
-\Omega_{\alpha(\beta1)}(B)+\frac{1}{\pi} h_\alpha(B)R_\beta(B,B).  
\label{calc2}
\end{equation}
Comparing (\ref{calc1}) and (\ref{calc2}) we find the relation
\begin{equation}
\Omega_{(\alpha1)\beta}(B)+\Omega_{\alpha(\beta1)}(B)=\frac{1}{\pi}
h_\alpha(B) h_\beta(B).  
\label{relati3a}
\end{equation}

\subsection{Relation among the physical quantities}

Let us define the integral equations, their solutions and boundary values
corresponding to the building blocks
\begin{equation}
\begin{split}
R_\alpha(\theta,R)&=\cosh n\theta,\\
R_\beta(\theta,R)&=\cosh m\theta,\\
R_\gamma(\theta,R)&=\sinh n\theta,\\
R_\delta(\theta,R)&=\sinh m\theta,
\end{split}
\qquad
\begin{split}
g_\alpha(\theta,R)&=\chi_n(\theta,B),\\
g_\beta(\theta,R)&=\chi_m(\theta,B),\\
g_\gamma(\theta,R)&=\psi_n(\theta,B),\\
g_\delta(\theta,R)&=\psi_m(\theta,B),
\end{split}
\qquad
\begin{split}
h_\alpha(R)&=\rho_n(B),\\
h_\beta(R)&=\rho_m(B),\\
h_\gamma(R)&=s_n(B),\\
h_\delta(R)&=s_m(B)
\end{split} 
\label{valtoz}
\end{equation}
and define the densities\footnote{By uniqueness of the solution, for $m=0$ the quantities $s_0$ and $T_{n,0}$ vanish, as the odd source in this case is identically zero $\sinh(0\cdot \theta) = 0$. Also, one should not confuse this notation for the boundary value with the constant in \eqref{s0const}.}
\begin{equation}
\Omega_{\alpha\beta}(B)=O_{n,m}(B),\qquad\quad  
\Omega_{\gamma\delta}(B)=T_{n,m}(B).  
\label{suru}
\end{equation}

We will apply the symmetry relation (\ref{relati1a}) and the identities
(\ref{relati2a}) and (\ref{relati3a}) to obtain the differential relations we
used in the main text.

Since there cannot be any ambiguity, from now on we will suppress the explicit
$B$ dependence. In terms of the variables (\ref{valtoz}) and the densities
(\ref{suru}) (\ref{relati2a}) and (\ref{relati3a}) read
\begin{equation}
\dot O_{n,m}=\frac{1}{\pi}\rho_n\rho_m,\qquad\quad  
\dot T_{n,m}=\frac{1}{\pi}s_ns_m,  
\label{relati2b}
\end{equation}
\begin{equation}
mO_{n,m}+nT_{n,m}=\frac{1}{\pi}\rho_n s_m.
\label{relati3b}  
\end{equation}
Writing (\ref{relati3b}) with $n,m$ and $m,n$ we can find the linear
combinations
\begin{equation}
(n^2-m^2)O_{n,m}=\frac{1}{\pi}\big[ns_n\rho_m-ms_m\rho_n\big],\qquad
(m^2-n^2)T_{n,m}=\frac{1}{\pi}\big[ms_n\rho_m-ns_m\rho_n\big].
\end{equation}

Next we take the $B$-derivative of (\ref{relati3b}), use (\ref{relati2b}) and
divide by $\rho_n s_m$ to obtain
\begin{equation}
\frac{\dot\rho_n}{\rho_n}-n\frac{s_n}{\rho_n}=
m\frac{\rho_m}{s_m}-\frac{\dot s_m}{s_m}=A.  
\end{equation}
This expression cannot depend on either $n$ or $m$. (Of course,
$A=\dot\rho_0/\rho_0$ still depends on~$B$.)

An alternative way to write the above relation is
\begin{equation}
\dot\rho_n-ns_n=A\rho_n,\qquad\quad
\dot s_m-m\rho_m=-As_m.  
\end{equation}
Excluding $s_n$ by using this pair of equations we can obtain a purely $\rho_n$
dependent formula\footnote{Here and throughout the paper ${ F }$ is not to be confused with the normalized version of the energy density in \eqref{physicals}.}
\begin{equation}\label{evenF}
\ddot\rho_n-n^2\rho_n={ F}\rho_n,\qquad { F}=A^2+\dot A.  
\end{equation}
Similarly\footnote{The function $G$ has also nothing to do with the normalized free energy density $G(\beta)$ defined in \eqref{eq:gbeta}.}
\begin{equation}\label{oddG}
\ddot s_n-n^2s_n={ G}s_n,\qquad { G}=A^2-\dot A.
\end{equation}
Let us now summarize the differential identities we obtained.

\subsubsection{Identities for the even quantities}

\begin{equation}
\begin{split}
\dot O_{n,m}&=\frac{1}{\pi}\rho_n\rho_m,\\
(n^2-m^2)O_{n,m}&=\frac{1}{\pi}\big[\rho_m\dot\rho_n-\rho_n\dot\rho_m\big],\\  
\ddot \rho_n-n^2\rho_n&={ F}\rho_n,\qquad { F}=A^2+\dot A,\qquad
A=\frac{\dot\rho_0}{\rho_0}.
\end{split}
\label{EVEN}
\end{equation}
These are the relations we actually need in our considerations in the main text.

\subsubsection{Identities for the odd quantities}

\begin{equation}
\begin{split}
\dot T_{n,m}&=\frac{1}{\pi}s_n s_m,\\
(n^2-m^2)T_{n,m}&=\frac{1}{\pi}\big[s_m\dot s_n-s_n\dot s_m\big],\\  
\ddot s_n-n^2s_n&={ G}s_n,\qquad { G}=A^2-\dot A.
\end{split}
\label{ODD}
\end{equation}

\subsubsection{Mixed relations}

\begin{equation}
\begin{split}
mO_{n,m}+nT_{n,m}&=\frac{1}{\pi}\rho_n s_m,\\
\dot\rho_n-ns_n&=A\rho_n,\\
\dot s_n-n\rho_n&=-As_n.  
\end{split}
\label{MIX}
\end{equation}

The mixed relations can be used to express the odd quantities in terms of the
even ones:
\begin{equation}\label{ODDFROMEVEN}
s_m=\frac{mO_{0,m}\pi}{\rho_0},\qquad\quad
T_{n,m}=\frac{m}{n}\left[\frac{\rho_nO_{0,m}}{\rho_0}-O_{n,m}\right].
\end{equation}

Finally we note that using our identities it is easy to derive a second order
differential equation satisfied by $O_{n,m}$:
\begin{equation}\label{secondordereven}
\ddot O_{n,m}-\frac{2\dot\rho_n}{\rho_n}\dot O_{n,m}=(m^2-n^2)O_{n,m}.  
\end{equation}

\section{Resurgence conventions and relations}

In this Appendix we collect the conventions and formulas from the resurgence theory we need.

\subsection{Borel resummation}
\label{AppResurgence}

The standard Borel resummation formulas are as follows. We write an asymptotic
series as
\begin{equation}
\Psi(v)=\Psi_{(-)}(v)+\Psi_{(+)}(v),\qquad
\Psi_{(-)}(v)=\sum_{n=-\infty}^0 \psi_n v^n,\quad  
\Psi_{(+)}(v)=\sum_{n=1}^\infty \psi_n v^n,  
\end{equation}
where it is assumed that only finitely many $\psi_n$, $n\leq0$, are
non-vanishing. The standard Borel transform is then
\begin{equation}\label{eq:Borel}
\hat\Psi(t)=\sum_{n=0}^\infty c_n t^n,\qquad c_n=\frac{\psi_{n+1}}{\Gamma(n+1)}  
\end{equation}
and lateral Borel resummations are given (after analytically continuing
$\hat\Psi$) by
\begin{equation}\label{eq:inverseBorel}
S^\pm(\Psi)(v)=\Psi_{(-)}(v)+\int_{{\cal C}_\pm}{\rm d}t\,{\rm e}^{-\frac{t}{v}}
\hat\Psi(t).  
\end{equation}
If the coefficients of the Borel transform behave (for large $n$) as
\begin{equation}
\begin{split}
c_n=&\big\{p^++\frac{p^+_0}{n}+\frac{p^+_1}{n(n-1)}+\frac{p^+_2}{n(n-1)(n-2)}
+\dots\big\}\\  
+(-1)^n &\big\{p^-+\frac{p^-_0}{n}+\frac{p^-_1}{n(n-1)}+\frac{p^-_2}{n(n-1)(n-2)}
+\dots\big\}
\end{split}
\end{equation}
then the Borel plane singularities are of logarithmic type and the alien
derivatives at $\pm1$ can be defined as
\begin{equation}\label{eq:aliender}
\Delta_{\pm1}\Psi(v)=\mp 2\pi i\big\{p^\pm\pm\sum_{m=0}^\infty(\pm1)^mp^\pm_mv^{m+1}
\big\}    
\end{equation}
and (the leading term of) the imaginary part of the lateral Borel resummation
is
\begin{equation}\label{eq:alienderSplus}
i{\rm Im}\,S^+(\Psi)(v)=-\frac{1}{2}{\rm e}^{-\frac{1}{v}}S^+(\Delta_1\Psi)(v). 
\end{equation}

We note the identity
\begin{equation}
S^\pm(v^\beta \Psi)(v)=v^\beta S^\pm(\Psi)(v).
\end{equation}

Let us consider the example (where $\gamma>-1$)
\begin{equation}
\Psi^{(\gamma)}_{(-)}(v)=0,\qquad \psi^{(\gamma)}_n=\Gamma(n+\gamma)\quad n\geq1.  
\label{exam}
\end{equation}
In this case
\begin{equation}
\hat\Psi^{(\gamma)}(t)=\frac{\Gamma(1+\gamma)}{(1-t)^{1+\gamma}}.
\end{equation}
It is easy to calculate that
\begin{equation}
{\rm Im}S^+(\Psi^{(\gamma)})(v)=
\pi v^{-\gamma}{\rm e}^{-\frac{1}{v}}.  
\label{impart}
\end{equation}

\subsection{Modified alien derivatives}
\label{sect6}


In this paper we will use modified alien derivatives. Starting from the
standard alien derivative associated to $z=1/v$ we define our version of
alien derivative by
\begin{equation}
\Delta_\omega={\rm e}^{\omega L}v^{-a\omega}\Delta_\omega^{({\rm st})}.
\end{equation}
The \lq\lq pointed'' alien derivative is then
\begin{equation}
{\overset{\circ}{\Delta}}_\omega={\rm e}^{-\omega z}\Delta_\omega^{({\rm st})}=
\nu^\omega\Delta_\omega.  
\end{equation}
Using the fact that the pointed alien derivative commutes with the ordinary
derivative,
\begin{equation}
\left[{\overset{\circ}{\Delta}}_\omega,\frac{{\rm d}}{{\rm d}z}\right]=0,
\end{equation}
we can derive the relation
\begin{equation}
\Delta_\omega \dot X=\dot{(\Delta_\omega X)}-2\omega(\Delta_\omega X).  
\end{equation}

We prove three alien derivative lemmas in appendix \ref{appB}, which will be
used in the main text.

\subsection{$S^{\pm}$ properties}
\label{ssectSplusprop}

We start from the observation that if $\psi$ and $\phi$ are $+$ type
asymptotic series then their Borel transforms satisfy the relation
\begin{equation}
\widehat{(\psi\cdot\phi)}=\hat\psi*\hat\phi,  
\end{equation}
where $*$ denotes convolution. Then for their Borel resummation (along the
ray $\omega$) we have
\begin{equation}
\begin{split}
S^\omega(\psi\cdot\phi)(v)&=\omega\int_0^\infty{\rm d}\tau{\rm e}^{-\omega
\frac{\tau}{v}}(\hat\psi*\hat\phi)(\omega\tau)\\
&=\omega^2\int_0^\infty{\rm d}\tau{\rm e}^{-\omega\frac{\tau}{v}}\int_0^\tau
{\rm d}w\hat\psi(\omega w)\hat\phi(\omega(\tau-w))=S^\omega(\psi)(v)\cdot
S^\omega(\phi)(v).
\end{split}  
\end{equation}
Thus we found the relation
\begin{equation}
S^\pm(\psi\cdot\phi)=S^\pm(\psi)\cdot S^\pm(\phi)
\end{equation}
for + type asymptotic series. Adding non-positive powers 
\begin{equation}
P(v)=\sum_{r=0}^M p_r v^{-r}  
\end{equation}
and using the identity
\begin{equation}
S^\pm(P\cdot\psi)=P\cdot S^\pm(\psi)  
\end{equation}
we can prove for generic asymptotic series
\begin{equation}
\psi_1=P+\psi,\qquad\qquad \phi_1=Q+\phi  
\end{equation}
the analogous relation
\begin{equation}
S^\pm(\psi_1\cdot\phi_1)=S^\pm(\psi_1)\cdot S^\pm(\phi_1).
\end{equation}
Next we study the case of full trans-series
\begin{equation}
\tilde{\cal A}=\sum_{r=0}^\infty A_r \nu^r,\qquad\qquad  
\tilde{\cal B}=\sum_{s=0}^\infty B_s \nu^s,  
\end{equation}
where $A_r$, $B_s$ are asymptotic series. Using $S^\pm(A_r\cdot B_s)=
S^\pm(A_r)\cdot S^\pm(B_s)$ it is easy to see that also for trans-series
\begin{equation}
S^\pm(\tilde{\cal A}\cdot\tilde{\cal B})=S^\pm(\tilde{\cal A})\cdot
S^\pm(\tilde{\cal B})
\end{equation}
holds. We can also prove easily that $S^\pm$ can be moved through the inverse,
multiple product, polynomial of trans-series. Finally, for \lq\lq small''
trans-series $\tilde{\cal A}$, $S^\pm$ can even be moved through analytic
functions:
\begin{equation}
S^\pm\left(f(\tilde{\cal A})\right)=f\left(S^\pm(\tilde{\cal A})\right),\qquad
f(\zeta)=\sum_{m=0}^\infty f_m\zeta^m.  
\end{equation}
(See Proposition 4.109 (p.108) in Costin's book \cite{costin2008asymptotics}.)

\subsection{$\mathfrak{S}^{-1/2}$ properties}
\label{ssectStokesprop}

Let ${\cal K}$ be a derivation (linear, with Leibniz property). Then, using the
binomial formula for higher derivatives,
\begin{equation}
{\cal K}^m (A\cdot B)=\sum_{r=0}^m \binom{m}{r} {\cal K}^r A\cdot{\cal K}^{m-r}B  
\end{equation}
it is easy to show that
\begin{equation}
{\rm e}^{\cal K}(A\cdot B)={\rm e}^{\cal K}A\cdot {\rm e}^{\cal K} B.  
\label{expK}
\end{equation}
Again, the exponentiated derivative ${\rm e}^{\cal K}$ can be moved through the
inverse, multiple product, polynomial and analytic function of trans-series.
In particular, for ${\cal K}=\frac{1}{2}{\cal R}$, (\ref{expK}) becomes
\begin{equation}
\mathfrak{S}^{-1/2}(A\cdot B)=\mathfrak{S}^{-1/2}A\cdot \mathfrak{S}^{-1/2}B.    
\end{equation}

\subsection{Composition of asymptotic series}
\label{compAsy}

In the following we will need the transformation rule of alien derivatives
under change of expansion variables. We will use (as before) the alien
derivative $\Delta_\omega^{\rm (st)}(v)$ in terms of the original expansion
parameter $v$, and denote by $D_\omega$ the alien derivative corresponding
to the new variable $\beta$ given by the asymptotic series 
\begin{equation}
\beta=\alpha(v)=2v+{\rm O}(v^2).  
\end{equation}
For a given asymptotic series $q(\beta)$ and the corresponding composite
function $C=q\circ \alpha$,
\begin{equation}
C(v)=q(\alpha(v)),  
\end{equation}
the relation between the two alien derivatives is
given \cite {delabaere1999resurgent} by
\begin{equation}
\left(\Delta^{\rm (st)}_\omega C\right)(v)
={\rm e}^{-\omega\left[\frac{2}{\alpha(v)}
-\frac{1}{v}\right]}(D_\omega q)(\alpha(v))+q^\prime(\alpha(v))\left(
\Delta^{\rm (st)}_\omega\alpha\right)(v).
\label{saee}
\end{equation}
(See eq. (5.7) in \cite{Abbott:2020qnl}.)

\section{Alien derivative lemmas}
\label{appB}

In the following considerations we will use the alien derivatives of the
relations (\ref{A1234}) and (\ref{V1234}). We will call the four relations of
the former (A1)-(A4) and call (V1)-(V4) those of the latter.

\subsection{Constant alien derivatives}

Let us assume that (as, for example, in the \mO{4} model
\cite{Bajnok:2021dri}) 
\begin{equation}
\Delta_1 A_{1,1}=C,  
\end{equation}
where $C$ is constant.
Then the alien derivative of (A1) immediately gives
\begin{equation}
\Delta_1 a_1=0  
\end{equation}
and (A3) (for $n=1$) gives then
\begin{equation}
\Delta_1 f=0.  
\end{equation}
Next we use (the alien derivative of) (A3) for generic $n$ to conclude that
\begin{equation}
\Delta_1 a_n=0.  
\end{equation}
This conclusion is based on the fact that the resulting equation has no
perturbative solution. Very similar considerations based on (V3) lead to
\begin{equation}
\Delta_1 s=0.  
\end{equation}
Next we consider the alien derivative of (A1) with generic $n$, $m$. We find
\begin{equation}
\Delta_1 A_{n,m}=0  
\label{Del1}
\end{equation}
for $n+m\not=2$. This conclusion is based again on the observation that the
equation obtained this way has no perturbative solution, unless $n+m=2$.
Alternatively, we can take the alien derivative of (A2) and conclude that
(\ref{Del1}) holds for generic $n$, $m$, except if $n=m$. Combining these
two conclusions we see that (\ref{Del1}) holds for all $n$, $m$, except of
course for $n=m=1$.

From (V2) we obtain
\begin{equation}
\Delta_1 V_n=0  
\end{equation}
and using (A4), (V4) we conclude that
\begin{equation}
\Delta_1 P_{n,-n}=0,\qquad \Delta_1 D_0=0.  
\end{equation}

Let us now study the consequences of the relation
\begin{equation}
\Delta_\kappa A_{1,1}=0,\qquad \kappa\not=1.  
\end{equation}
This holds for nearly all $\kappa$, since the points where the alien derivative
of $A_{1,1}$ does not vanish is a discrete set.

We can repeat the above considerations also in this case, except that (after
establishing $\Delta_\kappa a_1=0$, $\Delta_\kappa f=0$) from (A3) 
\begin{equation}
\Delta_\kappa a_n=0  
\label{ankappa}
\end{equation}
follows only for $n\not=\kappa$. For $n=\kappa$ we see that the alien derivative
of (A3) leads to a relation for $\Delta_\kappa a_\kappa$, which has a non-trivial
perturbative solution
\begin{equation}
\Delta_\kappa a_\kappa=X_0 a_{-\kappa},  
\end{equation}
where $X_0$ is some constant. Now combining (A4) and the alien derivative of
(A1) we arrive at
\begin{equation}
\dot{(\Delta_\kappa A_{\kappa,\kappa})}=2X_0(\dot{P}_{\kappa,-\kappa}+1),  
\end{equation}
which can be integrated to
\begin{equation}
\Delta_\kappa A_{\kappa,\kappa}=2X_0 P_{\kappa,-\kappa}+2BX_0+X_1,\qquad X_1
={\rm const.}  
\end{equation}
Since, as before, we require that the alien derivatives are perturbative, $X_0$
must vanish and finally we conclude that (\ref{ankappa}) holds for all $n$.

Next we establish
\begin{equation}
\Delta_\kappa s=0  
\end{equation}
and
\begin{equation}
{\rm from\ } (A1):\ \Delta_\kappa A_{n,m}=0,\quad n+m\not=2\kappa,\qquad
{\rm from\ } (A2):\ \Delta_\kappa A_{n,m}=0,\quad n\not=m.
\end{equation}
Combining the above two relations we see that
\begin{equation}
\Delta_\kappa A_{n,m}=0,\quad{\rm except\ }\Delta_\kappa A_{\kappa,\kappa}=X_1=
{\rm const.}  
\end{equation}
The same way as for $\Delta_1$ we can establish the relations
\begin{equation}
\Delta_\kappa V_n=0,\quad  
\Delta_\kappa P_{n,-n}=0,\quad  
\Delta_\kappa D_0=0.  
\end{equation}
To summarize, in this subsection we have shown that if $\Delta_1 A_{1,1}={\rm
const.}$ then $\Delta_1\equiv0$ on all other quantities and similarly if 
$\Delta_\kappa A_{1,1}=0$ then $\Delta_\kappa\equiv0$ on all other quantities
(except possibly on $A_{\kappa,\kappa}$).

\subsection{Alien derivative lemma}

Here the starting point is the relation (for some $\omega\not=1$)
\begin{equation}
\Delta_\omega A_{1,1}=\xi_\omega A_{1,-\omega}^2,\qquad \xi_\omega={\rm const.}  
\label{lemma0}
\end{equation}
Taking the alien derivative of (A1) immediately gives
\begin{equation}
\Delta_\omega a_1=\xi_\omega A_{1,-\omega} a_{-\omega}.  
\end{equation}
For $\Delta_\omega a_n$ with generic $n$ we take the ansatz
\begin{equation}
\Delta_\omega a_n=X_{n,\omega} a_{-\omega}  
\end{equation}
and use this ansatz in the alien derivative of (A3). The resulting equation
simplifies if we then use (A3) with $n=-\omega$. We obtain
\begin{equation}
2\dot{a}_{-\omega}[\dot{X}_{n,\omega}+(n-\omega)X_{n,\omega}]+a_{-\omega}
[\ddot{X}_{n,\omega}+(2n-4\omega)\dot{X}_{n,\omega}+4\omega(\omega-n)X_{n,\omega}]=
a_n\Delta_\omega f.
\label{Xnomega}
\end{equation}
This is a linear equation and (for $n\not=\omega$) we look for a solution
which is a sum of two terms:
\begin{equation}
X_{n,\omega}=\xi_\omega A_{n,-\omega}+Y_{n,\omega}.
\end{equation}
We see that $Y_{1,\omega}=0$.

By using (A1) repeatedly and (A2) the $A$-part of the equation simplifies
drastically and we find
\begin{equation}
4\xi_\omega a_n a_{-\omega}(\dot{a}_{-\omega}-\omega a_{-\omega})+
\{Y{\rm -part}\}=a_n\Delta_\omega f.  
\end{equation}
From the $n=1$ case (since in this case the $Y$-part vanishes) we conclude that
\begin{equation}
\Delta_\omega f=  
4\xi_\omega a_{-\omega}(\dot{a}_{-\omega}-\omega a_{-\omega}).
\end{equation}
It follows then that also for $n\not=1$ the $Y$-part must vanish and since
(for $n\not=\omega$) (\ref{Xnomega}) with zero right hand side has no
non-trivial perturbative solution,
\begin{equation}
Y_{n,\omega}=0.  
\end{equation}
This way we have shown that for all $n\not=\omega$
\begin{equation}
\Delta_\omega a_n=\xi_\omega A_{n,-\omega} a_{-\omega}.  
\end{equation}
A completely analogous calculation gives
\begin{equation}
\Delta_\omega s=\xi_\omega V_{-\omega} a_{-\omega}.  
\end{equation}
Similary to our considerations in the previous subsection, we first prove that
(for $n,m\not=\omega$)
\begin{equation}
\begin{split}
{\rm from\ (A1):}\quad\Delta_\omega A_{n,m}
&=\xi_\omega A_{n,-\omega} A_{m,-\omega},\quad n+m\not=2\omega,\\
{\rm from\ (A2):}\quad\Delta_\omega A_{n,m}
&=\xi_\omega A_{n,-\omega} A_{m,-\omega},\quad n\not=m.
\end{split}    
\end{equation}
Thus
\begin{equation}
\Delta_\omega A_{n,m}=\xi_\omega A_{n,-\omega} A_{m,-\omega},\quad n,m\not=\omega.
\label{lemma1}
\end{equation}
From (V2) it follows directly that
\begin{equation}
\Delta_\omega V_n=\xi_\omega A_{n,-\omega} V_{-\omega}
\end{equation}
and finally using (A4) and (V4) we get
\begin{equation}
\Delta_\omega P_{n,-n}=\xi_\omega A_{n,-\omega} A_{-n,-\omega},\quad n\not=\pm\omega
\end{equation}
and
\begin{equation}
\Delta_\omega D_0=\xi_\omega V_{-\omega}^2.
\end{equation}

\section{Recursive proof of the alien derivative formula}
\label{appC}


In this appendix we give a recursive proof of the relation (\ref{Deltaj}).
We start by combining the relations (\ref{M1}) and (\ref{psisub}) and write
\begin{equation}
{\rm Im}\,S^+(\Psi+M\nu)={\rm Im}\,S^+(\Psi_{\rm sub}).
\end{equation}
Next we introduce truncated quatities like
\begin{equation}
{\cal R}_m=\sum_{j=1}^m p^j\Delta_{jk},\qquad {\mathfrak{S}}_m^{-1/2}=
{\rm e}^{{\cal R}_m/2}.  
\end{equation}
For real $X$ we can write\footnote{\ \ $^*$ denotes complex conjugation.}
\begin{equation}
\left\{S^+({\mathfrak{S}}_m^{-1/2}X)\right\}^*=S^-({\mathfrak{S}}_m^{1/2}X)=  
S^+({\mathfrak{S}}^{-1}{\mathfrak{S}}_m^{1/2}X)=S^+({\mathfrak{S}}_m^{-1/2}X)
+{\rm O}(p^{m+1}).
\end{equation}
We see that $S^+({\mathfrak{S}}_m^{-1/2}X)$ is real up to order $m+1$,  
\begin{equation}
{\rm Im}\,S^+({\mathfrak{S}}_m^{-1/2}X)={\rm O}(p^{m+1}).  
\label{Im1}
\end{equation}
We also introduce the truncated version of $Y^{\rm T}{\cal L}Y$:
\begin{equation}
Y^{\rm T}{\cal L}Y=(Y^{\rm T}{\cal L}Y)_m+{\rm O}(p^{m+1}).  
\end{equation}
From our main conjecture the reality of $S^+(\hat\epsilon)$ follows. Its
truncated version is
\begin{equation}
{\rm Im}\,S^+(\Psi+M\nu+2(Y^{\rm T}{\cal L}Y)_m)={\rm O}(p^{m+1}),  
\end{equation}
which can also be written as
\begin{equation}
{\rm Im}\,S^+(\Psi_{\rm sub}+2(Y^{\rm T}{\cal L}Y)_m)={\rm O}(p^{m+1}).
\label{Im2}
\end{equation}

Now we assume that (as we have shown above to third order) (\ref{Deltaj}) holds
for $j=1,2,\dots,m-1$ and write
\begin{equation}
\frac{1}{2}{\cal R}_m\Psi_{\rm sub}=\frac{1}{2}{\cal R}_m\Psi=
\sum_{j=1}^{m-1}2iS_jp^jA_{1,-jk}^2+p^m\Delta_{mk}A_{1,1}=
\sum_{j=1}^m 2iS_jp^jA_{1,-jk}^2+p^m\xi_m,
\end{equation}
where
\begin{equation}
\xi_m=\Delta_{mk}A_{1,1}-2iS_mA_{1,-mk}^2
\end{equation}
and we have added and subtracted a term. Since $\Delta_{mk}$ does not occur in
higher powers of ${\cal R}_m$ up to ${\rm O}(p^m)$,
\begin{equation}
{\mathfrak{S}}_m^{-1/2}\Psi_{\rm sub}=\Psi_{\rm sub}+
2(Y^{\rm T}{\cal L}Y)_m+p^m\xi_m+{\rm O}(p^{m+1}).  
\end{equation}
Using (\ref{Im1}) and (\ref{Im2}) it follows that
\begin{equation}
{\rm Im}\,S^+(\xi_m)={\rm O}(p),
\end{equation}
but since $\xi_m$ is a purely imaginary asymptotic series,
\begin{equation}
\xi_m=0  
\end{equation}
must hold, which is the $j=m$ case of (\ref{Deltaj}).

\section{Proof of conjectures A and B in the \mO{4} model}
\label{appD}

In this appendix we check the validity of conjectures A and B (for the case of
the regular part of $\Omega$ in the \mO{4} model). We work up to
${\rm 2}^{\rm nd}$ order. We will see that the calculations are quite involved,
and are heavily based on special features of the particular case concerned.
It is possible that a general proof is less circular and will turn out to be
simpler. Nevertheless these calculations are useful because the results are
complicated enough to give some confidence in the validity of the conjectures.

\subsection{Conjecture A (reg part)}

We will use three important relations discussed in earlier sections. First from
section \ref{sect9} we recall (\ref{saee}), adapted to our case
here:
\begin{equation}
\Delta_\omega \left({\cal F}(b_0)\right)(v)=\left(\frac{{\rm e}b_0}{8v\bar S_0}
\right)^\omega(D_\omega{\cal F})\left(b_0(v)\right)+{\cal F}^\prime\left(b_0(v)
\right)(\Delta_\omega b_0)(v).
\label{conjectA1}
\end{equation}
Next from section \ref{sect6} we recall the rule for calculating the alien
derivative of a series, which is a $B$-derivative:
\begin{equation}
\Delta_\omega \dot X=\dot{(\Delta_\omega X)}-2\omega\Delta_\omega X.
\label{conjectA2}
\end{equation}
Finally we recall from section \ref{sect9} the defining relation of $b_0$ and
its $B$-derivative:
\begin{equation}
\frac{2}{b_0}+\ln b_0=2P+\frac{1}{v}+\ln(2v\bar S_0),
\end{equation}
\begin{equation}
\left(-\frac{2}{b_0^2}+\frac{1}{b_0}\right)\dot b_0=
2+\frac{(\dot{v\bar S}_0)}{v\bar S_0}=\frac{\bar S_1}{b_1}\dot b_0.
\label{conjectA3}
\end{equation}
We also recall that if
\begin{equation}
\tilde f={\mathfrak S}^{-1/2}f_0=f_0+f_1\nu^2+f_2\nu^4+\dots
\end{equation}
then the first two coefficients are
\begin{equation}
f_1=\frac{1}{2}\Delta_2 f_0,\qquad\quad
f_2=\frac{1}{2}\Delta_4 f_0+\frac{1}{8}\Delta_2^2 f_0
\end{equation}
and similarly, if  conjecture A is true:
\begin{equation}
{\cal G}_1=\frac{1}{2}D_2G_0,\qquad\quad
{\cal G}_2=\frac{1}{2}D_4 G_0+\frac{1}{8}D_2^2 G_0.
\label{true}
\end{equation}

Using (\ref{conjectA1}) with ${\cal F}=G_0$ we write
\begin{equation}
\Omega_1=\frac{1}{2}\Delta_2\Omega_0=\frac{1}{2}\left(\frac{{\rm e}b_0}{8}
\right)^2\frac{1}{(v\bar S_0)^2}(D_2G_0)(b_0)+G^\prime_0(b_0)b_1,
\label{Omga1}
\end{equation}
which, compared to (\ref{Omega012}), immediately gives
\begin{equation}
G_1(b_0)=\frac{1}{2}\left(\frac{{\rm e}b_0}{8}\right)^2(D_2 G_0)(b_0)
\end{equation}
and after removing the conventional prefactor
\begin{equation}
{\cal G}_1(\beta)=\frac{1}{2}(D_2 G_0)(\beta).
\end{equation}
This is the $1^{\rm st}$ order result, which we already obtained before.

Now using the $1^{\rm st}$ order result in (\ref{Omga1}) we can write 
\begin{equation}
\Delta_2\Omega_0=\frac{2G_1(b_0)}{(v\bar S_0)^2}+G^\prime_0(b_0)\Delta_2 b_0.
\end{equation}
We will use the above formula to calculate (for later purposes)
\begin{equation}
\Delta_2\dot\Omega_0=\Delta_2\left(G^\prime_0(b_0)\dot b_0\right)=
\Delta_2\left(G^\prime_0(b_0)\right)\dot b_0+
G^\prime_0(b_0)\Delta_2\dot b_0.
\end{equation}
We obtain
\begin{equation}
\begin{split}
\Delta_2\dot\Omega_0&=-4\Delta_2\Omega_0+\dot{(\Delta_2\Omega_0)}=
-4\Delta_2\Omega_0+\frac{2G^\prime_1(b_0)\dot b_0}{(v\bar S_0)^2}\\
&-\frac{4G_1(b_0)}{(v\bar S_0)^3}(\dot{v\bar S}_0)+G^{\prime\prime}_0(b_0)
\dot b_0\Delta_2b_0+G^\prime_0(b_0)\Delta_2\dot b_0+4G^\prime_0(b_0)\Delta_2b_0.
\end{split}
\end{equation}
Comparing the two expressions (for $\Delta_2\dot\Omega_0$) we get
\begin{equation}
\begin{split}
\Delta_2\left(G^\prime_0(b_0)\right)\dot b_0&=-\frac{8G_1(b_0)}{(v\bar S_0)^2}
+\frac{2G^\prime_1(b_0)\dot b_0}{(v\bar S_0)^2}-\frac{4G_1(b_0)}{(v\bar S_0)^3}
(\dot{v\bar S}_0)+G^{\prime\prime}_0(b_0)\dot b_0\Delta_2b_0\\
&=\dot b_0\Big\{G^{\prime\prime}_0(b_0)\Delta_2b_0+\frac{2G^\prime_1(b_0)}
{(v\bar S_0)^2}-\frac{4G_1(b_0)}{(v\bar S_0)^2}\frac{\bar S_1}{b_1}\Big\}.
\end{split}
\end{equation}
In the second line we used (\ref{conjectA3}). The purpose of the above exercise
was to calculate $\Delta_2(G^\prime_0(b_0))$, which is difficult to obtain
directly. Now we are able to calculate
\begin{equation}
\begin{split}
\Delta_2^2\Omega_0&=G^{\prime\prime}_0(b_0)(\Delta_2b_0)^2+\frac{2G^\prime_1(b_0)}
{(v\bar S_0)^2}\Delta_2 b_0-\frac{8G_1(b_0)\bar S_1}{(v\bar S_0)^2}+
G^\prime_0(b_0)\Delta_2^2b_0\\
&+\frac{2}{(v\bar S_0)^4}\Big(\frac{{\rm e}b_0}{8}\Big)^2(D_2G_1)(b_0)+
\frac{2}{(v\bar S_0)^2}G^\prime_1(b_0)\Delta_2b_0-\frac{4G_1(b_0)}{(v\bar S_0)^3}
\Delta_2(v\bar S_0),
\end{split}
\end{equation}
which can be somewhat simplified if we use
\begin{equation}
\Delta_2\bar S_0=2\bar S_0\bar S_1.
\end{equation}
At this point we have
\begin{equation}
\begin{split}
\Delta_2^2\Omega_0&=4G^{\prime\prime}_0(b_0)b_1^2+\frac{8G^\prime_1(b_0)}
{(v\bar S_0)^2}b_1-\frac{16G_1(b_0)}{(v\bar S_0)^2}\bar S_1\\
&+G^\prime_0(b_0)\big(8b_2-4\Delta_4b_0\big)+\frac{1}{(v\bar S_0)^4}\Big(
\frac{{\rm e}b_0}{8}\Big)^4(D_2^2G_0)(b_0).
\end{split}
\end{equation}
Finally we calculate
\begin{equation}
\begin{split}
\Omega_2&=\frac{1}{2}\Delta_4\Omega_0+\frac{1}{8}\Delta_2^2\Omega_0=
\frac{1}{(v\bar S_0)^4}\Big(\frac{{\rm e}b_0}{8}\Big)^4\big(\frac{1}{2}
D_4G_0+\frac{1}{8}D_2^2G_0\big)(b_0)\\
&+\frac{1}{2}G^{\prime\prime}_0(b_0)b_1^2+\frac{G_1(b_0)b_1}{(v\Bar S_0)^2}
-2\bar S_1\frac{G_1(b_0)}{(v\bar S_0)^2}+G^\prime_0(b_0)b_2.
\end{split}
\end{equation}
Comparison to (\ref{Omega012}) gives
\begin{equation}
G_2(b_0)=\left(\frac{{\rm e}b_0}{8}\right)^4\left(\frac{1}{2}D_4G_0
+\frac{1}{8}D_2^2G_0\right)(b_0)
\end{equation}
and finally after removing the prefactor
\begin{equation}
{\cal G}_2(\beta)=\left(\frac{1}{2}D_4G_0+\frac{1}{8}D_2^2G_0\right)(\beta),
\end{equation}
which verifies the conjecture up to $2^{\rm nd}$ order.

\subsection{Conjecture B (reg part)}

The proof is based on the following statement.
\begin{equation}
T^+({\cal F})\left(S^+(\alpha)(v)\right)=S^+\left({\cal F}(\alpha)\right)(v).
\label{Sauzin}
\end{equation}
This is Theorem 13.3 in \cite{Sauzin}. Here $\alpha(v)$ is an asymptotic
series in the variable $v$ and $S^+(\alpha)(v)$ is its lateral Borel sum.
Similarly ${\cal F}(a)$ is an asymptotic series in the variable $a$ and its
lateral Borel sum is denoted by $T^+({\cal F})(a)$.

We will also use the elementary relation
\begin{equation}
\left\{S^+(\psi)\right\}^\prime(v)=S^+(\psi^\prime)(v).
\label{derv}
\end{equation}
Similar relations hold also for higher derivatives.

Spelling out conjecture B in the new notation gives
\begin{equation}
S^+(\tilde\Omega)(v)=S^+\left(\tilde G(\tilde b)\right)(v)=
T^+(\tilde G)\left(S^+(\tilde b)(v)\right).
\end{equation}
Expanding the left hand side as trans-series up to $2^{\rm nd}$ order gives
\begin{equation}
S^+(\Omega_0)+S^+(\Omega_1)\nu^2+S^+(\Omega_2)\nu^4+\dots,
\end{equation}
whereas the right hand side is
\begin{equation}
T^+(G_0)(\beta)+T^+({\cal G}_1)(\beta){\rm e}^{-2x}
+T^+({\cal G}_2)(\beta){\rm e}^{-4x}+\dots,
\end{equation}
where
\begin{equation}
x=\frac{2}{\beta},\qquad \beta=S^+(\tilde b)=S^+(b_0)+S^+(b_1)\nu^2+
S^+(b_2)\nu^4+\dots
\end{equation}
Combining the prefactor with the exponent the right hand side becomes
\begin{equation}
T^+(G_0)(\beta)+T^+(G_1)(\beta)\left(\frac{\nu}{v(1+S)^2}\right)^2
+T^+(G_2)(\beta)\left(\frac{\nu}{v(1+S)^2}\right)^4+\dots
\end{equation}
and we recall that
\begin{equation}
\frac{1}{v(1+S)^2}=S^+\left(\frac{1}{v(1+\tilde S)^2}\right)=
S^+\left(\frac{1}{v\bar S_0}\left(1-\bar S_1\nu^2+\dots\right)\right).
\end{equation}
The first term on the right hand side can be expanded as
\begin{equation}
\begin{split}
T^+(G_0)\left(S^+(b_0)\right)
&+\left\{T^+(G_0)\right\}^\prime\left(S^+(b_0)\right)\cdot\left[S^+(b_1)\nu^2
+S^+(b_2)\nu^4\right]\\&+\frac{1}{2}
\left\{T^+(G_0)\right\}^{\prime\prime}\left(S^+(b_0)\right)\cdot
\left[S^+(b_1)\nu^2\right]^2+\dots
\end{split}
\end{equation}
Using the relations (\ref{derv}) and (\ref{Sauzin}) this becomes
\begin{equation}
S^+\left(G_0(b_0)\right)
+S^+\left(G^\prime_0(b_0)\right)\left[S^+(b_1)\nu^2+S^+(b_2)\nu^4\right]+
\frac{1}{2}S^+\left(G^{\prime\prime}_0(b_0)\right)S^+(b_1^2)\nu^4+\dots
\end{equation}
and using the properties of $S^+$ this can be further simplified to
\begin{equation}
S^+\left(G_0(b_0)\right)+
S^+\left(G^\prime_0(b_0)b_1\right)\nu^2+
S^+\left(G^\prime_0(b_0)b_2+\frac{1}{2}G^{\prime\prime}_0(b_0)b_1^2\right)\nu^4
+\dots
\end{equation}
Similarly
\begin{equation}
T^+(G_1)(\beta)=S^+\left(G_1(b_0)\right)
+S^+\left(G^\prime_1(b_0)b_1\right)\nu^2+\dots
\end{equation}
and the second contribution becomes
\begin{equation}
S^+\left(\frac{G_1(b_0)}{(v\bar S_0)^2}\right)\nu^2+
S^+\left(-2\bar S_1\frac{G_1(b_0)}{(v\bar S_0)^2}+\frac{G^\prime_1(b_0)b_1}
{(v\bar S_0)^2}\right)\nu^4+\dots
\end{equation}
The last contribution is
\begin{equation}
S^+\left(\frac{G_2(b_0)}{(v\bar S_0)^4}\right)\nu^4+\dots
\end{equation}
To summarize, the coefficients are
\begin{equation}
{\rm O}(1):\qquad
S^+\left(G_0(b_0)\right)=S^+(\Omega_0),
\end{equation}
\begin{equation}
{\rm O}(\nu^2):\qquad
S^+\left(G^\prime_0(b_0)b_1+\frac{G_1(b_0)}{(v\bar S_0)^2}\right)=S^+(\Omega_1),
\end{equation}
\begin{equation}
\begin{split}
{\rm O}(\nu^4):\qquad
S^+\Bigg(&G^\prime_0(b_0)b_2+\frac{1}{2}G^{\prime\prime}_0(b_0)b_1^2-2\bar S_1
\frac{G_1(b_0)}{(v\bar S_0)^2}\\
&+\frac{G^\prime_1(b_0)b_1}{(v\bar S_0)^2}+\frac{G_2(b_0)}{(v\bar S_0)^4}
\Bigg)=S^+(\Omega_2).
\end{split}
\end{equation}
Thus we have verified conjecture B to second order (in this special case).

\section{Proof of conjecture A in the \mO{3} model}
\label{AppO3proof}

Expanding both sides of \eqref{conjAsumm} in powers of $\nu$ we obtain the relations
\begin{equation}
\begin{split}
X_0&=G_0(b_0),\\
X_1&=G^\prime_0(b_0)b_1+G_1(b_0){\cal K}_0,\\
X_2+\frac{1}{2}\Delta_2 X_0&=G^\prime_0(b_0)\big(b_2+\frac{1}{2}\Delta_2 b_0
\big)+\frac{1}{2}G^{\prime\prime}_0(b_0)b_1^2+G_1(b_0){\cal K}_1\\
&+G^\prime_1(b_0)b_1{\cal K}_0+G_2(b_0){\cal K}_0^2,\\
X_3+\frac{1}{2}\Delta_2 X_1&=G^\prime_0(b_0)\big(b_3+\frac{1}{2}\Delta_2 b_1
\big)+G^{\prime\prime}_0(b_0)\big(b_2+\frac{1}{2}\Delta_2 b_0\big)b_1
+\frac{1}{6}G^{\prime\prime\prime}_0(b_0)b_1^3\\
&+G_1(b_0)\big({\cal K}_2+\frac{1}{2}
\Delta_2{\cal K}_0\big)+G_1^\prime(b_0)b_1{\cal K}_1+G^\prime_1(b_0){\cal K}_0
\big(b_2+\frac{1}{2}\Delta_2 b_0\big)\\
&+\frac{1}{2}G_1^{\prime\prime}(b_0)b_1^2{\cal K}_0+2G_2(b_0){\cal K}_0{\cal K}_1
+G_2^\prime(b_0)b_1{\cal K}^2_0+G_3(b_0){\cal K}^3_0.
\end{split}
\label{X0123}
\end{equation}
We can see that $G_0(\beta)$ and $G_1(\beta)$ are real but (because $\Delta_2$
is purely imaginary) $G_2(\beta)$ and $G_3(\beta)$ have both real and imaginary
parts:
\begin{equation}
G_s(\beta)=G_s^{({\rm r})}(\beta)+G_s^{({\rm i})}(\beta),\qquad s=2,3.
\end{equation}
Focussing on the imaginary parts first we write
\begin{equation}
\frac{1}{2}\Delta_2X_0=\frac{1}{2}G^\prime_0(b_0)\Delta_2 b_0
+G_2^{({\rm i})}(b_0){\cal K}^2_0  
\end{equation}
and comparing this to (\ref{saee}) (with $q=G_0$) we get
\begin{equation}
G_2^{({\rm i})}(b_0)=\frac{1}{2}\left(\frac{{\rm e}b_0^2}{32}\right)^2
(D_2G_0)(b_0)  
\end{equation}
and after removing the rescaling factor,
\begin{equation}
{\cal G}_2^{({\rm i})}(\beta)=\frac{1}{2}(D_2{\cal G}_0)(\beta).  
\end{equation}

The imaginary part of the $\nu^3$ relation is
\begin{equation}
\begin{split}
\frac{1}{2}\Delta_2X_1&=\frac{1}{2}G^\prime_0(b_0)\Delta_2b_1+\frac{1}{2}
G^{\prime\prime}_0(b_0)b_1\Delta_2 b_0+\frac{1}{2}G_1(b_0)\Delta_2{\cal K}_0\\
&+\frac{1}{2}G^\prime_1(b_0){\cal K}_0\Delta_2 b_0+2G_2^{({\rm i})}(b_0)
{\cal K}_0{\cal K}_1+G_2^{({\rm i})\prime}(b_0)b_1{\cal K}^2_0+G_3^{({\rm i})}
(b_0){\cal K}^3_0.
\end{split}
\end{equation}
On the other hand $X_1$ is given by the second line of (\ref{X0123}) and we can
explicitly calculate its alien derivative:
\begin{equation}
\begin{split}
\Delta_2X_1&=\Big(\frac{{\rm e}b_0^2}{32}{\cal K}_0\Big)^2(D_2G_1)(b_0){\cal K}_0
+G_1^\prime(b_0){\cal K}_0\Delta_2b_0+G_1(b_0)\Delta_2{\cal K}_0\\
&+G^\prime_0(b_0)\Delta_2b_1+b_1\Delta_2[G^\prime_0(b_0)].
\end{split}    
\end{equation}
Comparing the two expressions and using the above result for $G_2^{({\rm i})}$
many terms cancel and the relation determining $G_3^{({\rm i})}$ becomes
\begin{equation}
\begin{split}
\frac{1}{2}&b_1\Delta_2[G^\prime_0(b_0)]+\frac{1}{2}\Big(\frac{{\rm e}b_0^2}{32}
{\cal K}_0\Big)^2{\cal K}_0(D_2G_1)(b_0)\\
&=\frac{1}{2}G^{\prime\prime}_0(b_0)b_1\Delta_2b_0+G_3^{({\rm i})}(b_0){\cal K}^3_0
+\Big(\frac{{\rm e}b_0^2}{32}\Big)^2(D_2G_0)(b_0){\cal K}_0{\cal K}_1\\
&+\frac{1}{2}b_1{\cal K}_0^2\Big(\frac{{\rm e}b_0^2}{32}\Big)^2
(D_2G_0)^\prime(b_0)
+\frac{2}{b_0}\Big(\frac{{\rm e}b_0^2}{32}\Big)^2
(D_2G_0)(b_0)b_1{\cal K}^2_0.
\end{split}
\label{G3i}
\end{equation}
At this point we pause for calculating the alien derivative
$\Delta_2[G^\prime_0(b_0)]$. The trick is to calculate
$\Delta_2[\dot{\{G_0(b_0)\}}]$ in two different ways. On the one hand we write
\begin{equation}
\Delta_2[\dot{\{G_0(b_0)\}}]=\Delta_2\big[G_0^\prime(b_0)\dot b_0\big]=
\dot b_0\Delta_2[G^\prime_0(b_0)]+G^\prime_0(b_0)\dot{(\Delta_2 b_0)}-
4G^\prime_0(b_0)\Delta_2 b_0.
\end{equation}
Alternatively,
\begin{equation}
\begin{split}
\Delta_2&\left[\dot{\{G_0(b_0)\}}\right]=\frac{{\rm d}}{{\rm d}B}
{\big\{\Delta_2[G_0(b_0)]\big\}}-4\Delta_2[G_0(b_0)]\\
&=\Big(\frac{4\dot b_0}{b_0}+\frac{2\dot{\cal K}_0}{{\cal K}_0}\Big)
\big(\frac{{\rm e}b_0^2}{32}{\cal K}_0\big)^2(D_2G_0)(b_0)
+\dot b_0\big(\frac{{\rm e}b_0^2}{32}{\cal K}_0\big)^2(D_2G_0)^\prime(b_0)\\
&+G^{\prime\prime}_0(b_0)\dot b_0\Delta_2b_0
+G^\prime_0(b_0)\dot{(\Delta_2b_0)}-4G^\prime_0(b_0)\Delta_2b_0
-4\big(\frac{{\rm e}b_0^2}{32}{\cal K}_0\big)^2(D_2G_0)(b_0).
\end{split}
\end{equation}
Comparing this two representations, $\Delta_2[G^\prime_0(b_0)]$ can be expressed
as
\begin{equation}
\begin{split}
\Delta_2&[G^\prime_0(b_0)]=G^{\prime\prime}_0(b_0)\Delta_2b_0
+\big(\frac{{\rm e}b_0^2}{32}{\cal K}_0\big)^2(D_2G_0)^\prime(b_0)\\
&+\big(\frac{{\rm e}b_0^2}{32}{\cal K}_0\big)^2(D_2G_0)(b_0)
\Big[-\frac{4}{\dot b_0}+\frac{4}{b_0}+\frac{2\dot{\cal K}_0}
{{\cal K}_0\dot b_0}\Big]\\  
&=\frac{4}{b_0^2}\big(\frac{{\rm e}b_0^2}{32}{\cal K}_0\big)^2(D_2G_0)(b_0)
+G^{\prime\prime}_0(b_0)\Delta_2b_0
+\big(\frac{{\rm e}b_0^2}{32}{\cal K}_0\big)^2(D_2G_0)^\prime(b_0).
\end{split}
\end{equation}
In the last equality we used the second identity in (\ref{lat}).

Going back to the calculation of $G_3^{({\rm i})}$ and using the above formula
in (\ref{G3i}) we find
\begin{equation}
\begin{split}  
\frac{2b_1}{b_0^2}
&\big(\frac{{\rm e}b_0^2}{32}{\cal K}_0\big)^2(D_2G_0)(b_0)
+\frac{1}{2}
\big(\frac{{\rm e}b_0^2}{32}\big)^2{\cal K}_0^3(D_2G_1)(b_0)=
G_3^{({\rm i})}(b_0){\cal K}_0^3\\
&+\big(\frac{{\rm e}b_0^2}{32}{\cal K}_0\big)^2(D_2G_0)(b_0)
\big[\frac{{\cal K}_1}{{\cal K}_0}+\frac{2b_1}{b_0}\big].  
\end{split}
\end{equation}
Finally, using the first identity in (\ref{lat}) we find the simple result
\begin{equation}
G_3^{({\rm i})}(b_0)=\frac{1}{2}
\big(\frac{{\rm e}b_0^2}{32}\big)^2(D_2G_1)(b_0) 
\end{equation}
and after removing the rescalings
\begin{equation}
{\cal G}_3^{({\rm i})}(\beta)=\frac{1}{2}(D_2{\cal G}_1)(\beta). 
\end{equation}


\bibliographystyle{JHEP}
\bibliography{O3pap}

\end{document}